\documentclass[%
 reprint,
superscriptaddress,
 amsmath,amssymb,
 aps,showkeys,secnumarabic,
pra]{revtex4-2}

\newcommand{\inlinecomment}[1]{}
\usepackage{orcidlink}
\newcommand{\orcid}[1]{\orcidlink{#1}}

\usepackage{graphicx}
\usepackage{bm}
\usepackage{hyperref}
\usepackage{braket}
\usepackage{xcolor}
\usepackage[normalem]{ulem}

\definecolor{ReviewerBlue}{HTML}{1F4E79}
\definecolor{ReviewerGreen}{HTML}{38761D}
\definecolor{ReviewerRed}{HTML}{990000}
\definecolor{ReviewerPurple}{HTML}{6E2C91}

\usepackage{acro}
\usepackage{booktabs} 

\DeclareAcronym{DER}{
  short=DER,
  long=Distributed Energy Resource,
  }
  
\DeclareAcronym{FZJ}{
  short=FZJ,
  long=Forschungszentrum J{\"u}lich,
  }

\DeclareAcronym{RSA}{
  short=RSA,
  long=Rivest–Shamir–Adleman,
  }
\DeclareAcronym{ICE-1}{
  short=ICE-1,
  long=Institute of Climate and Energy Research:
Energy Systems Engineering,
  }
  
\DeclareAcronym{QKD}{
  short=QKD,
  long=Quantum key distribution,
  }  

\DeclareAcronym{IBM}{
  short=IBM,
  long=International Business Machines Corporation,
  }

\DeclareAcronym{RTDS}{
  short=RTDS,
  long=Real Time Digital Simulator,
  }

\DeclareAcronym{SCADA}{
  short=SCADA,
  long=Supervisory Control and Data Acquisition,
  }

\DeclareAcronym{MQTT}{
  short=MQTT,
  long=Message Queuing Telemetry Transport,
  }

\DeclareAcronym{EPB}{
  short=EPB,
  long=Electric Power Board,
  }

\DeclareAcronym{QUESS}{
  short=QUESS,
  long=Quantum Experiments at Space Scale,
  }

\begin{document}
\setcounter{secnumdepth}{3}
\renewcommand\thesection{\arabic{section}}
\renewcommand\thesubsection{\thesection.\arabic{subsection}}
\renewcommand\thesubsubsection{\thesubsection.\arabic{subsubsection}}

\title{Quantum Technologies and Edge Devices in Electrical Grids: Opportunities, Challenges, and Future Directions}

\author{Marjorie Hoegen\orcid{0009-0005-2265-8772}
}
\thanks{These authors contributed equally to this work.}
\affiliation{Institute for Automation of Complex Power Systems, RWTH Aachen University, 52074 Aachen, Germany}

\author{Ren\'e Glebke\orcid{0000-0001-7804-7536}
}
\thanks{These authors contributed equally to this work.}
\affiliation{Communication and Distributed Systems, RWTH Aachen University, 52074 Aachen, Germany}

\author{M. Sahnawaz Alam\orcid{0000-0001-6599-4964}}
\email[Corresponding author: ]{sahnawaz.alam@eonerc.rwth-aachen.de}
\affiliation{Institute for Automation of Complex Power Systems, RWTH Aachen University, 52074 Aachen, Germany}

\author{Alessandro David\orcid{0000-0002-3753-1396}}
\affiliation{Forschungszentrum Jülich GmbH, Peter Grünberg Institute,
Quantum Control (PGI-8), 52425 Jülich, Germany}
\author{Juan Navarro Arenas\orcid{0000-0002-9737-939X}}
\affiliation{Department for Quantum Technology, University of Münster, 48149 Münster, Germany}
\affiliation{Center for NanoTechnology -- CeNTech, 48149 Münster, Germany}
\author{Nikolaus Wirtz\orcid{0000-0003-0194-4452}}
\affiliation{Department of Digital Energy, Fraunhofer Institute for Applied Information Technology, Aachen 52074, Germany}

\author{Mario Albanese\orcid{0009-0003-6183-7090}}
\affiliation{ICE-1 Energy Systems Engineering, Forschungszentrum Jülich, 52428 Jülich, Germany}
\author{Daniele Carta\orcid{0000-0002-0182-8710}}
\affiliation{ICE-1 Energy Systems Engineering, Forschungszentrum Jülich, 52428 Jülich, Germany}

\author{Felix Motzoi}
\affiliation{Forschungszentrum Jülich GmbH, Peter Grünberg Institute,
Quantum Control (PGI-8), 52425 Jülich, Germany}
\affiliation{Institute for Theoretical Physics, University of Cologne, 50937 Köln, Germany}
\author{Antonello Monti\orcid{0000-0003-1914-9801}}
\affiliation{Institute for Automation of Complex Power Systems, RWTH Aachen University, 52074 Aachen, Germany}

\author{Carsten Schuck\orcid{0000-0002-9220-4021}}
\affiliation{Department for Quantum Technology, University of Münster, 48149 Münster, Germany}
\affiliation{Center for NanoTechnology -- CeNTech, 48149 Münster, Germany}

\author{Andrea Benigni\orcid{0000-0002-2475-7003}}
\affiliation{ICE-1 Energy Systems Engineering, Forschungszentrum Jülich, 52428 Jülich, Germany}
\affiliation{JARA-Energy, 52425 Jülich, Germany}

\author{Klaus Wehrle\orcid{0000-0001-7252-4186}}
\affiliation{Communication and Distributed Systems, RWTH Aachen University, 52074 Aachen, Germany}

\author{Ferdinanda Ponci\orcid{0000-0003-0431-9169}}
\affiliation{Institute for Automation of Complex Power Systems, RWTH Aachen University, 52074 Aachen, Germany}

\begin{abstract}
In modern power systems, edge devices serve as local hubs that collect data, perform on-site computing, sense electrical parameters, execute control actions, and communicate with neighboring edge devices as part of the larger grid. However, as the number of monitored nodes and control loops grows, traditional edge devices face serious limits. They can become overloaded by complex signal processing and decision tasks, causing delays and higher energy use. Standard sensors hit a noise floor that prevents them from detecting miniature changes, making it harder to spot early signs of faults or instability. Meanwhile, conventional communication links struggle with bandwidth limits, security risks, and rising encryption demands, which together slow down and weaken the transfer of critical grid information. Quantum technologies have the potential to overcome these challenges. Quantum computers can deliver exponential speed-ups for optimization and machine-learning tasks that ordinary processors cannot handle. Quantum sensors can sense signals with atomic precision, giving edge devices a more precise view of grid dynamics. Quantum communication techniques, including quantum key distribution, offer methods to achieve information-theoretic security and ensure that information arrives quickly and without tampering. We explore how quantum technologies can be integrated into edge devices, highlighting both opportunities and challenges..
\end{abstract}

\keywords{Quantum technologies; Edge Devices; Quantum computing; Quantum sensing; Quantum communication; Cybersecurity in power systems}
\maketitle

\date{\today}
\section{Introduction}
Power systems are large-scale, complex, nonlinear systems that have been transforming in recent years with the support of Internet of Things (IoT) and edge devices \cite{8281479,fi11040100}. The presence of renewable energy sources (RES) and power-electronic-interfaced resources have let to faster dynamics, requiring close monitoring and pervasive control of the resources. A fully centralized implementation of these monitoring and control functions is very demanding and hardly feasible. Consequently, distribution of intelligence and computational capacity among local devices capable of communication and cooperation is emerging as a more viable option. Edge devices are responsible for collecting data (sensing), performing computations, and communicating securely with other devices \cite{FENG2021100006, mehmood2021edge, edge-cloud-sg, 8877785}. However, as the scale and complexity of power-grid edge devices continue to grow, edge devices increasingly struggle to keep up the pace. First, as the number of monitored nodes increases, the computational burden for local signal processing and decision-making can quickly overwhelm the traditional micro-controllers, leading to high latency and energy consumption. Second, the sensitivity of conventional electromagnetic sensors impose hard limits on the atomistic perturbations in noisy environments, making it difficult to track the fine‐grained fluctuations that presage fault conditions or instability. Third, traditional bidirectional communication with the edge devices suffer from bandwidth constraints, interception vulnerability, and ever-rising demands for encryption, which together compromise both the timeliness and the integrity of grid‐state information \cite{bdcc3010008,IoT-Security-Survey-2, khanafer2017optimized}.

Quantum technologies offer a compelling path forward. In theory, quantum computing architectures can provide exponential speed-ups for specific classes of problems, such as machine‐learning tasks at the edge, enabling real-time route-finding for power flows and anomaly detection that are simply out of reach for current devices \cite{Feynman1982, Shor, Priyanka2024}. Quantum sensors, for example nitrogen-vacancy–center magnetometers in diamond, provide nanoscale spatial resolution in laboratory experiments \cite{Doherty2013TheDiamond,Barry2020}, and continued progress in these platforms could allow edge nodes to resolve much smaller variations in grid quantities. Finally, quantum communication methods, most notably Quantum Key Distribution (QKD), can provide information‐theoretic security for transferred data, ensuring that control commands and measurement data remain protected even against powerful adversaries~\cite{QKD-Sec-with-Imperfect-Devices}.

In this paper, we survey current challenges in edge devices applications in power grids, and describe how quantum technologies like quantum computing, sensing, and communication could address them. Next, we describe emerging quantum modalities, outline a roadmap for their integration into next-generation edge devices, and discuss the challenges that must be addressed to realize the so-called quantum-based edge intelligence in future power systems.

\begin{figure}[tb]
  \centering
  \includegraphics[width=0.95\columnwidth]{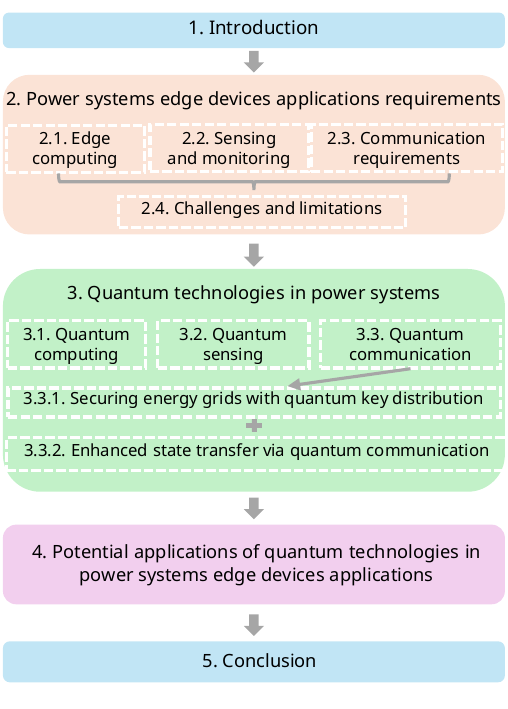}
  \caption{Structure of the paper.}
  \label{fig:layout}
\end{figure}

This overview paper is organized, as shown in
 Fig.~\ref{fig:layout}, as follows. In Sec.~\ref{sec:edge-requirements}, we describe several requirements of the edge device, including computing, sensing and communications applications, its challenges and limitations. Then, in Sec.~\ref{sec:quantum-technologies}, we introduce the relevant quantum technologies for the quantum-based edge device to address the challenges. In Sec.~\ref{sec:applications}, we discuss the applications and challenges of the quantum-based edge device. Finally, we conclude the paper in Sec.~\ref{sec:conclusions}. 

\begin{figure}[tb]
  \centering
  \includegraphics[width=0.5\columnwidth]{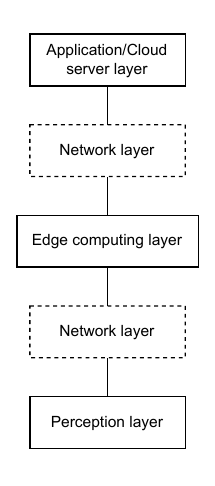}
  \caption{Four-layer architecture for edge computing systems. The three main layers (perception, computing and application) are in solid boxes and the network layer is presented in dotted boxes.}
  \label{fig:architectures}
\end{figure}

\section{Power systems edge devices applications requirements}\label{sec:edge-requirements}

\subsection{Edge computing systems}

The architecture of edge computing systems (Fig. ~\ref{fig:architectures}) usually involves three main layers: the perception or edge device layer, edge computing or edge server layer, and the application or cloud server layer (solid boxes in Fig.~\ref{fig:architectures}). These layers are connected via network layers (dotted boxes in Fig.~\ref{fig:architectures}) which allow data transmission between layers \cite{edge-computing-review2024, edge-computing-security, edge-cloud-sg}. 
The devices used in each layer have different characteristics, as each layer has a different purpose. The perception layer, closer to the edge, includes edge devices responsible for sensing and collecting data from the physical environment. In upper layers, edge computing or edge server devices have greater computational power to be able to preprocess data gathered from end devices and to perform some computational tasks \cite{edge-computing-review2024, mehmood2021edge}.

In \textcite{devices-review}, a survey of IoT devices, which could be used in edge computing layers, is executed. The devices are categorized in low, middle and high-end devices and their main specifications are listed and compared. Low-end devices are described as resource-constrained devices, used mostly for basic sensing and actuating applications, being part of the lower-level layers of an edge computing system. They usually have a clock speed in the range of MHz, and random access memory (RAM) and on-board storage in the order of kB. High-end IoT devices are devices with better resources, such as, powerful processing units, significant RAM and good on-board connectivity. In between, there are the middle-end devices, which have fewer resources than high-end devices but more processing capabilities than low-end devices. The clock speed of middle and high-end devices can reach ranges of GHz, and their RAM and on-board storage can be in the order of GB.
 
\subsection{Sensing and monitoring in edge devices}
The data used for monitoring and control of power grids is acquired through sensors, which are traditionally bulky devices usually restricted to substations, such as instrument transformers. However, there have been an increase in the demand for compact sensors. For efficient and real time monitoring and control, these sensors should enable real time data measurement and should be energy-efficient. Edge devices continuously measure local electrical parameters (voltage, current, frequency, etc.) and equipment status with high granularity. For instance, feeder sensors and smart meters at the edge can detect anomalies or power quality issues in real time, providing a fine-grained view of grid conditions \cite{Moghe2012}. This high-resolution monitoring underpins predictive maintenance and fast fault detection in modern grids. However, edge-level electrical sensors, namely instrument transformers (CTs and PTs), Hall effect devices, and Rogowski coils, exhibit accuracy classes typically in the 0.1\%--1\% range for CTs/PTs and around 0.5\% for Rogowski coils, while open-loop Hall effect sensors often incur about ±3--5\% full-scale error \cite{Xu2015,Metwally2010}. These devices can also degrade or fail over time due to core saturation, insulation breakdown, thermal drift, wiring faults, and inability to detect DC components \cite{Gazivoda2021,Schweitzer2018}.

\subsection{Communication requirements of power systems applications}
\label{subsec:communication-requirements}

\begin{center}
\small
\begin{table*}[tb]
    \begin{tabular}{||c|c|c|c||}
        \hline
        \textbf{Application} &\textbf{Ideal Latency} & \textbf{Data rate}& \textbf{Coverage range} \\ [0.5ex]
        \hline\hline
        Transmission line monitoring & 15 - 200 ms & 9.6 - 56 kbits/s & Up to 100 km\\ \hline
        Substation automation & 3 ms~\footnote{Trip signals for circuit breakers in protection systems} / 15 - 200 ms & 9.6 - 100 kbits/s & Up to 10 km\\ \hline
        Wide-area situational awareness & 15 - 200 ms & 600 - 1500 kbits/s & 100 km or more \\ \hline
        Distribution automation & 20 - 200 ms & 9.6 - 100 kbits/s & Up to 10 km \\ \hline
        Home energy management & 300 - 2000 ms & 9.6 - 56 kbits/s & Up to 100 m \\ \hline
        Flexibility provision & 500 ms - several minutes & 14 - 100 kbits/s & Up to 10 km \\ \hline
        Outage management & 2 s & 56 kbits/s & Up to 10 km \\ \hline
        Advanced metering infrastructure & 12 ms~\footnote{If used for real-time operations} - 15 s & 10 - 100 kbits/s & Up to 10 km \\ \hline
        Distributed energy resources / storage & 300 ms - 2 s & 0.6 - 56kbps & Up to 10 km \\ \hline
    \end{tabular}
    \caption{Communication requirements of fundamental smart grid applications, based on data from \cite{sg-review-2013, communication-sg-2019,8839117,sg-communication-survey-2013,3ms-for-trips}. Latencies are end-to-end, data rates are per individual node.}
    
    \label{tab:smartgrid-comm-requirements}    
\end{table*}
\end{center}

Edge devices process data near its source and can reduce the reliance on distant cloud servers and control centers enabling faster response times and avoiding network burdens \cite{Yldrm2025}. Smart grids require bidirectional communication between different types of components connected to the generation, transmission and distribution parts of the grid. The increase in the utilization of edge devices also leads to the communication infrastructure and its requirements to have a fundamental role in smart grids, as this large flow of information should be maintained via reliable and secure channels.

The smart grid communication network can be represented in a hierarchical multi-layered architecture that includes short-range, medium-range and long-range networks. Short-range networks are in customer premises and include home area network, building area network and industrial area network having a coverage range of up to 100 meters. An example of a smart grid application that use this type of network is home or building automation systems. Medium-range networks include neighbour area network and field area network and have a coverage range of up to $10$~kilometres. Meter reading and distribution automation can be considered examples of applications in this type of network. Long-range networks are wide area networks (WANs), which have a coverage range of $100$~kilometres or more(\textit{e.g.,} applications that involve synchrophasors use WANs \cite{communication-sg-2013}).

Typical analyzed requirements related to the smart grids’ applications in the different communication networks layers are latency, data rate, reliability, and security, which have been reviewed in Refs. like \textcite{communication-sg-2013}, \textcite{sg-review-2013}, \textcite{communication-sg-2019}. Latency and data rate requirements for some applications are summarized in Table \ref{tab:smartgrid-comm-requirements}. Latency requirements can range from 12 milliseconds to some minutes, depending on the application. For real-time sensing purposes, latency is critical and should be between $12$ to $20$ milliseconds. Wide-area situational awareness systems and distribution automation systems also require low latency in the range of milliseconds. Applications where latency is not that critical and can be around $2$ to $15$ seconds are advanced metering infrastructure (AMI) and home energy management (HEMs). Data rate requirements vary between $9.6$~kbit/s to $10$~Mbit/s. Wide area situational awareness systems require high data rate values to achieve accurate and reliable data transfers. Data rates for other applications, such as AMI, home and building automation, and distribution automation can be lower, usually less than $100$~kbit/s.

Smart grids are highly dependent on information and communication technologies (ICT) and can be very vulnerable to cyberattacks, as each network connection is a potential attack target \cite{etsi-report}. Considering the interconnection of these systems and that they contain sensitive information of end-users and power systems components, cyberattacks in smart grids can lead to serious consequences, such as power outages. Therefore, providing reliable end-to-end security is a high priority for almost all smart grid applications.

\subsection{Challenges and limitations}

Although modern developments of the edge device shows significant improvements, recent review highlights several bottlenecks that must be addressed \cite{Yldrm2025}. The complexity of smart grids creates challenges in designing their computation and communication infrastructures and as well as sensing protocols. 

The limitations of IoT devices regarding their computational and storage characteristics influence their capabilities of running complex algorithms \cite{iot-aided-sg, edge-intelligence-sg}. Yet, there is an increasing demand for the use of artificial intelligence (AI) algorithms in various grid-related applications such as demand response~\cite{AI-Demand-Response}, protection~\cite{AI-Grid-Protection}, among several other fields~\cite{AI-SmartGrid-Survey, edge-intelligence-sg}. However, the practical feasibility of applying AI models directly on the edge is low, due to an overall weak computational power of edge devices \cite{edge-intelligence-sg}. The power consumption related to the computational performance to train AI models is also non-negligible, so that constraints may arise in situations in which edge devices are required to perform the training themselves. Distributed AI schemes such as Federated Learning~\cite{federated-learning-overview} may reduce the workload of individual devices by introducing global models and learning procedures coordinated by a central instance, so that a single device may only be required to train on an individual subset of the data. Such collaborative approaches enable the devices to essentially trade computational or power consumption burdens against communication requirements. This however renders the reliable and secure communication between the involved entities all the more important, as not all data for informed decision-making may be readily available locally.

In terms of the sensing requirements of the edge devices, smart grids operate under varying environmental conditions, which can result in noisy and less reliable data. However, high reliability and availability of the collected data in  high precision collected data in practical environment conditions are desired \cite{iot-aided-sg}.

Security has been increasingly relevant to guarantee the reliability of the systems as well as data privacy and can be a challenge in edge devices, as the number of attack entry points increase with the number of devices and the resource-constraints can also play a role in limiting the security \cite{edge-computing-security}. Data sensing and communications in smart grids may introduce more and more points of intrusions and attacks, increasing the importance of cybersecurity. For an appropriate functioning of modern power grids, many systems need to be interconnected, which creates new possibilities and threats in cybersecurity \cite{sg-communication-survey-2013}. Considering the use of IoT and edge devices in smart grids (such as smart meters, controllers of PV systems, battery or vehicle charging systems, and equipment in electrical substations), the majority of such devices is typically resource-constrained, with consequences on the complexity of the security mechanisms they can implement~\cite{IoT-Security-Survey-1}. The number of IoT malware attacks has been increasing, with a 400\% growth in IoT-related cyberattacks from 2022 to 2023, according to \textcite{iot-analytics}. The maintainability of these devices and hence the ability of owners or operators to react to newly-emerging threats without physically accessing the devices also creates a challenge in guaranteeing cybersecurity ~\cite{IoT-Firmware-Updates-1, IoT-Firmware-Updates-2}. New vulnerabilities arise as traditional encryption algorithms used to secure systems are mostly not effective against quantum computing attacks \cite{ahn_toward_2022}. This means that eventually, not only confidential personal or business data may be decrypted and used without consent, but also control data exchanged between the devices for grid operation may be intercepted or altered in manners that may endanger grid stability or safety.

\section{Quantum technologies in power systems}\label{sec:quantum-technologies}
Modern technologies built on classical physics capture the behavior of everyday macroscopic systems, but they fail in the microscopic realm; there, quantum principles succeed in describing these atomic-scale systems and feed back into ever faster, more accurate, and more efficient macroscopic engineering. This quantum world is defined by fundamental principles such as wave-particle duality, the uncertainty principle, and the quantization of physical properties. In this domain, particles such as electrons and photons display both wave-like and particle-like behavior, and their actions can only be described in terms of probabilities \cite{Feynman1963Lectures}. This marks a clear shift from the deterministic nature of classical physics to the probabilistic framework of quantum physics, which has significantly changed our understanding of the universe \cite{Halliday2014Fundamentals,Feynman1963Lectures}. Quantum physics also forms the basis of many modern technologies by explaining how matter and energy behave at very small scales. This early understanding led to the first quantum revolution, which enabled the development of technologies such as lasers, transistors, and semiconductors \cite{Kleppner2000OneHundred}. More recent advancements, often referred to as the second quantum revolution, involve the manipulation of quantum systems and have led to emerging applications in quantum computing, sensing and communication \cite{Dowling2003Quantum, Wiseman2009QuantumMeasurement, Deutsch2020SecondQuantumRevolution}. 

Building on this foundation and its success, recent research explores the utilization of its potential in broader applications, particularly in highly complex engineered infrastructures such as power systems. While large-scale power grids operate firmly in the macroscopic regime, several subproblems within planning, operation, sensing, and communication expose computational, accuracy, or security limitations where quantum technologies may offer advantages. In particular, developments arising from the second quantum revolution have enabled early investigations into the use of quantum computing for optimisation, learning, and simulation tasks in power-system applications \cite{Bryn2022, Priyanka2024, Priyanka2025}. In parallel, quantum sensing technologies have demonstrated atomic-scale sensitivity to electric and magnetic fields, enabling high-resolution current, voltage, and field measurements relevant for grid monitoring and diagnostics \cite{Doherty2013TheDiamond,Hatano2022EV}, which are beyond the reach of classical sensors. Furthermore, quantum communication protocols, notably quantum key distribution, have been proposed to enhance the security and resilience of communication links in smart grids and future power systems \cite{ahn_toward_2022,kong2022qkd}. Although many of these approaches remain at an early or experimental stage, with limited functionality, they collectively indicate a growing research effort to assess where quantum technologies can realistically complement modern power systems, as discussed in greater detail in the following sections. Table~\ref{tab:smartgrid-quantum-directions} provides a brief overview of the research directions and its challenges.

\begin{center}
\small
\begin{table*}[tb]
    \begin{tabular}{||p{2.2cm}|p{3cm}|p{5cm}|p{6cm}||}
        \hline
        \textbf{Application} &\textbf{Research direction} & \textbf{Relevance for power systems}& \textbf{Research challenges} \\ [0.5ex]
        \hline\hline
        Quantum computing & (i) Improved optimization and global search for hard problems. (ii) Tools for cryptography breaking & (i) Potential applications for grid operation and planning, quantum-based ML. (ii) Potential threat for encryption, and thus for secure operation of energy systems. & (i) Translating information to quantum states counterbalances computational speed-ups direct use of quantum measurements suffers from measurement errors and decoherence. (ii) Long-term secure communication in vast power grids despite the quantum threat\\ \hline
        Quantum sensing & High-precision measurements & Measurement of currents via magnetic fields, temperature measurement; potential use in edge devices due to miniaturization & NV already outperform classical sensors in high-precision applications\\ \hline
        Quantum communication & (i) Quantum key distribution. (ii) Enhanced state transfer  & (i) Securing critical power-system communication channels, e.g., between control centers and field devices. (ii) Quantum teleportation towards grid control mechanisms; quantum SDN / on-path signal handling for achieving consensus and robustness in distributed control. & (i) Deployment complexity in real smart grids; robust, energy- and cost-efficient systems at the edge; combination with classical channels to avoid denial-of-service by impaired QKD during attacks. (ii) Practical implementations of control strategies require additional communication channels and imply added complexity\\ \hline
    \end{tabular}
    \caption{Overview on research directions and challenges for quantum in power systems.}
    
    \label{tab:smartgrid-quantum-directions}    
\end{table*}
\end{center}

\subsection{Quantum computing}

Quantum computing harnesses quantum phenomena like superposition and entanglement to execute computations that, for certain problems, are exponentially faster than those achievable by classical computers. This capability opens up advancements in areas such as cryptography, optimization, quantum simulation, and machine learning \cite{Chae2024,Q-CompCommSurvey}.

Since the introduction of quantum computation in 1982 \cite{Feynman1982, Shor}, many new quantum algorithms have been proposed since the 1990s and the fields of interest vary from quantum chemistry to finance \cite{Dalzell2023}. We can broadly distinguish them in three main categories: (i) algorithms for more efficient simulation of quantum mechanics, introduced mainly to lead to breakthrough scientific discoveries and improved theoretical models as well as to optimal pharmaceutical design and knowledge of molecular chemistry; (ii) algorithms for improved optimization and global search for hard problems, leveraging the superposition of quantum states (this includes general algorithm like Grover search or amplitude amplification, variational quantum solvers, accelerated linear algebra and quantum machine learning); and (iii) tools for cryptography breaking. Of these three categories we envision the second one (ii) as the right direction to look for applications in power systems. Category (iii) poses threats to the secure operation of the systems, and we consider this aspect later in Section~\ref{sec:q-communication}.

In category (ii), quantum machine learning (QML) exploits quantum parallelism over classical machine learning techniques to potentially achieve speed-ups in data-intensive machine learning tasks \cite{Biamonte2017}. Variational quantum algorithms, including quantum support vector machines, quantum neural networks, and related models, form the cornerstone of current QML research, bridging theory and practice on the noisy intermediate-scale quantum (NISQ) devices available today. A comparison of some important classical and quantum time complexities of various machine learning algorithms as functions of dataset size \(N\) as reported in Ref.~\cite{Fedorov2022} and references therein, is presented in Table~\ref{tab:ml-algo}. This list of quantum algorithms for linear algebra and QML that might have an important impact for the analysis and optimization of energy grids. An overview of the potential and applications of quantum computing and QML in power systems, in domains such as system monitoring, control, optimization, and protection in presented in \textcite{Priyanka2024}.

\begin{table}[tb]
    \centering
    \begin{tabular}{||l|c|c||}
    \hline

      \textbf{Algorithm} & \textbf{Classical} & \textbf{Quantum} \\
    \hline\hline

      Linear regression & $\mathcal{O}(N)$ & $\mathcal{O}(\log N)$ \\    \hline

      Gaussian process regression & $\mathcal{O}(N)$ & $\mathcal{O}(\log N)$ \\    \hline

      Ensemble methods & $\mathcal{O}(N)$ & $\mathcal{O}(\sqrt{N})$ \\    \hline

      Support vector machines & $\mathcal{O}(N^2\!-\!N^3)$ & $\mathcal{O}(\log N)$ \\    \hline

      Bayesian networks & $\mathcal{O}(N)$ & $\mathcal{O}(\sqrt{N})$ \\    \hline

      k-Means clustering & $\mathcal{O}(kN)$ & $\mathcal{O}(k \log N )$ \\    \hline

      Principal component analysys & $\mathcal{O}(N)$ & $\mathcal{O}(\log N)$ \\    \hline

      Persistent homology & $\mathcal{O}(\exp N)$ & $\mathcal{O}(N^2)$ \\    \hline

      Gaussian mixture models & $\mathcal{O}(\log N)$ & $\mathcal{O}(\mathrm{polylog}\,N)$ \\    \hline

      Convolutional neural networks & $\mathcal{O}(N)$ & $\mathcal{O}(\log N)$ \\    \hline

      Bayesian deep learning & $\mathcal{O}(N)$ & $\mathcal{O}(\sqrt{N})$ \\    \hline

      Generative adversarial networks & $\mathcal{O}(N)$ & $\mathcal{O}(\mathrm{polylog}\,N)$ \\    \hline

      Boltzmann machines & $\mathcal{O}(N)$ & $\mathcal{O}(\sqrt{N})$ \\    \hline

    \end{tabular}
    \caption{Comparison of classical and quantum time complexities of various machine learning algorithms as functions of dataset size \(N\).}
    \label{tab:ml-algo}
\end{table}

When looking at the performance improvement offered by quantum computing, there is an important caveat to keep in mind (see Ref.~\cite{Aaronson2015}): in most cases the algorithmic speedup is guaranteed only in terms of processing the quantum state for the problem at hand, but usually an initial quantum state needs to be prepared to appropriately represent the data and it also needs to be read out. Especially if the problem uses classical data, the loading process into a quantum state introduces linear (for vectors) and quadratic (for matrices) overheads counterbalancing the benefits of quadratic speedups \cite{Hoefler2023}. For this reason the most promising algorithms are those involving cubic, quartic or even higher speedups. On the output/read out side, often the solutions are encoded into too many probability amplitudes leading to the efficient extraction of only global properties, e.g., evenness or oddness or confirmation of identical solutions. On the input side, the performance can be improved when the data to be processed can be generated algorithmically or when the data is already encoded in a quantum state. In this sense it would be worthwhile investigating the use of sensors that collect the state of the power grid and encode it in a quantum information carrier (photons, spins, etc.). Depending on the scenario, quantum states encoded by such sensors may however need to be collected from different locations within the grid, and then transferred over to some quantum computing equipment for further processing. As this may happen tens to hundreds of kilometers away, the quantum states need to be transferred by either classical means (i.e., sending photons through a fiber or through free-space), or via quantum teleportation (i.e., via entanglement effects in combination with classical communication~\cite{Q-Teleportation}). Both methods are prone to measurement errors and decoherence, an effect which intensifies with longer distances and time spans~\cite{Q-Decoherence}. The positioning of quantum-enabled sensors and their integration into quantum algorithmic procedures hence needs to be carefully planned to avoid prohibitive error rates on the inputs for the computations, even if advances in quantum communication (cf. Sec.~\ref{sec:q-communication}) may mitigate such problems.

In this context, hybrid quantum–classical architectures offer a pragmatic way to leverage emerging quantum capabilities while avoiding many of the bottlenecks associated with full quantum-state preparation, long-distance quantum transmission, and noisy readout. These architectures integrate quantum subroutines within classical workflows that handle data acquisition, large-scale computation, and real-time control. Despite the challenges outlined above, early experimental work shows that such hybrid approaches can already yield practical benefits in energy and IoT applications. For example, a recent demonstration machine-learning model for 5G base stations and solar PVs in which a quantum annealer performs the combinatorial siting step while a classical optimizer (DEA) manages operational planning \cite{Xu2025}. Similarly, a report from 2025 shows a hybrid workflow for electric-bus fleet management where quantum routines train a risk classifier offline, and the resulting compressed model is deployed as a classical neural network on IoT edge controllers for real-time charging decisions \cite{Moniruzzaman2025Hybrid}.

Despite this progress, demonstrating a clear quantum advantage in machine learning remains an open problem, largely because present-day hardware is noisy and error-prone \cite{Zimborás2024QCmyths}.  Current NISQ devices face errors of diverse origins, ranging from gate infidelities to measurement errors, spurious environmental couplings and cross-talk \cite{patel_experimental_2020}. Although a roadmap to full fault-tolerant quantum chips is known already \cite{fowler_surface_2012}, a scalable architecture capable of supporting the required number of physical qubits has not yet been demonstrated. Alternative approaches include accepting the current error rates and applying error mitigation techniques to promptly improve the quality of the results \cite{cai_quantum_2023}. Nevertheless, since edge devices have limited computational resources, integrating quantum computing, and QML in particular, into edge environments is still a significant challenge \cite{Herbst2024}.

\subsection{Quantum sensing}

Quantum sensing leverages quantum systems or phenomena to measure physical quantities with high precision, enabling the detection of minute changes in fields or forces that are often imperceptible to classical sensors \cite{Degen2017QuantumSensing}. This ability has broad societal applications; for example, detecting magnetic field fluctuations produced by current-carrying devices can reveal anomalies in electric systems through the electromagnetic fields generated by these disturbances. For power system applications, however, it is crucial to employ quantum systems that operate at room temperature. Quantum sensors such as negatively charged nitrogen-vacancy (NV) centers in diamond are sensitive to magnetic fields, electric fields, temperature, and pressure, and are used to detect subtle variations in these quantities with unprecedented precision \cite{Doherty2013TheDiamond, Doherty2014ElectronicPressure, Fang2024, Oh2024} and can be operated in room temperature. NV centers have been employed to probe condensed matter systems, studying static and dynamic magnetic textures and current distributions \cite{Casola2018ProbingCondensed}. Additionally, NV centers are utilized in nanoscale thermometry to measure miniature temperature variations \cite{Hayashi2018OptimizationEnsemble} and microscopic variations in crystal deformations or strains \cite{Doherty2014ElectronicPressure,Barry2020, Alam2024Determining}. Notably, NV centers can simultaneously detect magnetic fields and temperature variations \cite{Hatano2021Simultaneous}. These capabilities make quantum sensing one of the few quantum technologies that can already be deployed for power system applications under real hardware limitations at the power-system edge.

Certain properties of power systems, such as the charge and discharge currents in batteries, can be determined by accurately estimating current flows. Precise determination of current can be achieved by measuring the magnetic fields produced by these systems. Conventional systems can determine current with an accuracy on the order of milliamperes; however, for more precise determination of power flow, even in noisy environments, quantum sensors like NV centers offer a significant advantage \cite{Hatano2022EV}. Recent developments have enabled NV sensor implementations to achieve remarkable miniaturization, with devices as small as $0.42\,\text{cm}^3$ \cite{Pogorzelski2024Compact}, making them promising candidates for integration into edge devices, providing enhanced performance over current technologies based on classical sensors.

\subsection{Quantum communication}
\label{sec:q-communication}
Quantum communication employs quantum-physical effects to enhance data exchanges between two or multiple parties. By encoding information within quantum objects and employing quantum system-specific phenomena such as state superposition and entanglement, quantum communication allows the sharing of information in ways beyond what is possible with classical systems. The ongoing developments touch two important aspects of power system applications: (data) security and enhanced state transfer, both of which we elaborate on in the following sections.

\subsubsection{Securing energy grids with quantum key distribution}
\label{subsec:qkd}

Addressing data security within energy grids, the work in~\cite{ahn_toward_2022} reviews the vulnerabilities of current distributed energy resource (DER) systems to quantum cyberattacks. It highlights the potential timeline for quantum security threats, and notes that IBM's roadmap for quantum computing~\cite{IBM-QC-Roadmap} suggests a $10^6$-qubit quantum computer to be available somewhen after 2026. Having evolved from chips with only tens of qubits in less than a decade, such systems would be just one order of magnitude shy of qubit counts that allow breaking officially endorsed encryption methods such as \ac{RSA} with key lengths of 2048~\cite{Nist-2048bit-RSA} or 3000~\cite{BSI-3000bit-RSA} bits within hours~\cite{rsa-2024-in-8-hours}, while recent achievements using classical methods require thousands of core years on currently-available classical hardware for key lengths of only 829 bits~\cite{RSA-Thousands-of-Core-Years}. However, depending on the use case, the encryption needs in energy grids primarily concern data with short required secrecy duration and any sensible encryption setup would use individual keys for each pair of communicating entities. This means that some of the possibly daunting cryptographic conditions arising from the new quantum computing capabilities are alleviated; attacks would also usually need to be repeated for each individual connection within a power system, thus requiring significant effort by potential attackers to gain full control. Nevertheless, given the potential vulnerabilities of energy grids as critical infrastructure, there is an urgent demand for proactive solutions against future threats. 

Quantum key distribution (QKD) presents one such solution. It establishes provably secure~\cite{QKD-Sec-with-Imperfect-Devices} communication between spatially separated parties in a network via the joint generation of symmetric keys relying on fundamental principles of practical quantum mechanics such as the Heisenberg uncertainty principle, the no-cloning theorem and photon polarization. In QKD settings, adversaries, even those with unlimited computing power, inevitably disturb the state of transmitted qubits in ways that reveal  attempts to eavesdrop on a quantum channel, thus allowing the parties that wish to communicate privately to discard the key that is established over this channel before it is being used to encode private information \cite{pirandola}. In practice, QKD so far has been implemented in two main modalities: fiber‐optic implementations for up to hundreds of kilometer distances \cite{Sharma2021,Cao2022} and free-space realizations for long-haul/global key distribution, e.g. via satellites \cite{10713122, Bedington2017}.
While both variants have been considered for securing communication in smart grids \cite{10845578}, the fiber optic variants are particularly well suited for power-grid applications because the typical distances between substations and control centers match well with state-of-the-art utility fiber-based QKD implementations over $10-100$~km.

The communication demands in smart grid infrastructure, as introduced in Sec.~\ref{subsec:communication-requirements} generally compare favorably to the capabilities already attainable with state-of-the-art QKD systems. While smart grid communication systems must support a wide range of applications, the latency, bandwidth, and link-distance requirements are relatively modest as compared to telecommunication standards (see Table~\ref{tab:smartgrid-comm-requirements} for reference). Modern fiber-based QKD can reliably produce key rates ranging from several kbit/s ($>200$~km) up to a hundred Mbps ($<10$~km)~\cite{Li2023}, depending on distance and hardware configuration. Novel photonic integrated QKD transmitters achieved key rates of $1.82$~Mbps over $50$~km and allowed for maintaining secrete key generation over up to $250$~km distance \cite{Dolphin2023}, thus covering the overwhelming part of communication needs in smart power grids. Additionally, fiber-based QKD introduces negligible transmission delay and allows key generation to be decoupled from application-layer timing, making it highly compatible with latency-sensitive grid functions.

QKD hardware can already be considered technologically mature and field-ready, especially compared to e.g., quantum processors. It thus offers a practical near-term enhancement for securing critical power-system communication channels, such as those used in Wide-Area Monitoring, Protection, and Control (WAMPAC) systems, by strengthening the confidentiality and authenticity of measurements exchanged between field devices and control centers, as well as the integrity of the control signals sent back to the grid.
Realized example applications with regard to power grids include the work by \textcite{alshowkan2022authentication} which demonstrated the distribution of quantum keys over links of $3.4$~km between a distribution center and a substation within a fiber-optic utility network. The demonstrator was isolated from the operational network, relied on commercial QKD hardware, and used the publish-subscribe Message Queuing Telemetry Transport (MQTT) protocol in upper layers of the network stack. Similarly, QKD was also deployed in a control system for distributed energy resources on a live electrical grid, demonstrating enhanced resilience against spoofing and interception \cite{9405393}. A satellite-based QKD variant was implemented to secure power grid communications in a Chinese Grid~\cite{Mozi_Satellite}. As part of the \ac{QUESS} mission, it aims to demonstrate long-range quantum security and moves toward a potential quantum internet. In Refs.~\cite{tang_quantum_2021} and \cite{tang_quantum-secure_2020}, communication via a simulated QKD protocol between a control center and a microgrid is recreated in a real-time environment simulated by the commercial \ac{RTDS}.

While these advancements and examples are encouraging, the research on QKD applications in smart grids and power systems remains limited~\cite{kong2022qkd}, and no large-scale real-world deployment has yet been achieved. 
Securing communication in larger networks whose layouts follow that of a smart grid also presents additional challenges. Besides the cost and availability of dark fiber links for connecting devices via quantum channels in a smart grid \cite{kong2022qkd, 10852309}, the diverse range of existing energy control devices make it difficult to achieve scalable QKD integration in a heterogeneous network. Instead of general purpose QKD implementations, smart grids would thus further benefit from customized solutions that are optimized for the specific use cases relevant to communication between substations and control centers of energy systems. Such application specific QKD solutions must ensure compatibility of hardware subsystems, protocols, and interfaces, to ensure optimal performance and security of the overall system \cite{Diamanti2016}. Deployment complexity can be reduced by considering the electrical interrelations and control structures of grid segments during the planning of QKD systems. The work in Ref.~\cite{gado_upgrading_2024} for example explores the concept of \emph{semi}-QKD in power systems, where only a subset of nodes requires full quantum capabilities. This study aims to optimize resource allocation while maintaining security, and it proposes a greedy algorithm for placing quantum servers in a power grid and validates it on the IEEE $14$-bus system.

For practical large-scale deployments permeating also the edges of networks, the size, power consumption and costs of current QKD equipment as well as the environmental conditions under which such systems operate may require further improvements. In this regard, chip-based photonic integrated QKD transceivers benefit the miniaturization of cryptographic modules and provide exciting possibilities that are accompanied by rapidly improving key generation rates and link distances. Superconducting nanowire single photon detectors (SNSPDs) have here revolutionized QKD receiver technology because they achieve near-unity detection efficiency, picosecond timing jitter, and count rates approaching the GHz range \cite{Beutel2021}, vastly outperforming legacy avalanche photodiodes (APDs). Together, these advances promise high-rate, cost-effective fiber optic-QKD systems that can scale to the size and reliability demands of future smart grids.

It shall be noted that while QKD does provide convincing additional security for the transferred data when employed, it does not preclude attacks on the encryption mechanics that can negatively affect operational performance. Especially, eavesdropping on a quantum channel opens vectors for denial-of-service attacks~\cite{QKD-Eavesdropping-DOS-1, QKD-Eavesdropping-DOS-2}, as the quantum-physical laws dictate that eavesdropping on transmissions introduces disturbances in the communication between the original communication partners, so that keys potentially cannot be established or updated during an attack. In systems relying on the continuous delivery or updating of key material (as e.g., na\"{i}ve implementations of one-time pads do), service may hence be impaired until the eavesdropping event has ended. 
It is consequently important to identify the time horizons under which specific parts of power systems are able to operate with possibly unchanged key pairs (i.e., temporarily weakened protection), as well as the implications of potential successful attacks. As an example, applications issuing emergency commands in teleprotection scenarios can be considered highly critical because of the large-scale impacts that misuse of the signalling mechanisms may cause. Yet, such signals are likely to be unaffected by sudden attacks, as the IEC 68150 GOOSE / R-GOOSE signalling mechanism employed for e.g., for tripping signals, employs a bursty signalling mechanism when an event arises (i.e., it sends messages in quick succession) but resorts to slow periodic repetitions when no changes occur~\cite{R-GOOSE}. Eavesdrop-free times before an event can be thus used to establish enough key material in advance to cater for the burst of signals that arises in an emergency situation, while the key might be updated with some delay after such an event. Similarly, the rapid data updates of advanced metering infrastructure might at first sight require frequent re-keying and hence the reliance on a steady availability of QKD resources. The large-scale and distributed nature of the measurements would however render attacks to such applications more difficult because of the higher number of systems to target physically, and methods such as those for state estimation under incomplete observability~\cite{Buechel} may be used to remedy missing data caused by a potential intermittent unavailability of new quantum key material. Moreover, modern data-driven intrusion detection techniques, such as the Bayesian GAN–based false data injection detection scheme proposed in~\cite{Xie}, can be used to identify compromised measurement data.  
Assessing the degree and length of compromise until an application or control scenario fails might help to evaluate which guarantees a cryptographic protection mechanism needs to fulfill, irrespective of whether that mechanism is based on quantum effects or on classical technology. In this regard, standards developed by the European Telecommunications Standards Institute (ETSI), in particular its specifications for QKD key-management interfaces~\cite{etsi-qkd-interfaces} and network architectures~\cite{etsi-qkd-architectures}, allow QKD to be integrated into an IEC~61850 setting as a source of high-assurance symmetric keys for existing IEC~62351-based protection profiles (e.g., TLS- or IPsec-protected client/server traffic and group keys for GOOSE / R-GOOSE and Sampled Values), rather than as a separate security silo, and can be dimensioned in accordance with the criticality and time horizon of the underlying applications.

\subsubsection{Enhanced state transfer via quantum communication}
\label{subsec:enhanced-state-transfer}
Besides providing necessary enhancements of cryptographic functionality, quantum communication can also aid in the development of novel control approaches. Leveraging the effect that groups of pre-entangled qubits can be employed to create duplicates of the states of these qubits remotely (``quantum teleportation'')~\cite{Q-Teleportation}, quantum state information can be transferred between different locations without the need of physical transfer of a quantum information carrier (such as a photon) at that point in time. While such teleportations still require the communication of additional information after the necessary measurement processes on the sender side and hence are not faster than the speed of light~\cite{No-SOL-Q-Communication, No-Communication}, the possibility of theoretically lossless and (to a certain degree) quantum channel state-independent communication of quantum information opens novel possibilities with regard to sharing information between different regions of power grids.

The work in Refs.~\cite{MG-Control-With-Q-Teleportation, MG-Control-With-Q-Teleportation-2} proposes leveraging quantum teleportation to achieve consensus between \ac{DER} system controllers placed along an energy grid, and simulates microgrid applications of this mechanism for AC frequency- and DC voltage control. However, a recent review of distributed control strategies for microgrids by \textcite{Distributed-Control-MG-Review} found only limited work involving quantum communication beyond these approaches, yet employing entangled and/or teleported qubits may yield further research opportunities.

The application of especially quantum teleportation to grid control appears appealing, however some fundamental principles of quantum physics such as the no-cloning theorem~\cite{No-Cloning-1, No-Cloning2} and the no-communication theorem~\cite{No-Communication} preclude some scenarios, respectively complicate the application of quantum communication in other scenarios. State information residing with a sender is unavoidably and irrevocably destroyed within the process of teleportation, so that information cannot be copied using this method, only \emph{transferred}. This has implications on the working mechanisms of control algorithms as well as communication protocols as a shared state (in the sense of the same information being available on both ends) cannot be achieved with the means of quantum teleportation only. The requirement of an additional channel for meta information, as elaborated in Sec.~\ref{subsec:qkd}, adds another level of complexity to control mechanisms as both channels need to operate in conjunction to achieve benefits.

\section{Potential applications of quantum technologies in power systems edge devices applications}\label{sec:applications}

Quantum technologies have been successfully tried to tackle some of the challenges and limitations of modern power grids and to support the efforts in building and maintaining reliable and secure power grids. Most theoretical quantum computing models present better computational speed over classical models \cite{qc-power-ystems-zhou} and there are investigations that suggest quantum computing has a lower energy consumption in comparison to classical computing \cite{qc-power-systems-golestan}. Quantum computing, quantum machine learning and quantum key distribution have been used in power systems applications and can be found in several reviews such by \textcite{qc-power-systems-golestan}, \textcite{qgrid-eskandarpour}, \textcite{qc-power-ystems-zhou}, \textcite{Priyanka2024}, etc. Existing applications include mainly power system problems in grid analytics, optimization, operations and security, such as power flow, state estimation, contingency analysis, EMT simulation, stability assessment, fault diagnosis and management, unit commitment, facility location-allocation, PMU placement, power system control, cybersecurity, and forecasting. In this paper, we explore the possible integration of quantum technologies in edge devices.

Although large-scale quantum processors are still limited by qubit coherence, noise and physical size, several power system tasks can already benefit from quantum technologies in their current form. Firstly, shallow-circuit variational quantum algorithms enable small-scale optimisation and machine learning tasks, such as reduced-order classification or local optimisation. In \cite{qc-power-ystems-zhou} a detailed review of quantum algorithms that remain feasible under NISQ constraints is provided, highlighting their specific relevance to energy-system optimisation. 
Secondly, room-temperature NV-centre quantum sensors can make high-precision measurements of magnetic fields, currents, temperature and strain. This enables improved anomaly detection and equipment diagnostics at the edge. 
Thirdly, quantum key distribution (QKD), which has already been demonstrated in utility fibre networks, offers practical, quantum-secure communication for protection, control and distributed energy resource (DER) coordination. In \cite{alshowkan2022authentication} a working QKD deployment over 3.4 km of operational utility fibre, securing and MQTT-based SCADA communication between a substation and a distribution centre, has been shown. Similarly, in \cite{9405393}, the integration of QKD into a three-node system for a provider has been demonstrated using dark fiber-based communication. 
Fourthly, limited forms of quantum communication (e.g. entanglement-assisted synchronisation) can support distributed coordination and failover functions, even without fully error-corrected networks. For example, it has been demonstrated in \cite{MG-Control-With-Q-Teleportation} how teleportation-based information sharing can support distributed microgrid controllers. This is achieved by deploying a quantum distributed controller (QDC) with quantum computers at each node, as well as quantum communication between the nodes. The same authors have shown, in \cite{MG-Control-With-Q-Teleportation-2}, how interacting qubits can synchronize DER controllers under realistic noise conditions. 
Finally, post-quantum cryptography (PQC) can be used immediately for edge devices with limited resources. As shown in \cite{liu2018postquantumiot}, lattice-based PQC algorithms are well-suited to low-power embedded hardware and can be incorporated into IoT security architectures and microcontrollers. In \cite{ebrahimi2019cryptoprocessor}, two different architectures are presented that allow for PQC in powerful edge IoT devices, as well as in resource-constrained IoT nodes.
Together, these applications represent realistic use cases for quantum technologies within the constraints of current hardware.

In terms of sensors, NV center–based quantum sensors can measure magnetic, electric, temperature, and pressure fields with extremely high precision and spatial resolution at room temperature \cite{Pogorzelski2024Compact, Doherty2014ElectronicPressure}. These devices are solid-state and compact and can be operated in noisy environments \cite{Pogorzelski2024Compact}. Such miniature, low-power sensors can be embedded directly in grid edge hardware (e.g. on power lines, transformers, or IoT devices) to provide rich real-time monitoring of local currents, fields, and environmental conditions. Emerging research also explores distributed quantum sensing, the use of entangled particles in sensor networks, to achieve collective measurements that surpass classical precision limits. For example, continuous‑variable multipartite entangled networks have been demonstrated to reach Heisenberg‑limit scaling in sensitivity, outperforming separable schemes \cite{Zhuang2018}. Beyond sensing, lightweight quantum algorithms are being designed for deployment across distributed edge networks, with quantum communication links proposed to synchronize sensing tasks, ensure secure coordination between edge devices via entanglement‑assisted protocols, and perform entanglement‑assisted data fusion and decision‑making \cite{Li2023}. 

In terms of communication technologies for edge devices, QKD is increasingly being explored for direct integration with quantum-enabled sensors and processors at the grid edge. Advances in integrated photonics have enabled the miniaturization of QKD systems, making it feasible to deploy compact, low-power QKD modules directly within edge hardware such as smart meters, transformers, and IoT sensors. These integrated photonic QKD devices leverage on-chip sources, detectors, and modulators to generate and distribute quantum keys securely over existing fiber or free-space links with minimal footprint and cost \cite{Diamanti2016}. Recent demonstrations in photonic integration and packaging have highlighted promising routes to scale QKD deployment across edge networks, potentially enabling real-time, quantum-secure communication among edge nodes without the need for extensive specialized infrastructure \cite{Beutel2021}. 

Alongside QKD, post-quantum cryptography (PQC) algorithms offer complementary, software-based security solutions that can be implemented on resource-constrained edge devices to protect data against quantum-enabled attacks. Studies by \textcite{liu2018postquantumiot}, \textcite{ebrahimi2019cryptoprocessor}, and \textcite{schöffel2022secureiotq} demonstrate that lattice- and code-based PQC algorithms can be optimized for low-power IoT environments. As edge devices evolve into quantum-enabled sensors and processors, robust and scalable quantum communication infrastructures become essential for their coordination \cite{Li2023}. Such edge devices will incorporate the necessary physical components to establish and maintain entangled links across distributed nodes: entangled photon sources, quantum repeaters, and quantum routers. By leveraging integrated photonics, quantum communication technologies can be cost-efficiently scaled across distributed grid edge environments.

The potentially large amount of information required in grid control centers in combination with the real-time requirements for control signals gives rise to problems should these control centers experience disruptions in their own operation (e.g., hardware problems). If the grid data cannot be properly processed at one location anymore, the current state of the control system needs to be transferred to a standby backup control center as quickly as possible to resume control from there. Classical parts of this information can well be distributed among different locations even without the intervention of the senders, e.g., via Software Defined Networking (SDN)-based approaches as suggested for smart grids in e.g.,~\cite{SDN-FastFailover, SDN-for-PMUs}. In such scenarios, the data may either be directly copied in the network and sent both to the main control center and to the backup station, or can be re-routed to the backup if the main facility experiences problems. Problematic situations triggering the re-routing may even be detected by the network hardware itself and hence with minimal delay; recent hardware developments allow the collection of the necessary statistics (e.g., packets delivered from and to the control center) or calculations on the transmitted data (e.g., threshold analyses) at close to line speed as long as these calculations simple enough~\cite{COMSYS-In-Network-Computing}. In combination with QKD, such hardware also opens up the possibility of executing specific cryptographic functionality on the data while in-transit, alleviating edge devices from having to implement such functionality altogether. One-time pads (which may employ QKD to generate the keys) e.g., require only exclusive-OR operations between the data and the keys for en- and decryption, which programmable network equipment is capable of at high speeds. Such options may be viable if the edge equipment cannot be retrofitted or where retrofitting may not be viable for economic reasons (e.g., smart meters that have been installed recently, or small-scale equipment) and where the protection/authentication of individual data items on the ``first mile'' from prosumers to the grid operator may be less important than securing (possibly aggregated) data within the core of the grid.

For the reasons elaborated in Sec.~\ref{subsec:enhanced-state-transfer}, proactively providing copies of a specific quantum state is however, not possible. Thus, quantum algorithms running in a specific location can only have their state transferred to another location upon specific notification. Teleportation of entangled qubits may provide an opportunity for doing so without losses, but the solely reactive character of such failover scenarios needs to be considered in algorithm design. Another possibility may be to have quantum sensors send their respective readings to quantum algorithm copies at different locations in, e.g., a round-robin fashion, so that for $n$ copies of an algorithm, each copy can work with a copy of the data that is at most $n$ sensing iterations old. In case of failures, backup control could thus be established with some minor incongruences with regard to the current system state, but without having to initiate a full state teleportation first. As with classical data, such warm/hot standby strategies need not be implemented by the quantum sensors or algorithm copies themselves but can be realized within the quantum network itself. Especially when point-to-point quantum channels are replaced by switched quantum networks with the help of quantum repeaters~\cite{Quantum-Repeaters}, these repeaters could facilitate a ``quantum SDN''. To which degree quantum information may be used to control the switching processes, and the impact of the resulting essentially subsampling-based control strategies on grid operation, have yet to be determined. Methods akin to this strategy may however provide interesting ways for remedying some of the less intuitive effects of integrating distributed quantum data into control scenarios.

The adoption of quantum technologies in real-world smart grids requires interoperable and adaptable classical systems. These are not only required for the quick integration of quantum technologies, but also for parallel and coordinated operation of quantum-based and traditional applications, the switch between those and the use of fallback solutions. One example are modular approaches to control systems such as SOGNO~\cite{SOGNO} and MCCS~\cite{MCCS}, which simplify the integration of new, also quantum-based, functionalities as services and therefore serve as enabling technology.

Lastly, we considered the practical engineering constraints that shape current perspectives on the feasibility of deploying quantum capabilities at the edge of power and energy systems. Large dilution refrigerators exemplify the challenge of quantum integration. Cryostats can be several meters tall and weigh hundreds of kilograms, requiring dedicated multi-phase power and floor space. On the other hand, Josephson junctions in superconducting circuits are extremely sensitive to magnetic fields, necessitating heavy shielding. Trapped-ion systems similarly require ultra-high-vacuum pumps, stabilized lasers, and vibration-isolation platforms, all of which can sometimes exceed the environmental and infrastructural tolerances of grid’s substations \cite{ESQMalevannaya}. Even room-temperature diamond-based NV platforms require tight environmental control, requiring costly packaging, temperature stabilization and $\sim$100 mW optical pumping lasers. Representative commercial units draw $\sim$17 W, substantially higher than classical magnetic or current sensors \cite{Le2025}. 

These physical constraints create non-trivial barriers for embedding quantum elements directly into PMUs, protection relays, or distributed controllers. Quantum sensors and communication modules currently incur capital costs driven by specialized optics, precise thermal control, and low-volume manufacturing. Operational expenditures are also elevated: calibration intervals are shorter; photon sources, detectors, and optical couplers degrade measurably over time; and cryogenic or laser-based components require periodic servicing, realignment, or replacement. Consequently, their total cost of ownership remains significantly higher than that of conventional edge devices (typical 10–20-year utility lifecycles). Nevertheless, the stationary nature of substations, access to utility-grade power, and the possibility of hybrid architectures that isolate quantum modules suggest that, relative to other edge-computing use cases, power-grid applications may eventually be among the first in which these constraints become technically and economically manageable.

\section{Conclusion}\label{sec:conclusions}

The possible integration of quantum technologies such as quantum computing, sensing, and communication into edge devices could represent a new paradigm in power grids. In this paper, we introduce possible ways in which quantum-enabled edge devices can overcome key limitations of current edge devices capabilities. Quantum computing utilizes the principles of superposition and entanglement at its core and can, at the grid edge, tackle large-scale complex optimization problems and update the control system in real time—for example, in real-time power flow optimization, which can become impossible with classical processors. Similarly, quantum sensing can measure physical quantities such as electromagnetic signals with high precision and enables the detection of even minute changes in the fields that are often undetected by classical sensor devices, even in noisy environments. Utilizing these quantum sensing protocols enables monitoring of grid conditions and provides early warning of faults or disturbances beyond the sensitivity of conventional classical sensors. Additionally, quantum communication use quantum systems to transmit information, faster and more securely, for example, QKD enables fundamentally secure information exchange that is impossible in classical encryption schemes. Thus, these quantum technologies promise to elevate the capabilities of edge devices into a new class of fast, responsive, and ultra-secure nodes, which will amplify the overall resilience and efficiency of future power systems. In the near term, quantum sensing, QKD, PQC and small-scale NISQ-compatible algorithms are the most realistic and beneficial quantum tools for application in power-system edge devices given the current constraints of the hardware.

Despite these prospective opportunities, significant hurdles need to be overcome before quantum-based edge devices can be practically be realized in electrical grids. For example, current quantum hardware is constrained by resource-intensive requirements and is noisy. Similarly, NV-based miniature quantum sensors must also be calibrated to be noise-free to maintain accuracy. Additionally, system interoperability remains an open issue: quantum components need to interface seamlessly with classical grid infrastructure and communication protocols. Without well-defined standards and architectures for hybrid quantum-classical operation, the benefits of quantum edge devices cannot be fully realized.

Building on these challenges, several emerging research directions can help guide the practical evolution of quantum-enabled edge devices. First, substantial progress is required in miniaturizing and quantum hardware, as current systems may remain incompatible with the environmental and operational demands of substations; advances in integrated photonics, compact cryogenics, and application-specific quantum sensors represent promising paths \cite{Crawford2021QuantumReview}. Second, a qualitative cost–lifecycle assessment is urgently needed, as quantum components currently incur higher acquisition, calibration, and maintenance costs than classical equivalents, shaping the economic feasibility of field deployment. Third, as future implementations will rely on hybrid quantum–classical architectures that combine quantum sensors and PQC/QKD modules with NISQ-era processors, new software frameworks, orchestration mechanisms, and interoperability standards must be developed.

Addressing these challenges through continued research will help in realizing quantum-based edge devices in practice. Several national initiatives already recognize the importance of quantum technologies in power systems, including the QuGrids project, which aims to pursue quantum solutions to build quantum-based energy infrastructure for better resilience and intelligence of future electric power systems. In a nutshell, realizing this quantum-based edge device vision is not only a technical challenge but a strategic necessity to enable secure, resilient, and intelligent next-generation smart grids that can meet the evolving challenges in energy system landscapes.

\begin{acknowledgments}
We are grateful for the support from the Ministry of Culture and Science of North Rhine-Westphalia (MKW NRW) as part of the “Profilbildung 2022" program within the framework of the project “Quantum based Energy Grids (QuGrids)".

\end{acknowledgments}

\appendix

\section{Basic Quantum Physics and its Applications}\label{appendix:qauntum-physics-foundations}

Quantum physics describes nature at the smallest scales, and quantum systems are governed by probabilities, complex amplitudes, and measurement-induced randomness. In the second quantum revolution, the primary applications of quantum physics are quantum computing, sensing, and communication. Quantum superposition principle and entanglement are the foundation of these applications. We briefly describe them in this section and show the distinction from classical physics.

\

\subsection{Quantum superposition principle and entanglement}
The superposition principle states that a quantum system can exist in multiple classical states at once and any linear combination of valid quantum states is itself a valid quantum state. Physically, this implies that a quantum-bit (qubit) can exist in a coherent mixture of $|0\rangle$ and $|1\rangle$ rather than occupying only one ($0$ or $1$) classical configuration. Mathematically, it can be represented as
\begin{equation}
\ket{\psi} = \alpha \ket{0} + \beta \ket{1}.
\end{equation}
where $|\alpha|^2$ and $|\beta|^2$ are probabilities of measuring the system in states $\ket{0}$ and $\ket{1}$. Unlike a classical bit, which is either 0 or 1, a qubit stores probability amplitudes for both simultaneously.

Whereas, quantum entanglement arises when the quantum state of a composite system cannot be written as a tensor product of its subsystems. For example, a two-qubit state $|\Psi\rangle \in \mathbb{C}^2 \otimes \mathbb{C}^2$ is entangled if
\begin{equation}
|\Psi\rangle \neq |\psi_1\rangle \otimes |\psi_2\rangle.
\end{equation}
A canonical example is the Bell state
\begin{equation}
|\Phi^+\rangle = \frac{|00\rangle + |11\rangle}{\sqrt{2}}.
\end{equation}
This state exhibits perfect correlations: a measurement on one qubit fixes the outcome on the other, regardless of distance. Entanglement is a fundamental resource for quantum algorithms, communication, and sensing.

\subsection{Quantum Computing}
\label{appendix:qauntum-computing}

Quantum computing exploits quantum states and unitary evolution to perform computations beyond the capabilities of classical systems. The fundamental unit of information is the qubit, a two-level quantum system described by the state
\begin{equation}
\ket{\psi} = \alpha\ket{0} + \beta\ket{1},
\end{equation}
where the complex amplitudes satisfy $|\alpha|^2 + |\beta|^2 = 1$. Computation is implemented through quantum gates, each represented by a unitary operator $U$ obeying
\begin{equation}
U^\dagger U = I,
\end{equation}
which guarantees the reversibility of quantum evolution. A canonical example is the Hadamard gate,
\begin{equation}
H = \frac{1}{\sqrt{2}}
\begin{pmatrix}
1 & 1 \\
1 & -1
\end{pmatrix},
\end{equation}
which generates superposition by acting on the computational basis as
\begin{align}
H|0\rangle &= \frac{|0\rangle + |1\rangle}{\sqrt{2}}, \\
H|1\rangle &= \frac{|0\rangle - |1\rangle}{\sqrt{2}}.
\end{align}
For an $n$-qubit register initialized in the state $|0\rangle^{\otimes n}$, applying the operator $H^{\otimes n}$ produces a uniform superposition over all $2^n$ computational basis states,
\begin{equation}\label{eqn:hadamardsuperposition}
H^{\otimes n} |0\rangle^{\otimes n}
= \frac{1}{\sqrt{2^n}} \sum_{x \in \{0,1\}^n} |x\rangle,
\end{equation}
which constitutes the structural origin of quantum parallelism. A quantum computation is therefore described as a sequence of unitary operations arranged in a circuit,
\begin{equation}
\ket{\psi_{\text{out}}} = U_n U_{n-1} \cdots U_1 \ket{\psi_{\text{in}}},
\end{equation}
followed by measurement, which probabilistically converts the final quantum state amplitudes into a classical output.

\subsubsection{Unitary encoding of a function}
Any function acting on $n$ input bits and resulting in $m$ output bits can be decomposed into basic boolean operations (\emph{and}, \emph{or} and \emph{not}). These boolean operations cannot be directly implemented by a quantum system because they are not reversible, i.e. their inputs cannot be reconstructed from their outputs. Since quantum evolution is unitary and, therefore, reversible we have to use a logical gate that reflect this property. This is represented by the \emph{Toffoli gate}, which has three inputs and three outputs,
\begin{equation}
    T(a, b, c) = \{a, b, (a \, \mathrm{and} \, b) \, \mathrm{xor} \, c\},
\end{equation}
where \emph{xor} is the \emph{exclusive-or}, or \emph{difference}, operator. This gate is capable of reversibly implement any basic boolean operation.

It has been demonstrated that the Toffoli gate can be further decomposed into elementary quantum operations acting on single qubits and two qubits at a time \cite{barenco_elementary_1995}. This constitutes the theoretical background to implement a function $f$ as a sequence of simple quantum gates producing a total unitary gate $U_f$. The unitary $U_f$ computes the value of $f$ for a given input value encoded as a quantum state made of multiple qubits.

Focusing on functions with $n$ input bits and a single output bit, $f : \{0, 1\}^{\times n} \to \{0, 1\}$ (extension to multiple outputs is obtained through vectorization), the unitary $U_f$ has the following behavior,
\begin{equation}
U_f \ket{x}\ket{y}
=
\ket{x}\ket{y \oplus f(x)},
\quad x \in \{0,1\}^n.
\end{equation}
Combining this action with the linearity of unitary gates, we see the true expression of quantum parallelism when $U_f$ acts on superposition of states generated like in Eq.~\eqref{eqn:hadamardsuperposition},
\begin{equation}
\begin{split}
U_f \Bigg(
\frac{1}{\sqrt{2^n}}
\sum_{x \in \{0,1\}^n}
\ket{x}
\Bigg) \otimes \ket{y}
\\
=
\frac{1}{\sqrt{2^n}}
\sum_{x \in \{0,1\}^n}
\ket{x}
\otimes
\ket{y \oplus f(x)}.
\end{split}
\end{equation}
The function $f$ is evaluated on all possible inputs simultaneously. The result is an entangle (non-separable) state whose probability amplitudes encode all the outputs of $f$. Decoding $f$ entirely is highly inefficient, as measuring the qubits leads at most $n + 1$ bits of information at a time and requires a number of repetitions growing exponentially with $n$. It is generally more convenient to extract global properties of $f$ by leveraging constructive and destructive interference like in the Deutsch-Jozsa algorithm \cite{deutsch_rapid_1992}.


\subsection{Quantum Sensing}
\label{appendix:quantum-sensing}
Quantum sensing exploits controlled quantum evolution to measure physical parameters with precision reaching fundamental quantum limits. Information about an unknown parameter is encoded in the system dynamics through its Hamiltonian; specifically, if a physical parameter $\theta$ couples to a generator $G$, the Hamiltonian can be written as
\begin{equation}
H = \theta G,
\end{equation}
such that the system evolves according to
\begin{equation}
\ket{\psi(\theta)} = e^{-i \theta G t} \ket{\psi(0)}.
\end{equation}
Quantum sensing protocols proceed through repeated cycles of state preparation, signal-dependent evolution, and measurement. The sensor is initialized in a known reference state $\lvert 0 \rangle$ and transformed via coherent control into a sensitive superposition state $\lvert \psi_0 \rangle$, which maximizes the response to the parameter of interest. During an interrogation time $t$, the system evolves under the signal-dependent Hamiltonian, imprinting the parameter $\theta$ as a relative phase or population shift in the quantum state. A final control operation maps this encoded information onto a measurement basis, after which projective measurement produces classical outcomes with probabilities determined by the evolved state. Repeating this sequence and statistically averaging the outcomes yields a signal-dependent probability $p(t)$, from which an estimate of the physical parameter is inferred with a precision limited by quantum noise and the available interrogation resources.

\subsection{Quantum Communication}
\label{appendix:qauntum-communication}
Quantum communication uses quantum states and entanglement to transmit information via a quantum channel. Quantum channels generalize the classical communication channels to systems that carry qubits instead of classical bits. In this framework, information is encoded in nonorthogonal quantum states whose transmission and manipulation obeyed by quantum physics, enabling communication protocols with capabilities in classical systems. A central application of quantum communication is quantum key distribution, which allows two parties to establish a shared secret key whose security is guaranteed by fundamental physical principles: any attempt at eavesdropping disturbs the transmitted quantum states, thereby revealing the presence of the eavesdropper. This security is further underpinned by the no-cloning theorem, which prohibits the creation of identical copies of an unknown quantum state and ensures that quantum information cannot be intercepted, duplicated, or forwarded without detection. Together, these properties enable intrinsically secure communication and form the foundation of quantum networks and long-distance quantum information transfer. The no-cloning theory is elaborated further below:

Assume an arbitrary pure quantum state $\ket{\psi}$ and a known blank state $\ket{0}$. It can be proven~\cite{No-Cloning-1, No-Cloning2} that there exists no unitary operation $U$ (as introduced in~Appendix~\ref{appendix:qauntum-computing}) that allows the perfect cloning of $\ket{\psi}$:

\begin{equation}
\nexists\, U\ \text{unitary}, \ket{0}:\quad U(\ket{\psi} \otimes \ket{0}) = \ket{\psi} \otimes \ket{\psi}\quad \forall\,\ket{\psi}    
\end{equation}

In contrast to the classical case, it is hence impossible to retain copies of quantum information that have been transferred to some other location.
This implies that it is both not possible to work on the same quantum information in two places at the same time, and also that whenever quantum information is disturbed by external factors during or after transmission (e.g., by interaction with the environment), it cannot be retransmitted by the original sender.
Quantum information hence needs to be secured by other mechanisms.
These include quantum error correction codes (QECCs)~\cite{QECC}, which essentially distribute the information of a logical qubit among several physical qubits, allowing for the detection and correction of specific disturbances within that specific QECC's design scope, at the expense of requiring the storage and transmission of multiple physical qubits for a single piece of ``real'' quantum information.
A related approach, entanglement purification (or distillation)~\cite{Entanglement-Purification}, similarly employs multiple imperfectly entangled pairs, local operations on the entangled particles, as well as classical communication, to probabilistically keep entangled pairs with higher fidelity (i.e., closer to a maximally entangled state), which are better suited for quantum information transfer.


\begin{thebibliography}{133}%
\makeatletter
\providecommand \@ifxundefined [1]{%
 \@ifx{#1\undefined}
}%
\providecommand \@ifnum [1]{%
 \ifnum #1\expandafter \@firstoftwo
 \else \expandafter \@secondoftwo
 \fi
}%
\providecommand \@ifx [1]{%
 \ifx #1\expandafter \@firstoftwo
 \else \expandafter \@secondoftwo
 \fi
}%
\providecommand \natexlab [1]{#1}%
\providecommand \enquote  [1]{``#1''}%
\providecommand \bibnamefont  [1]{#1}%
\providecommand \bibfnamefont [1]{#1}%
\providecommand \citenamefont [1]{#1}%
\providecommand \href@noop [0]{\@secondoftwo}%
\providecommand \href [0]{\begingroup \@sanitize@url \@href}%
\providecommand \@href[1]{\@@startlink{#1}\@@href}%
\providecommand \@@href[1]{\endgroup#1\@@endlink}%
\providecommand \@sanitize@url [0]{\catcode `\\12\catcode `\$12\catcode
  `\&12\catcode `\#12\catcode `\^12\catcode `\_12\catcode `\%12\relax}%
\providecommand \@@startlink[1]{}%
\providecommand \@@endlink[0]{}%
\providecommand \url  [0]{\begingroup\@sanitize@url \@url }%
\providecommand \@url [1]{\endgroup\@href {#1}{\urlprefix }}%
\providecommand \urlprefix  [0]{URL }%
\providecommand \Eprint [0]{\href }%
\providecommand \doibase [0]{https://doi.org/}%
\providecommand \selectlanguage [0]{\@gobble}%
\providecommand \bibinfo  [0]{\@secondoftwo}%
\providecommand \bibfield  [0]{\@secondoftwo}%
\providecommand \translation [1]{[#1]}%
\providecommand \BibitemOpen [0]{}%
\providecommand \bibitemStop [0]{}%
\providecommand \bibitemNoStop [0]{.\EOS\space}%
\providecommand \EOS [0]{\spacefactor3000\relax}%
\providecommand \BibitemShut  [1]{\csname bibitem#1\endcsname}%
\let\auto@bib@innerbib\@empty
\bibitem [{\citenamefont {Bedi}\ \emph {et~al.}(2018)\citenamefont {Bedi},
  \citenamefont {Venayagamoorthy}, \citenamefont {Singh}, \citenamefont
  {Brooks},\ and\ \citenamefont {Wang}}]{8281479}%
  \BibitemOpen
  \bibfield  {author} {\bibinfo {author} {\bibfnamefont {G.}~\bibnamefont
  {Bedi}}, \bibinfo {author} {\bibfnamefont {G.~K.}\ \bibnamefont
  {Venayagamoorthy}}, \bibinfo {author} {\bibfnamefont {R.}~\bibnamefont
  {Singh}}, \bibinfo {author} {\bibfnamefont {R.~R.}\ \bibnamefont {Brooks}},\
  and\ \bibinfo {author} {\bibfnamefont {K.-C.}\ \bibnamefont {Wang}},\
  }\bibfield  {title} {\bibinfo {title} {Review of internet of things (iot) in
  electric power and energy systems},\ }\href
  {https://doi.org/10.1109/JIOT.2018.2802704} {\bibfield  {journal} {\bibinfo
  {journal} {IEEE Internet of Things Journal}\ }\textbf {\bibinfo {volume}
  {5}},\ \bibinfo {pages} {847} (\bibinfo {year} {2018})}\BibitemShut {NoStop}%
\bibitem [{\citenamefont {Capra}\ \emph {et~al.}(2019)\citenamefont {Capra},
  \citenamefont {Peloso}, \citenamefont {Masera}, \citenamefont {Ruo~Roch},\
  and\ \citenamefont {Martina}}]{fi11040100}%
  \BibitemOpen
  \bibfield  {author} {\bibinfo {author} {\bibfnamefont {M.}~\bibnamefont
  {Capra}}, \bibinfo {author} {\bibfnamefont {R.}~\bibnamefont {Peloso}},
  \bibinfo {author} {\bibfnamefont {G.}~\bibnamefont {Masera}}, \bibinfo
  {author} {\bibfnamefont {M.}~\bibnamefont {Ruo~Roch}},\ and\ \bibinfo
  {author} {\bibfnamefont {M.}~\bibnamefont {Martina}},\ }\bibfield  {title}
  {\bibinfo {title} {Edge computing: A survey on the hardware requirements in
  the internet of things world},\ }\bibfield  {journal} {\bibinfo  {journal}
  {Future Internet}\ }\textbf {\bibinfo {volume} {11}},\ \href
  {https://doi.org/10.3390/fi11040100} {10.3390/fi11040100} (\bibinfo {year}
  {2019})\BibitemShut {NoStop}%
\bibitem [{\citenamefont {Feng}\ \emph {et~al.}(2021)\citenamefont {Feng},
  \citenamefont {Wang}, \citenamefont {Chen}, \citenamefont {Ding},
  \citenamefont {Strbac},\ and\ \citenamefont {Kang}}]{FENG2021100006}%
  \BibitemOpen
  \bibfield  {author} {\bibinfo {author} {\bibfnamefont {C.}~\bibnamefont
  {Feng}}, \bibinfo {author} {\bibfnamefont {Y.}~\bibnamefont {Wang}}, \bibinfo
  {author} {\bibfnamefont {Q.}~\bibnamefont {Chen}}, \bibinfo {author}
  {\bibfnamefont {Y.}~\bibnamefont {Ding}}, \bibinfo {author} {\bibfnamefont
  {G.}~\bibnamefont {Strbac}},\ and\ \bibinfo {author} {\bibfnamefont
  {C.}~\bibnamefont {Kang}},\ }\bibfield  {title} {\bibinfo {title} {Smart grid
  encounters edge computing: opportunities and applications},\ }\href
  {https://doi.org/https://doi.org/10.1016/j.adapen.2020.100006} {\bibfield
  {journal} {\bibinfo  {journal} {Advances in Applied Energy}\ }\textbf
  {\bibinfo {volume} {1}},\ \bibinfo {pages} {100006} (\bibinfo {year}
  {2021})}\BibitemShut {NoStop}%
\bibitem [{\citenamefont {Mehmood}\ \emph {et~al.}(2021)\citenamefont
  {Mehmood}, \citenamefont {Oad}, \citenamefont {Abrar}, \citenamefont {Munir},
  \citenamefont {Hasan}, \citenamefont {Muqeet},\ and\ \citenamefont
  {Golilarz}}]{mehmood2021edge}%
  \BibitemOpen
  \bibfield  {author} {\bibinfo {author} {\bibfnamefont {M.~Y.}\ \bibnamefont
  {Mehmood}}, \bibinfo {author} {\bibfnamefont {A.}~\bibnamefont {Oad}},
  \bibinfo {author} {\bibfnamefont {M.}~\bibnamefont {Abrar}}, \bibinfo
  {author} {\bibfnamefont {H.~M.}\ \bibnamefont {Munir}}, \bibinfo {author}
  {\bibfnamefont {S.~F.}\ \bibnamefont {Hasan}}, \bibinfo {author}
  {\bibfnamefont {H.~A.~u.}\ \bibnamefont {Muqeet}},\ and\ \bibinfo {author}
  {\bibfnamefont {N.~A.}\ \bibnamefont {Golilarz}},\ }\bibfield  {title}
  {\bibinfo {title} {Edge computing for iot-enabled smart grid},\ }\href@noop
  {} {\bibfield  {journal} {\bibinfo  {journal} {Security and communication
  networks}\ }\textbf {\bibinfo {volume} {2021}},\ \bibinfo {pages} {5524025}
  (\bibinfo {year} {2021})}\BibitemShut {NoStop}%
\bibitem [{\citenamefont {Li}\ \emph {et~al.}(2022)\citenamefont {Li},
  \citenamefont {Gu}, \citenamefont {Xiang},\ and\ \citenamefont
  {Li}}]{edge-cloud-sg}%
  \BibitemOpen
  \bibfield  {author} {\bibinfo {author} {\bibfnamefont {J.}~\bibnamefont
  {Li}}, \bibinfo {author} {\bibfnamefont {C.}~\bibnamefont {Gu}}, \bibinfo
  {author} {\bibfnamefont {Y.}~\bibnamefont {Xiang}},\ and\ \bibinfo {author}
  {\bibfnamefont {F.}~\bibnamefont {Li}},\ }\bibfield  {title} {\bibinfo
  {title} {Edge-cloud computing systems for smart grid: State-of-the-art,
  architecture, and applications},\ }\href
  {https://doi.org/10.35833/MPCE.2021.000161} {\bibfield  {journal} {\bibinfo
  {journal} {Journal of Modern Power Systems and Clean Energy}\ }\textbf
  {\bibinfo {volume} {10}},\ \bibinfo {pages} {805} (\bibinfo {year}
  {2022})}\BibitemShut {NoStop}%
\bibitem [{\citenamefont {Gedeon}\ \emph {et~al.}(2019)\citenamefont {Gedeon},
  \citenamefont {Brandherm}, \citenamefont {Egert}, \citenamefont {Grube},\
  and\ \citenamefont {Mühlhäuser}}]{8877785}%
  \BibitemOpen
  \bibfield  {author} {\bibinfo {author} {\bibfnamefont {J.}~\bibnamefont
  {Gedeon}}, \bibinfo {author} {\bibfnamefont {F.}~\bibnamefont {Brandherm}},
  \bibinfo {author} {\bibfnamefont {R.}~\bibnamefont {Egert}}, \bibinfo
  {author} {\bibfnamefont {T.}~\bibnamefont {Grube}},\ and\ \bibinfo {author}
  {\bibfnamefont {M.}~\bibnamefont {Mühlhäuser}},\ }\bibfield  {title}
  {\bibinfo {title} {What the fog? edge computing revisited: Promises,
  applications and future challenges},\ }\href
  {https://doi.org/10.1109/ACCESS.2019.2948399} {\bibfield  {journal} {\bibinfo
   {journal} {IEEE Access}\ }\textbf {\bibinfo {volume} {7}},\ \bibinfo {pages}
  {152847} (\bibinfo {year} {2019})}\BibitemShut {NoStop}%
\bibitem [{\citenamefont {Hussain}\ and\ \citenamefont
  {Beg}(2019)}]{bdcc3010008}%
  \BibitemOpen
  \bibfield  {author} {\bibinfo {author} {\bibfnamefont {M.~M.}\ \bibnamefont
  {Hussain}}\ and\ \bibinfo {author} {\bibfnamefont {M.~S.}\ \bibnamefont
  {Beg}},\ }\bibfield  {title} {\bibinfo {title} {Fog computing for internet of
  things (iot)-aided smart grid architectures},\ }\bibfield  {journal}
  {\bibinfo  {journal} {Big Data and Cognitive Computing}\ }\textbf {\bibinfo
  {volume} {3}},\ \href {https://doi.org/10.3390/bdcc3010008}
  {10.3390/bdcc3010008} (\bibinfo {year} {2019})\BibitemShut {NoStop}%
\bibitem [{\citenamefont {Sun}\ \emph {et~al.}(2024)\citenamefont {Sun},
  \citenamefont {Shen}, \citenamefont {Wan}, \citenamefont {Wu}, \citenamefont
  {Fang},\ and\ \citenamefont {Gao}}]{IoT-Security-Survey-2}%
  \BibitemOpen
  \bibfield  {author} {\bibinfo {author} {\bibfnamefont {P.}~\bibnamefont
  {Sun}}, \bibinfo {author} {\bibfnamefont {S.}~\bibnamefont {Shen}}, \bibinfo
  {author} {\bibfnamefont {Y.}~\bibnamefont {Wan}}, \bibinfo {author}
  {\bibfnamefont {Z.}~\bibnamefont {Wu}}, \bibinfo {author} {\bibfnamefont
  {Z.}~\bibnamefont {Fang}},\ and\ \bibinfo {author} {\bibfnamefont {X.-Z.}\
  \bibnamefont {Gao}},\ }\bibfield  {title} {\bibinfo {title} {{A Survey of IoT
  Privacy Security: Architecture, Technology, Challenges, and Trends}},\ }\href
  {https://doi.org/10.1109/JIOT.2024.3372518} {\bibfield  {journal} {\bibinfo
  {journal} {IEEE Internet of Things Journal}\ }\textbf {\bibinfo {volume}
  {11}},\ \bibinfo {pages} {34567} (\bibinfo {year} {2024})}\BibitemShut
  {NoStop}%
\bibitem [{\citenamefont {Khanafer}\ \emph {et~al.}(2017)\citenamefont
  {Khanafer}, \citenamefont {Al-Anbagi},\ and\ \citenamefont
  {Mouftah}}]{khanafer2017optimized}%
  \BibitemOpen
  \bibfield  {author} {\bibinfo {author} {\bibfnamefont {M.}~\bibnamefont
  {Khanafer}}, \bibinfo {author} {\bibfnamefont {I.}~\bibnamefont
  {Al-Anbagi}},\ and\ \bibinfo {author} {\bibfnamefont {H.~T.}\ \bibnamefont
  {Mouftah}},\ }\bibfield  {title} {\bibinfo {title} {An optimized wsn design
  for latency-critical smart grid applications},\ }\href@noop {} {\bibfield
  {journal} {\bibinfo  {journal} {Journal of Sensors}\ }\textbf {\bibinfo
  {volume} {2017}},\ \bibinfo {pages} {5274715} (\bibinfo {year}
  {2017})}\BibitemShut {NoStop}%
\bibitem [{\citenamefont {Feynman}(1982)}]{Feynman1982}%
  \BibitemOpen
  \bibfield  {author} {\bibinfo {author} {\bibfnamefont {R.~P.}\ \bibnamefont
  {Feynman}},\ }\bibfield  {title} {\bibinfo {title} {Simulating physics with
  computers},\ }\href {https://doi.org/10.1007/bf02650179} {\bibfield
  {journal} {\bibinfo  {journal} {International Journal of Theoretical
  Physics}\ }\textbf {\bibinfo {volume} {21}},\ \bibinfo {pages} {467–488}
  (\bibinfo {year} {1982})}\BibitemShut {NoStop}%
\bibitem [{\citenamefont {Shor}(1994)}]{Shor}%
  \BibitemOpen
  \bibfield  {author} {\bibinfo {author} {\bibfnamefont {P.}~\bibnamefont
  {Shor}},\ }\bibfield  {title} {\bibinfo {title} {Algorithms for quantum
  computation: discrete logarithms and factoring},\ }in\ \href
  {https://doi.org/10.1109/SFCS.1994.365700} {\emph {\bibinfo {booktitle}
  {Proceedings 35th {Annual} {Symposium} on {Foundations} of {Computer}
  {Science}}}}\ (\bibinfo {year} {1994})\ pp.\ \bibinfo {pages}
  {124--134}\BibitemShut {NoStop}%
\bibitem [{\citenamefont {Ganeshamurthy}\ \emph
  {et~al.}(2024{\natexlab{a}})\citenamefont {Ganeshamurthy}, \citenamefont
  {Ghosh}, \citenamefont {O'Meara}, \citenamefont {Cortiana},\ and\
  \citenamefont {Schiefelbein-Lach}}]{Priyanka2024}%
  \BibitemOpen
  \bibfield  {author} {\bibinfo {author} {\bibfnamefont {P.~A.}\ \bibnamefont
  {Ganeshamurthy}}, \bibinfo {author} {\bibfnamefont {K.}~\bibnamefont
  {Ghosh}}, \bibinfo {author} {\bibfnamefont {C.}~\bibnamefont {O'Meara}},
  \bibinfo {author} {\bibfnamefont {G.}~\bibnamefont {Cortiana}},\ and\
  \bibinfo {author} {\bibfnamefont {J.}~\bibnamefont {Schiefelbein-Lach}},\
  }\href {https://doi.org/10.48550/ARXIV.2408.02432} {\bibinfo {title}
  {Bridging the gap to next generation power system planning and operation with
  quantum computation}} (\bibinfo {year} {2024}{\natexlab{a}})\BibitemShut
  {NoStop}%
\bibitem [{\citenamefont {Doherty}\ \emph {et~al.}(2013)\citenamefont
  {Doherty}, \citenamefont {Manson}, \citenamefont {Delaney}, \citenamefont
  {Jelezko}, \citenamefont {Wrachtrup},\ and\ \citenamefont
  {Hollenberg}}]{Doherty2013TheDiamond}%
  \BibitemOpen
  \bibfield  {author} {\bibinfo {author} {\bibfnamefont {M.~W.}\ \bibnamefont
  {Doherty}}, \bibinfo {author} {\bibfnamefont {N.~B.}\ \bibnamefont {Manson}},
  \bibinfo {author} {\bibfnamefont {P.}~\bibnamefont {Delaney}}, \bibinfo
  {author} {\bibfnamefont {F.}~\bibnamefont {Jelezko}}, \bibinfo {author}
  {\bibfnamefont {J.}~\bibnamefont {Wrachtrup}},\ and\ \bibinfo {author}
  {\bibfnamefont {L.~C.}\ \bibnamefont {Hollenberg}},\ }\bibfield  {title}
  {\bibinfo {title} {{The nitrogen-vacancy colour centre in diamond}},\ }\href
  {https://doi.org/10.1016/j.physrep.2013.02.001} {\bibfield  {journal}
  {\bibinfo  {journal} {Phys. Rep.}\ }\textbf {\bibinfo {volume} {528}},\
  \bibinfo {pages} {1} (\bibinfo {year} {2013})}\BibitemShut {NoStop}%
\bibitem [{\citenamefont {Barry}\ \emph {et~al.}(2020)\citenamefont {Barry},
  \citenamefont {Schloss}, \citenamefont {Bauch}, \citenamefont {Turner},
  \citenamefont {Hart}, \citenamefont {Pham},\ and\ \citenamefont
  {Walsworth}}]{Barry2020}%
  \BibitemOpen
  \bibfield  {author} {\bibinfo {author} {\bibfnamefont {J.~F.}\ \bibnamefont
  {Barry}}, \bibinfo {author} {\bibfnamefont {J.~M.}\ \bibnamefont {Schloss}},
  \bibinfo {author} {\bibfnamefont {E.}~\bibnamefont {Bauch}}, \bibinfo
  {author} {\bibfnamefont {M.~J.}\ \bibnamefont {Turner}}, \bibinfo {author}
  {\bibfnamefont {C.~A.}\ \bibnamefont {Hart}}, \bibinfo {author}
  {\bibfnamefont {L.~M.}\ \bibnamefont {Pham}},\ and\ \bibinfo {author}
  {\bibfnamefont {R.~L.}\ \bibnamefont {Walsworth}},\ }\bibfield  {title}
  {\bibinfo {title} {Sensitivity optimization for nv-diamond magnetometry},\
  }\bibfield  {journal} {\bibinfo  {journal} {Reviews of Modern Physics}\
  }\textbf {\bibinfo {volume} {92}},\ \href
  {https://doi.org/10.1103/revmodphys.92.015004} {10.1103/revmodphys.92.015004}
  (\bibinfo {year} {2020})\BibitemShut {NoStop}%
\bibitem [{\citenamefont {Gottesman}\ \emph {et~al.}(2004)\citenamefont
  {Gottesman}, \citenamefont {Lo}, \citenamefont {L\"{u}tkenhaus},\ and\
  \citenamefont {Preskill}}]{QKD-Sec-with-Imperfect-Devices}%
  \BibitemOpen
  \bibfield  {author} {\bibinfo {author} {\bibfnamefont {D.}~\bibnamefont
  {Gottesman}}, \bibinfo {author} {\bibfnamefont {H.-K.}\ \bibnamefont {Lo}},
  \bibinfo {author} {\bibfnamefont {N.}~\bibnamefont {L\"{u}tkenhaus}},\ and\
  \bibinfo {author} {\bibfnamefont {J.}~\bibnamefont {Preskill}},\ }\bibfield
  {title} {\bibinfo {title} {Security of quantum key distribution with
  imperfect devices},\ }\href@noop {} {\bibfield  {journal} {\bibinfo
  {journal} {Quantum Info. Comput.}\ }\textbf {\bibinfo {volume} {4}},\
  \bibinfo {pages} {325–360} (\bibinfo {year} {2004})}\BibitemShut {NoStop}%
\bibitem [{\citenamefont {Liang}\ \emph {et~al.}(2024)\citenamefont {Liang},
  \citenamefont {Jin},\ and\ \citenamefont {Chen}}]{edge-computing-review2024}%
  \BibitemOpen
  \bibfield  {author} {\bibinfo {author} {\bibfnamefont {S.}~\bibnamefont
  {Liang}}, \bibinfo {author} {\bibfnamefont {S.}~\bibnamefont {Jin}},\ and\
  \bibinfo {author} {\bibfnamefont {Y.}~\bibnamefont {Chen}},\ }\bibfield
  {title} {\bibinfo {title} {A review of edge computing technology and its
  applications in power systems},\ }\bibfield  {journal} {\bibinfo  {journal}
  {Energies}\ }\textbf {\bibinfo {volume} {17}},\ \href
  {https://doi.org/10.3390/en17133230} {10.3390/en17133230} (\bibinfo {year}
  {2024})\BibitemShut {NoStop}%
\bibitem [{\citenamefont {Xiao}\ \emph {et~al.}(2019)\citenamefont {Xiao},
  \citenamefont {Jia}, \citenamefont {Liu}, \citenamefont {Cheng},
  \citenamefont {Yu},\ and\ \citenamefont {Lv}}]{edge-computing-security}%
  \BibitemOpen
  \bibfield  {author} {\bibinfo {author} {\bibfnamefont {Y.}~\bibnamefont
  {Xiao}}, \bibinfo {author} {\bibfnamefont {Y.}~\bibnamefont {Jia}}, \bibinfo
  {author} {\bibfnamefont {C.}~\bibnamefont {Liu}}, \bibinfo {author}
  {\bibfnamefont {X.}~\bibnamefont {Cheng}}, \bibinfo {author} {\bibfnamefont
  {J.}~\bibnamefont {Yu}},\ and\ \bibinfo {author} {\bibfnamefont
  {W.}~\bibnamefont {Lv}},\ }\bibfield  {title} {\bibinfo {title} {Edge
  computing security: State of the art and challenges},\ }\href
  {https://doi.org/10.1109/JPROC.2019.2918437} {\bibfield  {journal} {\bibinfo
  {journal} {Proceedings of the IEEE}\ }\textbf {\bibinfo {volume} {107}},\
  \bibinfo {pages} {1608} (\bibinfo {year} {2019})}\BibitemShut {NoStop}%
\bibitem [{\citenamefont {Ojo}\ \emph {et~al.}(2018)\citenamefont {Ojo},
  \citenamefont {Giordano}, \citenamefont {Procissi},\ and\ \citenamefont
  {Seitanidis}}]{devices-review}%
  \BibitemOpen
  \bibfield  {author} {\bibinfo {author} {\bibfnamefont {M.~O.}\ \bibnamefont
  {Ojo}}, \bibinfo {author} {\bibfnamefont {S.}~\bibnamefont {Giordano}},
  \bibinfo {author} {\bibfnamefont {G.}~\bibnamefont {Procissi}},\ and\
  \bibinfo {author} {\bibfnamefont {I.~N.}\ \bibnamefont {Seitanidis}},\
  }\bibfield  {title} {\bibinfo {title} {A review of low-end, middle-end, and
  high-end iot devices},\ }\href {https://doi.org/10.1109/ACCESS.2018.2879615}
  {\bibfield  {journal} {\bibinfo  {journal} {IEEE Access}\ }\textbf {\bibinfo
  {volume} {6}},\ \bibinfo {pages} {70528} (\bibinfo {year}
  {2018})}\BibitemShut {NoStop}%
\bibitem [{\citenamefont {Moghe}\ \emph {et~al.}(2012)\citenamefont {Moghe},
  \citenamefont {Lambert},\ and\ \citenamefont {Divan}}]{Moghe2012}%
  \BibitemOpen
  \bibfield  {author} {\bibinfo {author} {\bibfnamefont {R.}~\bibnamefont
  {Moghe}}, \bibinfo {author} {\bibfnamefont {F.~C.}\ \bibnamefont {Lambert}},\
  and\ \bibinfo {author} {\bibfnamefont {D.}~\bibnamefont {Divan}},\ }\bibfield
   {title} {\bibinfo {title} {Smart “stick-on” sensors for the smart
  grid},\ }\href {https://doi.org/10.1109/tsg.2011.2166280} {\bibfield
  {journal} {\bibinfo  {journal} {IEEE Transactions on Smart Grid}\ }\textbf
  {\bibinfo {volume} {3}},\ \bibinfo {pages} {241} (\bibinfo {year}
  {2012})}\BibitemShut {NoStop}%
\bibitem [{\citenamefont {Xu}\ \emph {et~al.}(2015)\citenamefont {Xu},
  \citenamefont {Liu}, \citenamefont {Zhang}, \citenamefont {Xu},\ and\
  \citenamefont {Yang}}]{Xu2015}%
  \BibitemOpen
  \bibfield  {author} {\bibinfo {author} {\bibfnamefont {C.}~\bibnamefont
  {Xu}}, \bibinfo {author} {\bibfnamefont {J.-G.}\ \bibnamefont {Liu}},
  \bibinfo {author} {\bibfnamefont {Q.}~\bibnamefont {Zhang}}, \bibinfo
  {author} {\bibfnamefont {C.}~\bibnamefont {Xu}},\ and\ \bibinfo {author}
  {\bibfnamefont {Y.}~\bibnamefont {Yang}},\ }\bibfield  {title} {\bibinfo
  {title} {Investigation of the thermal drift of open-loop hall effect current
  sensor and its improvement},\ }in\ \href
  {https://doi.org/10.1109/amps.2015.7312732} {\emph {\bibinfo {booktitle}
  {2015 IEEE International Workshop on Applied Measurements for Power Systems
  (AMPS)}}}\ (\bibinfo  {publisher} {IEEE},\ \bibinfo {year} {2015})\ pp.\
  \bibinfo {pages} {1--6}\BibitemShut {NoStop}%
\bibitem [{\citenamefont {Metwally}(2010)}]{Metwally2010}%
  \BibitemOpen
  \bibfield  {author} {\bibinfo {author} {\bibfnamefont {I.}~\bibnamefont
  {Metwally}},\ }\bibfield  {title} {\bibinfo {title} {Self-integrating
  rogowski coil for high-impulse current measurement},\ }\href
  {https://doi.org/10.1109/tim.2009.2023821} {\bibfield  {journal} {\bibinfo
  {journal} {IEEE Transactions on Instrumentation and Measurement}\ }\textbf
  {\bibinfo {volume} {59}},\ \bibinfo {pages} {353} (\bibinfo {year}
  {2010})}\BibitemShut {NoStop}%
\bibitem [{\citenamefont {Gazivoda}\ \emph {et~al.}(2021)\citenamefont
  {Gazivoda}, \citenamefont {Ziger}, \citenamefont {Novko},\ and\ \citenamefont
  {Zupan}}]{Gazivoda2021}%
  \BibitemOpen
  \bibfield  {author} {\bibinfo {author} {\bibfnamefont {D.}~\bibnamefont
  {Gazivoda}}, \bibinfo {author} {\bibfnamefont {I.}~\bibnamefont {Ziger}},
  \bibinfo {author} {\bibfnamefont {I.}~\bibnamefont {Novko}},\ and\ \bibinfo
  {author} {\bibfnamefont {T.}~\bibnamefont {Zupan}},\ }\bibfield  {title}
  {\bibinfo {title} {Method for internal fault testing of instrument
  transformers with sectioned active parts},\ }\href
  {https://doi.org/10.1109/access.2021.3121429} {\bibfield  {journal} {\bibinfo
   {journal} {IEEE Access}\ }\textbf {\bibinfo {volume} {9}},\ \bibinfo {pages}
  {143295} (\bibinfo {year} {2021})}\BibitemShut {NoStop}%
\bibitem [{\citenamefont {Schweitzer}\ and\ \citenamefont
  {Kasztenny}(2018)}]{Schweitzer2018}%
  \BibitemOpen
  \bibfield  {author} {\bibinfo {author} {\bibfnamefont {E.~O.}\ \bibnamefont
  {Schweitzer}}\ and\ \bibinfo {author} {\bibfnamefont {B.}~\bibnamefont
  {Kasztenny}},\ }\bibfield  {title} {\bibinfo {title} {Distance protection:
  Why have we started with a circle, does it matter, and what else is out
  there?},\ }in\ \href {https://doi.org/10.1109/cpre.2018.8349791} {\emph
  {\bibinfo {booktitle} {2018 71st Annual Conference for Protective Relay
  Engineers (CPRE)}}}\ (\bibinfo  {publisher} {IEEE},\ \bibinfo {year} {2018})\
  pp.\ \bibinfo {pages} {1--19}\BibitemShut {NoStop}%
\bibitem [{\citenamefont {Gungor}\ \emph {et~al.}(2013)\citenamefont {Gungor},
  \citenamefont {Sahin}, \citenamefont {Kocak}, \citenamefont {Ergut},
  \citenamefont {Buccella}, \citenamefont {Cecati},\ and\ \citenamefont
  {Hancke}}]{sg-review-2013}%
  \BibitemOpen
  \bibfield  {author} {\bibinfo {author} {\bibfnamefont {V.~C.}\ \bibnamefont
  {Gungor}}, \bibinfo {author} {\bibfnamefont {D.}~\bibnamefont {Sahin}},
  \bibinfo {author} {\bibfnamefont {T.}~\bibnamefont {Kocak}}, \bibinfo
  {author} {\bibfnamefont {S.}~\bibnamefont {Ergut}}, \bibinfo {author}
  {\bibfnamefont {C.}~\bibnamefont {Buccella}}, \bibinfo {author}
  {\bibfnamefont {C.}~\bibnamefont {Cecati}},\ and\ \bibinfo {author}
  {\bibfnamefont {G.~P.}\ \bibnamefont {Hancke}},\ }\bibfield  {title}
  {\bibinfo {title} {A survey on smart grid potential applications and
  communication requirements},\ }\href
  {https://doi.org/10.1109/TII.2012.2218253} {\bibfield  {journal} {\bibinfo
  {journal} {IEEE Transactions on Industrial Informatics}\ }\textbf {\bibinfo
  {volume} {9}},\ \bibinfo {pages} {28} (\bibinfo {year} {2013})}\BibitemShut
  {NoStop}%
\bibitem [{\citenamefont {Ghorbanian}\ \emph
  {et~al.}(2019{\natexlab{a}})\citenamefont {Ghorbanian}, \citenamefont
  {Dolatabadi}, \citenamefont {Masjedi},\ and\ \citenamefont
  {Siano}}]{communication-sg-2019}%
  \BibitemOpen
  \bibfield  {author} {\bibinfo {author} {\bibfnamefont {M.}~\bibnamefont
  {Ghorbanian}}, \bibinfo {author} {\bibfnamefont {S.~H.}\ \bibnamefont
  {Dolatabadi}}, \bibinfo {author} {\bibfnamefont {M.}~\bibnamefont
  {Masjedi}},\ and\ \bibinfo {author} {\bibfnamefont {P.}~\bibnamefont
  {Siano}},\ }\bibfield  {title} {\bibinfo {title} {Communication in smart
  grids: A comprehensive review on the existing and future communication and
  information infrastructures},\ }\href
  {https://doi.org/10.1109/JSYST.2019.2928090} {\bibfield  {journal} {\bibinfo
  {journal} {IEEE Systems Journal}\ }\textbf {\bibinfo {volume} {13}},\
  \bibinfo {pages} {4001} (\bibinfo {year} {2019}{\natexlab{a}})}\BibitemShut
  {NoStop}%
\bibitem [{\citenamefont {Ghorbanian}\ \emph
  {et~al.}(2019{\natexlab{b}})\citenamefont {Ghorbanian}, \citenamefont
  {Dolatabadi}, \citenamefont {Masjedi},\ and\ \citenamefont
  {Siano}}]{8839117}%
  \BibitemOpen
  \bibfield  {author} {\bibinfo {author} {\bibfnamefont {M.}~\bibnamefont
  {Ghorbanian}}, \bibinfo {author} {\bibfnamefont {S.~H.}\ \bibnamefont
  {Dolatabadi}}, \bibinfo {author} {\bibfnamefont {M.}~\bibnamefont
  {Masjedi}},\ and\ \bibinfo {author} {\bibfnamefont {P.}~\bibnamefont
  {Siano}},\ }\bibfield  {title} {\bibinfo {title} {Communication in smart
  grids: A comprehensive review on the existing and future communication and
  information infrastructures},\ }\href
  {https://doi.org/10.1109/JSYST.2019.2928090} {\bibfield  {journal} {\bibinfo
  {journal} {IEEE Systems Journal}\ }\textbf {\bibinfo {volume} {13}},\
  \bibinfo {pages} {4001} (\bibinfo {year} {2019}{\natexlab{b}})}\BibitemShut
  {NoStop}%
\bibitem [{\citenamefont {Yan}\ \emph {et~al.}(2013)\citenamefont {Yan},
  \citenamefont {Qian}, \citenamefont {Sharif},\ and\ \citenamefont
  {Tipper}}]{sg-communication-survey-2013}%
  \BibitemOpen
  \bibfield  {author} {\bibinfo {author} {\bibfnamefont {Y.}~\bibnamefont
  {Yan}}, \bibinfo {author} {\bibfnamefont {Y.}~\bibnamefont {Qian}}, \bibinfo
  {author} {\bibfnamefont {H.}~\bibnamefont {Sharif}},\ and\ \bibinfo {author}
  {\bibfnamefont {D.}~\bibnamefont {Tipper}},\ }\bibfield  {title} {\bibinfo
  {title} {A survey on smart grid communication infrastructures: Motivations,
  requirements and challenges},\ }\href
  {https://doi.org/10.1109/SURV.2012.021312.00034} {\bibfield  {journal}
  {\bibinfo  {journal} {{IEEE Communications Surveys \& Tutorials}}\ }\textbf
  {\bibinfo {volume} {15}},\ \bibinfo {pages} {5} (\bibinfo {year}
  {2013})}\BibitemShut {NoStop}%
\bibitem [{\citenamefont {Myoung}\ \emph {et~al.}(2010)\citenamefont {Myoung},
  \citenamefont {Kim},\ and\ \citenamefont {Lee}}]{3ms-for-trips}%
  \BibitemOpen
  \bibfield  {author} {\bibinfo {author} {\bibfnamefont {N.}~\bibnamefont
  {Myoung}}, \bibinfo {author} {\bibfnamefont {Y.}~\bibnamefont {Kim}},\ and\
  \bibinfo {author} {\bibfnamefont {S.}~\bibnamefont {Lee}},\ }\bibfield
  {title} {\bibinfo {title} {{The Design of Communication Infrastructures for
  Smart DAS and AMI}},\ }in\ \href {https://doi.org/10.1109/ICTC.2010.5674796}
  {\emph {\bibinfo {booktitle} {2010 International Conference on Information
  and Communication Technology Convergence (ICTC)}}}\ (\bibinfo {year} {2010})\
  pp.\ \bibinfo {pages} {461--462}\BibitemShut {NoStop}%
\bibitem [{\citenamefont {Yıldırım}\ \emph {et~al.}(2025)\citenamefont
  {Yıldırım}, \citenamefont {Yalman}, \citenamefont {Bayındır},\ and\
  \citenamefont {Terciyanlı}}]{Yldrm2025}%
  \BibitemOpen
  \bibfield  {author} {\bibinfo {author} {\bibfnamefont {F.}~\bibnamefont
  {Yıldırım}}, \bibinfo {author} {\bibfnamefont {Y.}~\bibnamefont {Yalman}},
  \bibinfo {author} {\bibfnamefont {K.~u.}\ \bibnamefont {Bayındır}},\ and\
  \bibinfo {author} {\bibfnamefont {E.}~\bibnamefont {Terciyanlı}},\
  }\bibfield  {title} {\bibinfo {title} {Comprehensive review of edge computing
  for power systems: State of the art, architecture, and applications},\ }\href
  {https://doi.org/10.3390/app15084592} {\bibfield  {journal} {\bibinfo
  {journal} {Applied Sciences}\ }\textbf {\bibinfo {volume} {15}},\ \bibinfo
  {pages} {4592} (\bibinfo {year} {2025})}\BibitemShut {NoStop}%
\bibitem [{\citenamefont {Kuzlu}\ and\ \citenamefont
  {Pipattanasomporn}(2013)}]{communication-sg-2013}%
  \BibitemOpen
  \bibfield  {author} {\bibinfo {author} {\bibfnamefont {M.}~\bibnamefont
  {Kuzlu}}\ and\ \bibinfo {author} {\bibfnamefont {M.}~\bibnamefont
  {Pipattanasomporn}},\ }\bibfield  {title} {\bibinfo {title} {Assessment of
  communication technologies and network requirements for different smart grid
  applications},\ }in\ \href {https://doi.org/10.1109/ISGT.2013.6497873} {\emph
  {\bibinfo {booktitle} {2013 IEEE PES Innovative Smart Grid Technologies
  Conference (ISGT)}}}\ (\bibinfo {year} {2013})\ pp.\ \bibinfo {pages}
  {1--6}\BibitemShut {NoStop}%
\bibitem [{\citenamefont {ETSI}(2010)}]{etsi-report}%
  \BibitemOpen
  \bibfield  {author} {\bibinfo {author} {\bibnamefont {ETSI}},\ }\href@noop {}
  {\emph {\bibinfo {title} {Draft TR 102 935, Version: 0.1.3, Machine to
  Machine Applicability of M2M Architecture to Smart Grid Networks}}},\
  \bibinfo {type} {Tech. Rep.}\ (\bibinfo  {institution} {ETSI},\ \bibinfo
  {year} {2010})\BibitemShut {NoStop}%
\bibitem [{\citenamefont {Saleem}\ \emph {et~al.}(2019)\citenamefont {Saleem},
  \citenamefont {Crespi}, \citenamefont {Rehmani},\ and\ \citenamefont
  {Copeland}}]{iot-aided-sg}%
  \BibitemOpen
  \bibfield  {author} {\bibinfo {author} {\bibfnamefont {Y.}~\bibnamefont
  {Saleem}}, \bibinfo {author} {\bibfnamefont {N.}~\bibnamefont {Crespi}},
  \bibinfo {author} {\bibfnamefont {M.~H.}\ \bibnamefont {Rehmani}},\ and\
  \bibinfo {author} {\bibfnamefont {R.}~\bibnamefont {Copeland}},\ }\bibfield
  {title} {\bibinfo {title} {Internet of things-aided smart grid: Technologies,
  architectures, applications, prototypes, and future research directions},\
  }\href {https://doi.org/10.1109/ACCESS.2019.2913984} {\bibfield  {journal}
  {\bibinfo  {journal} {IEEE Access}\ }\textbf {\bibinfo {volume} {7}},\
  \bibinfo {pages} {62962} (\bibinfo {year} {2019})}\BibitemShut {NoStop}%
\bibitem [{\citenamefont {Gooi}\ \emph {et~al.}(2023)\citenamefont {Gooi},
  \citenamefont {Wang},\ and\ \citenamefont {Tang}}]{edge-intelligence-sg}%
  \BibitemOpen
  \bibfield  {author} {\bibinfo {author} {\bibfnamefont {H.~B.}\ \bibnamefont
  {Gooi}}, \bibinfo {author} {\bibfnamefont {T.}~\bibnamefont {Wang}},\ and\
  \bibinfo {author} {\bibfnamefont {Y.}~\bibnamefont {Tang}},\ }\bibfield
  {title} {\bibinfo {title} {Edge intelligence for smart grid: A survey on
  application potentials},\ }\href
  {https://doi.org/10.17775/CSEEJPES.2022.02210} {\bibfield  {journal}
  {\bibinfo  {journal} {CSEE Journal of Power and Energy Systems}\ }\textbf
  {\bibinfo {volume} {9}},\ \bibinfo {pages} {1623} (\bibinfo {year}
  {2023})}\BibitemShut {NoStop}%
\bibitem [{\citenamefont {Khan}\ \emph {et~al.}(2023)\citenamefont {Khan},
  \citenamefont {Saleh}, \citenamefont {Waseem},\ and\ \citenamefont
  {Sajjad}}]{AI-Demand-Response}%
  \BibitemOpen
  \bibfield  {author} {\bibinfo {author} {\bibfnamefont {M.~A.}\ \bibnamefont
  {Khan}}, \bibinfo {author} {\bibfnamefont {A.~M.}\ \bibnamefont {Saleh}},
  \bibinfo {author} {\bibfnamefont {M.}~\bibnamefont {Waseem}},\ and\ \bibinfo
  {author} {\bibfnamefont {I.~A.}\ \bibnamefont {Sajjad}},\ }\bibfield  {title}
  {\bibinfo {title} {{Artificial Intelligence Enabled Demand Response:
  Prospects and Challenges in Smart Grid Environment}},\ }\href
  {https://doi.org/10.1109/ACCESS.2022.3231444} {\bibfield  {journal} {\bibinfo
   {journal} {IEEE Access}\ }\textbf {\bibinfo {volume} {11}},\ \bibinfo
  {pages} {1477} (\bibinfo {year} {2023})}\BibitemShut {NoStop}%
\bibitem [{\citenamefont {Bakkar}\ \emph {et~al.}(2022)\citenamefont {Bakkar},
  \citenamefont {Bogarra}, \citenamefont {Córcoles}, \citenamefont
  {Aboelhassan}, \citenamefont {Wang},\ and\ \citenamefont
  {Iglesias}}]{AI-Grid-Protection}%
  \BibitemOpen
  \bibfield  {author} {\bibinfo {author} {\bibfnamefont {M.}~\bibnamefont
  {Bakkar}}, \bibinfo {author} {\bibfnamefont {S.}~\bibnamefont {Bogarra}},
  \bibinfo {author} {\bibfnamefont {F.}~\bibnamefont {Córcoles}}, \bibinfo
  {author} {\bibfnamefont {A.}~\bibnamefont {Aboelhassan}}, \bibinfo {author}
  {\bibfnamefont {S.}~\bibnamefont {Wang}},\ and\ \bibinfo {author}
  {\bibfnamefont {J.}~\bibnamefont {Iglesias}},\ }\bibfield  {title} {\bibinfo
  {title} {{Artificial Intelligence-Based Protection for Smart Grids}},\
  }\bibfield  {journal} {\bibinfo  {journal} {Energies}\ }\textbf {\bibinfo
  {volume} {15}},\ \href {https://doi.org/10.3390/en15134933}
  {10.3390/en15134933} (\bibinfo {year} {2022})\BibitemShut {NoStop}%
\bibitem [{\citenamefont {Omitaomu}\ and\ \citenamefont
  {Niu}(2021)}]{AI-SmartGrid-Survey}%
  \BibitemOpen
  \bibfield  {author} {\bibinfo {author} {\bibfnamefont {O.~A.}\ \bibnamefont
  {Omitaomu}}\ and\ \bibinfo {author} {\bibfnamefont {H.}~\bibnamefont {Niu}},\
  }\bibfield  {title} {\bibinfo {title} {Artificial intelligence techniques in
  smart grid: A survey},\ }\href {https://www.mdpi.com/2624-6511/4/2/29}
  {\bibfield  {journal} {\bibinfo  {journal} {Smart Cities}\ }\textbf {\bibinfo
  {volume} {4}},\ \bibinfo {pages} {548} (\bibinfo {year} {2021})}\BibitemShut
  {NoStop}%
\bibitem [{\citenamefont {Khan}\ \emph {et~al.}(2021)\citenamefont {Khan},
  \citenamefont {Saad}, \citenamefont {Han}, \citenamefont {Hossain},\ and\
  \citenamefont {Hong}}]{federated-learning-overview}%
  \BibitemOpen
  \bibfield  {author} {\bibinfo {author} {\bibfnamefont {L.~U.}\ \bibnamefont
  {Khan}}, \bibinfo {author} {\bibfnamefont {W.}~\bibnamefont {Saad}}, \bibinfo
  {author} {\bibfnamefont {Z.}~\bibnamefont {Han}}, \bibinfo {author}
  {\bibfnamefont {E.}~\bibnamefont {Hossain}},\ and\ \bibinfo {author}
  {\bibfnamefont {C.~S.}\ \bibnamefont {Hong}},\ }\bibfield  {title} {\bibinfo
  {title} {{Federated Learning for Internet of Things: Recent Advances,
  Taxonomy, and Open Challenges}},\ }\href
  {https://doi.org/10.1109/COMST.2021.3090430} {\bibfield  {journal} {\bibinfo
  {journal} {{IEEE Communications Surveys \& Tutorials}}\ }\textbf {\bibinfo
  {volume} {23}},\ \bibinfo {pages} {1759} (\bibinfo {year}
  {2021})}\BibitemShut {NoStop}%
\bibitem [{\citenamefont {Adam}\ \emph {et~al.}(2024)\citenamefont {Adam},
  \citenamefont {Hammoudeh}, \citenamefont {Alrawashdeh},\ and\ \citenamefont
  {Alsulaimy}}]{IoT-Security-Survey-1}%
  \BibitemOpen
  \bibfield  {author} {\bibinfo {author} {\bibfnamefont {M.}~\bibnamefont
  {Adam}}, \bibinfo {author} {\bibfnamefont {M.}~\bibnamefont {Hammoudeh}},
  \bibinfo {author} {\bibfnamefont {R.}~\bibnamefont {Alrawashdeh}},\ and\
  \bibinfo {author} {\bibfnamefont {B.}~\bibnamefont {Alsulaimy}},\ }\bibfield
  {title} {\bibinfo {title} {{A Survey on Security, Privacy, Trust, and
  Architectural Challenges in IoT Systems}},\ }\href
  {https://doi.org/10.1109/ACCESS.2024.3382709} {\bibfield  {journal} {\bibinfo
   {journal} {IEEE Access}\ }\textbf {\bibinfo {volume} {12}},\ \bibinfo
  {pages} {57128} (\bibinfo {year} {2024})}\BibitemShut {NoStop}%
\bibitem [{\citenamefont {Analytics}(2024)}]{iot-analytics}%
  \BibitemOpen
  \bibfield  {author} {\bibinfo {author} {\bibfnamefont {I.}~\bibnamefont
  {Analytics}},\ }\href
  {https://iot-analytics.com/number-connected-iot-devices/} {\bibinfo {title}
  {State of iot 2024: Number of connected iot devices growing 13
  billion globally}} (\bibinfo {year} {2024})\BibitemShut {NoStop}%
\bibitem [{\citenamefont {Kolehmainen}(2018)}]{IoT-Firmware-Updates-1}%
  \BibitemOpen
  \bibfield  {author} {\bibinfo {author} {\bibfnamefont {A.}~\bibnamefont
  {Kolehmainen}},\ }\bibfield  {title} {\bibinfo {title} {{Secure Firmware
  Updates for IoT: A Survey}},\ }in\ \href
  {https://doi.org/10.1109/Cybermatics_2018.2018.00051} {\emph {\bibinfo
  {booktitle} {2018 IEEE iThings / GreenCom / CPSCom / SmartData}}}\ (\bibinfo
  {year} {2018})\ pp.\ \bibinfo {pages} {112--117}\BibitemShut {NoStop}%
\bibitem [{\citenamefont {Arakadakis}\ \emph {et~al.}(2021)\citenamefont
  {Arakadakis}, \citenamefont {Charalampidis}, \citenamefont {Makrogiannakis},\
  and\ \citenamefont {Fragkiadakis}}]{IoT-Firmware-Updates-2}%
  \BibitemOpen
  \bibfield  {author} {\bibinfo {author} {\bibfnamefont {K.}~\bibnamefont
  {Arakadakis}}, \bibinfo {author} {\bibfnamefont {P.}~\bibnamefont
  {Charalampidis}}, \bibinfo {author} {\bibfnamefont {A.}~\bibnamefont
  {Makrogiannakis}},\ and\ \bibinfo {author} {\bibfnamefont {A.}~\bibnamefont
  {Fragkiadakis}},\ }\bibfield  {title} {\bibinfo {title} {{Firmware
  Over-the-air Programming Techniques for IoT Networks - A Survey}},\
  }\bibfield  {journal} {\bibinfo  {journal} {ACM Comput. Surv.}\ }\textbf
  {\bibinfo {volume} {54}},\ \href {https://doi.org/10.1145/3472292}
  {10.1145/3472292} (\bibinfo {year} {2021})\BibitemShut {NoStop}%
\bibitem [{\citenamefont {Ahn}\ \emph {et~al.}(2022)\citenamefont {Ahn},
  \citenamefont {Kwon}, \citenamefont {Ahn}, \citenamefont {Park},
  \citenamefont {Kim}, \citenamefont {Lee}, \citenamefont {Kim},\ and\
  \citenamefont {Chung}}]{ahn_toward_2022}%
  \BibitemOpen
  \bibfield  {author} {\bibinfo {author} {\bibfnamefont {J.}~\bibnamefont
  {Ahn}}, \bibinfo {author} {\bibfnamefont {H.-Y.}\ \bibnamefont {Kwon}},
  \bibinfo {author} {\bibfnamefont {B.}~\bibnamefont {Ahn}}, \bibinfo {author}
  {\bibfnamefont {K.}~\bibnamefont {Park}}, \bibinfo {author} {\bibfnamefont
  {T.}~\bibnamefont {Kim}}, \bibinfo {author} {\bibfnamefont {M.-K.}\
  \bibnamefont {Lee}}, \bibinfo {author} {\bibfnamefont {J.}~\bibnamefont
  {Kim}},\ and\ \bibinfo {author} {\bibfnamefont {J.}~\bibnamefont {Chung}},\
  }\bibfield  {title} {\bibinfo {title} {Toward quantum secured distributed
  energy resources: Adoption of post-quantum cryptography (pqc) and quantum key
  distribution (qkd)},\ }\href {https://doi.org/10.3390/en15030714} {\bibfield
  {journal} {\bibinfo  {journal} {Energies}\ }\textbf {\bibinfo {volume}
  {15}},\ \bibinfo {pages} {714} (\bibinfo {year} {2022})}\BibitemShut
  {NoStop}%
\bibitem [{\citenamefont {Feynman}\ \emph {et~al.}(1963)\citenamefont
  {Feynman}, \citenamefont {Leighton},\ and\ \citenamefont
  {Sands}}]{Feynman1963Lectures}%
  \BibitemOpen
  \bibfield  {author} {\bibinfo {author} {\bibfnamefont {R.~P.}\ \bibnamefont
  {Feynman}}, \bibinfo {author} {\bibfnamefont {R.~B.}\ \bibnamefont
  {Leighton}},\ and\ \bibinfo {author} {\bibfnamefont {M.}~\bibnamefont
  {Sands}},\ }\href@noop {} {\emph {\bibinfo {title} {The Feynman Lectures on
  Physics, Vol. I: Mainly Mechanics, Radiation, and Heat}}}\ (\bibinfo
  {publisher} {Addison-Wesley},\ \bibinfo {year} {1963})\BibitemShut {NoStop}%
\bibitem [{\citenamefont {Halliday}\ \emph {et~al.}(2014)\citenamefont
  {Halliday}, \citenamefont {Resnick},\ and\ \citenamefont
  {Walker}}]{Halliday2014Fundamentals}%
  \BibitemOpen
  \bibfield  {author} {\bibinfo {author} {\bibfnamefont {D.}~\bibnamefont
  {Halliday}}, \bibinfo {author} {\bibfnamefont {R.}~\bibnamefont {Resnick}},\
  and\ \bibinfo {author} {\bibfnamefont {J.}~\bibnamefont {Walker}},\
  }\href@noop {} {\emph {\bibinfo {title} {Fundamentals of Physics}}},\
  \bibinfo {edition} {10th}\ ed.\ (\bibinfo  {publisher} {Wiley},\ \bibinfo
  {year} {2014})\BibitemShut {NoStop}%
\bibitem [{\citenamefont {Kleppner}\ and\ \citenamefont
  {Jackiw}(2000)}]{Kleppner2000OneHundred}%
  \BibitemOpen
  \bibfield  {author} {\bibinfo {author} {\bibfnamefont {D.}~\bibnamefont
  {Kleppner}}\ and\ \bibinfo {author} {\bibfnamefont {R.}~\bibnamefont
  {Jackiw}},\ }\bibfield  {title} {\bibinfo {title} {One hundred years of
  quantum physics},\ }\href {https://doi.org/10.1126/science.289.5481.893}
  {\bibfield  {journal} {\bibinfo  {journal} {Science}\ }\textbf {\bibinfo
  {volume} {289}},\ \bibinfo {pages} {893–898} (\bibinfo {year}
  {2000})}\BibitemShut {NoStop}%
\bibitem [{\citenamefont {Dowling}\ and\ \citenamefont
  {Milburn}(2003)}]{Dowling2003Quantum}%
  \BibitemOpen
  \bibfield  {author} {\bibinfo {author} {\bibfnamefont {J.~P.}\ \bibnamefont
  {Dowling}}\ and\ \bibinfo {author} {\bibfnamefont {G.~J.}\ \bibnamefont
  {Milburn}},\ }\bibfield  {title} {\bibinfo {title} {Quantum technology: the
  second quantum revolution},\ }\href {https://doi.org/10.1098/rsta.2003.1227}
  {\bibfield  {journal} {\bibinfo  {journal} {Philosophical Transactions of the
  Royal Society of London. Series A: Mathematical, Physical and Engineering
  Sciences}\ }\textbf {\bibinfo {volume} {361}},\ \bibinfo {pages}
  {1655–1674} (\bibinfo {year} {2003})}\BibitemShut {NoStop}%
\bibitem [{\citenamefont {Wiseman}\ and\ \citenamefont
  {Milburn}(2009)}]{Wiseman2009QuantumMeasurement}%
  \BibitemOpen
  \bibfield  {author} {\bibinfo {author} {\bibfnamefont {H.~M.}\ \bibnamefont
  {Wiseman}}\ and\ \bibinfo {author} {\bibfnamefont {G.~J.}\ \bibnamefont
  {Milburn}},\ }\href {https://doi.org/10.1017/cbo9780511813948} {\emph
  {\bibinfo {title} {Quantum Measurement and Control}}}\ (\bibinfo  {publisher}
  {Cambridge University Press},\ \bibinfo {year} {2009})\BibitemShut {NoStop}%
\bibitem [{\citenamefont {Deutsch}(2020)}]{Deutsch2020SecondQuantumRevolution}%
  \BibitemOpen
  \bibfield  {author} {\bibinfo {author} {\bibfnamefont {I.~H.}\ \bibnamefont
  {Deutsch}},\ }\bibfield  {title} {\bibinfo {title} {Harnessing the power of
  the second quantum revolution},\ }\bibfield  {journal} {\bibinfo  {journal}
  {PRX Quantum}\ }\textbf {\bibinfo {volume} {1}},\ \href
  {https://doi.org/10.1103/prxquantum.1.020101} {10.1103/prxquantum.1.020101}
  (\bibinfo {year} {2020})\BibitemShut {NoStop}%
\bibitem [{\citenamefont {Sævarsson}\ \emph {et~al.}(2022)\citenamefont
  {Sævarsson}, \citenamefont {Chatzivasileiadis}, \citenamefont
  {Jóhannsson},\ and\ \citenamefont {Østergaard}}]{Bryn2022}%
  \BibitemOpen
  \bibfield  {author} {\bibinfo {author} {\bibfnamefont {B.}~\bibnamefont
  {Sævarsson}}, \bibinfo {author} {\bibfnamefont {S.}~\bibnamefont
  {Chatzivasileiadis}}, \bibinfo {author} {\bibfnamefont {H.}~\bibnamefont
  {Jóhannsson}},\ and\ \bibinfo {author} {\bibfnamefont {J.}~\bibnamefont
  {Østergaard}},\ }\href {https://doi.org/10.48550/ARXIV.2204.14028} {\bibinfo
  {title} {Quantum computing for power flow algorithms: Testing on real quantum
  computers}} (\bibinfo {year} {2022})\BibitemShut {NoStop}%
\bibitem [{\citenamefont {Ganeshamurthy}\ \emph
  {et~al.}(2024{\natexlab{b}})\citenamefont {Ganeshamurthy}, \citenamefont
  {Ghosh}, \citenamefont {O'Meara}, \citenamefont {Cortiana}, \citenamefont
  {Schiefelbein-Lach},\ and\ \citenamefont {Monti}}]{Priyanka2025}%
  \BibitemOpen
  \bibfield  {author} {\bibinfo {author} {\bibfnamefont {P.~A.}\ \bibnamefont
  {Ganeshamurthy}}, \bibinfo {author} {\bibfnamefont {K.}~\bibnamefont
  {Ghosh}}, \bibinfo {author} {\bibfnamefont {C.}~\bibnamefont {O'Meara}},
  \bibinfo {author} {\bibfnamefont {G.}~\bibnamefont {Cortiana}}, \bibinfo
  {author} {\bibfnamefont {J.}~\bibnamefont {Schiefelbein-Lach}},\ and\
  \bibinfo {author} {\bibfnamefont {A.}~\bibnamefont {Monti}},\ }\href
  {https://doi.org/10.48550/ARXIV.2411.09123} {\bibinfo {title} {Quantum
  multi-output gaussian processes based machine learning for line parameter
  estimation in electrical grids}} (\bibinfo {year}
  {2024}{\natexlab{b}})\BibitemShut {NoStop}%
\bibitem [{\citenamefont {Hatano}\ \emph {et~al.}(2022)\citenamefont {Hatano},
  \citenamefont {Shin}, \citenamefont {Tanigawa}, \citenamefont {Shigenobu},
  \citenamefont {Nakazono}, \citenamefont {Sekiguchi}, \citenamefont {Onoda},
  \citenamefont {Ohshima}, \citenamefont {Arai}, \citenamefont {Iwasaki},\ and\
  \citenamefont {Hatano}}]{Hatano2022EV}%
  \BibitemOpen
  \bibfield  {author} {\bibinfo {author} {\bibfnamefont {Y.}~\bibnamefont
  {Hatano}}, \bibinfo {author} {\bibfnamefont {J.}~\bibnamefont {Shin}},
  \bibinfo {author} {\bibfnamefont {J.}~\bibnamefont {Tanigawa}}, \bibinfo
  {author} {\bibfnamefont {Y.}~\bibnamefont {Shigenobu}}, \bibinfo {author}
  {\bibfnamefont {A.}~\bibnamefont {Nakazono}}, \bibinfo {author}
  {\bibfnamefont {T.}~\bibnamefont {Sekiguchi}}, \bibinfo {author}
  {\bibfnamefont {S.}~\bibnamefont {Onoda}}, \bibinfo {author} {\bibfnamefont
  {T.}~\bibnamefont {Ohshima}}, \bibinfo {author} {\bibfnamefont
  {K.}~\bibnamefont {Arai}}, \bibinfo {author} {\bibfnamefont {T.}~\bibnamefont
  {Iwasaki}},\ and\ \bibinfo {author} {\bibfnamefont {M.}~\bibnamefont
  {Hatano}},\ }\bibfield  {title} {\bibinfo {title} {High-precision robust
  monitoring of charge/discharge current over a wide dynamic range for electric
  vehicle batteries using diamond quantum sensors},\ }\bibfield  {journal}
  {\bibinfo  {journal} {Scientific Reports}\ }\textbf {\bibinfo {volume}
  {12}},\ \href {https://doi.org/10.1038/s41598-022-18106-x}
  {10.1038/s41598-022-18106-x} (\bibinfo {year} {2022})\BibitemShut {NoStop}%
\bibitem [{\citenamefont {Kong}(2022)}]{kong2022qkd}%
  \BibitemOpen
  \bibfield  {author} {\bibinfo {author} {\bibfnamefont {P.-Y.}\ \bibnamefont
  {Kong}},\ }\bibfield  {title} {\bibinfo {title} {A review of quantum key
  distribution protocols in the perspective of smart grid communication
  security},\ }\href {https://doi.org/10.1109/JSYST.2020.3024956} {\bibfield
  {journal} {\bibinfo  {journal} {IEEE Systems Journal}\ }\textbf {\bibinfo
  {volume} {16}},\ \bibinfo {pages} {41} (\bibinfo {year} {2022})}\BibitemShut
  {NoStop}%
\bibitem [{\citenamefont {Chae}\ \emph {et~al.}(2024)\citenamefont {Chae},
  \citenamefont {Choi},\ and\ \citenamefont {Kim}}]{Chae2024}%
  \BibitemOpen
  \bibfield  {author} {\bibinfo {author} {\bibfnamefont {E.}~\bibnamefont
  {Chae}}, \bibinfo {author} {\bibfnamefont {J.}~\bibnamefont {Choi}},\ and\
  \bibinfo {author} {\bibfnamefont {J.}~\bibnamefont {Kim}},\ }\bibfield
  {title} {\bibinfo {title} {An elementary review on basic principles and
  developments of qubits for quantum computing},\ }\bibfield  {journal}
  {\bibinfo  {journal} {Nano Convergence}\ }\textbf {\bibinfo {volume} {11}},\
  \href {https://doi.org/10.1186/s40580-024-00418-5}
  {10.1186/s40580-024-00418-5} (\bibinfo {year} {2024})\BibitemShut {NoStop}%
\bibitem [{\citenamefont {Yang}\ \emph {et~al.}(2023)\citenamefont {Yang},
  \citenamefont {Zolanvari},\ and\ \citenamefont {Jain}}]{Q-CompCommSurvey}%
  \BibitemOpen
  \bibfield  {author} {\bibinfo {author} {\bibfnamefont {Z.}~\bibnamefont
  {Yang}}, \bibinfo {author} {\bibfnamefont {M.}~\bibnamefont {Zolanvari}},\
  and\ \bibinfo {author} {\bibfnamefont {R.}~\bibnamefont {Jain}},\ }\bibfield
  {title} {\bibinfo {title} {{A Survey of Important Issues in Quantum Computing
  and Communications}},\ }\href {https://doi.org/10.1109/COMST.2023.3254481}
  {\bibfield  {journal} {\bibinfo  {journal} {{IEEE Communications Surveys \&
  Tutorials}}\ }\textbf {\bibinfo {volume} {25}},\ \bibinfo {pages} {1059}
  (\bibinfo {year} {2023})}\BibitemShut {NoStop}%
\bibitem [{\citenamefont {Dalzell}\ \emph {et~al.}(2023)\citenamefont
  {Dalzell}, \citenamefont {McArdle}, \citenamefont {Berta}, \citenamefont
  {Bienias}, \citenamefont {Chen}, \citenamefont {Gilyén}, \citenamefont
  {Hann}, \citenamefont {Kastoryano}, \citenamefont {Khabiboulline},
  \citenamefont {Kubica}, \citenamefont {Salton}, \citenamefont {Wang},\ and\
  \citenamefont {Brandão}}]{Dalzell2023}%
  \BibitemOpen
  \bibfield  {author} {\bibinfo {author} {\bibfnamefont {A.~M.}\ \bibnamefont
  {Dalzell}}, \bibinfo {author} {\bibfnamefont {S.}~\bibnamefont {McArdle}},
  \bibinfo {author} {\bibfnamefont {M.}~\bibnamefont {Berta}}, \bibinfo
  {author} {\bibfnamefont {P.}~\bibnamefont {Bienias}}, \bibinfo {author}
  {\bibfnamefont {C.-F.}\ \bibnamefont {Chen}}, \bibinfo {author}
  {\bibfnamefont {A.}~\bibnamefont {Gilyén}}, \bibinfo {author} {\bibfnamefont
  {C.~T.}\ \bibnamefont {Hann}}, \bibinfo {author} {\bibfnamefont {M.~J.}\
  \bibnamefont {Kastoryano}}, \bibinfo {author} {\bibfnamefont {E.~T.}\
  \bibnamefont {Khabiboulline}}, \bibinfo {author} {\bibfnamefont
  {A.}~\bibnamefont {Kubica}}, \bibinfo {author} {\bibfnamefont
  {G.}~\bibnamefont {Salton}}, \bibinfo {author} {\bibfnamefont
  {S.}~\bibnamefont {Wang}},\ and\ \bibinfo {author} {\bibfnamefont {F.~G.
  S.~L.}\ \bibnamefont {Brandão}},\ }\href
  {https://doi.org/10.48550/ARXIV.2310.03011} {\bibinfo {title} {Quantum
  algorithms: A survey of applications and end-to-end complexities}} (\bibinfo
  {year} {2023})\BibitemShut {NoStop}%
\bibitem [{\citenamefont {Biamonte}\ \emph {et~al.}(2017)\citenamefont
  {Biamonte}, \citenamefont {Wittek}, \citenamefont {Pancotti}, \citenamefont
  {Rebentrost}, \citenamefont {Wiebe},\ and\ \citenamefont
  {Lloyd}}]{Biamonte2017}%
  \BibitemOpen
  \bibfield  {author} {\bibinfo {author} {\bibfnamefont {J.}~\bibnamefont
  {Biamonte}}, \bibinfo {author} {\bibfnamefont {P.}~\bibnamefont {Wittek}},
  \bibinfo {author} {\bibfnamefont {N.}~\bibnamefont {Pancotti}}, \bibinfo
  {author} {\bibfnamefont {P.}~\bibnamefont {Rebentrost}}, \bibinfo {author}
  {\bibfnamefont {N.}~\bibnamefont {Wiebe}},\ and\ \bibinfo {author}
  {\bibfnamefont {S.}~\bibnamefont {Lloyd}},\ }\bibfield  {title} {\bibinfo
  {title} {Quantum machine learning},\ }\href
  {https://doi.org/10.1038/nature23474} {\bibfield  {journal} {\bibinfo
  {journal} {Nature}\ }\textbf {\bibinfo {volume} {549}},\ \bibinfo {pages}
  {195–202} (\bibinfo {year} {2017})}\BibitemShut {NoStop}%
\bibitem [{\citenamefont {Fedorov}\ \emph {et~al.}(2022)\citenamefont
  {Fedorov}, \citenamefont {Gisin}, \citenamefont {Beloussov},\ and\
  \citenamefont {Lvovsky}}]{Fedorov2022}%
  \BibitemOpen
  \bibfield  {author} {\bibinfo {author} {\bibfnamefont {A.~K.}\ \bibnamefont
  {Fedorov}}, \bibinfo {author} {\bibfnamefont {N.}~\bibnamefont {Gisin}},
  \bibinfo {author} {\bibfnamefont {S.~M.}\ \bibnamefont {Beloussov}},\ and\
  \bibinfo {author} {\bibfnamefont {A.~I.}\ \bibnamefont {Lvovsky}},\ }\href
  {https://doi.org/10.48550/ARXIV.2203.17181} {\bibinfo {title} {Quantum
  computing at the quantum advantage threshold: a down-to-business review}}
  (\bibinfo {year} {2022})\BibitemShut {NoStop}%
\bibitem [{\citenamefont {Aaronson}(2015)}]{Aaronson2015}%
  \BibitemOpen
  \bibfield  {author} {\bibinfo {author} {\bibfnamefont {S.}~\bibnamefont
  {Aaronson}},\ }\bibfield  {title} {\bibinfo {title} {Read the fine print},\
  }\href {https://doi.org/10.1038/nphys3272} {\bibfield  {journal} {\bibinfo
  {journal} {Nature Physics}\ }\textbf {\bibinfo {volume} {11}},\ \bibinfo
  {pages} {291–293} (\bibinfo {year} {2015})}\BibitemShut {NoStop}%
\bibitem [{\citenamefont {Hoefler}\ \emph {et~al.}(2023)\citenamefont
  {Hoefler}, \citenamefont {H\"{a}ner},\ and\ \citenamefont
  {Troyer}}]{Hoefler2023}%
  \BibitemOpen
  \bibfield  {author} {\bibinfo {author} {\bibfnamefont {T.}~\bibnamefont
  {Hoefler}}, \bibinfo {author} {\bibfnamefont {T.}~\bibnamefont {H\"{a}ner}},\
  and\ \bibinfo {author} {\bibfnamefont {M.}~\bibnamefont {Troyer}},\
  }\bibfield  {title} {\bibinfo {title} {Disentangling hype from practicality:
  On realistically achieving quantum advantage},\ }\href
  {https://doi.org/10.1145/3571725} {\bibfield  {journal} {\bibinfo  {journal}
  {Communications of the ACM}\ }\textbf {\bibinfo {volume} {66}},\ \bibinfo
  {pages} {82–87} (\bibinfo {year} {2023})}\BibitemShut {NoStop}%
\bibitem [{\citenamefont {Bennett}\ \emph {et~al.}(1993)\citenamefont
  {Bennett}, \citenamefont {Brassard}, \citenamefont {Cr\'epeau}, \citenamefont
  {Jozsa}, \citenamefont {Peres},\ and\ \citenamefont
  {Wootters}}]{Q-Teleportation}%
  \BibitemOpen
  \bibfield  {author} {\bibinfo {author} {\bibfnamefont {C.~H.}\ \bibnamefont
  {Bennett}}, \bibinfo {author} {\bibfnamefont {G.}~\bibnamefont {Brassard}},
  \bibinfo {author} {\bibfnamefont {C.}~\bibnamefont {Cr\'epeau}}, \bibinfo
  {author} {\bibfnamefont {R.}~\bibnamefont {Jozsa}}, \bibinfo {author}
  {\bibfnamefont {A.}~\bibnamefont {Peres}},\ and\ \bibinfo {author}
  {\bibfnamefont {W.~K.}\ \bibnamefont {Wootters}},\ }\bibfield  {title}
  {\bibinfo {title} {{Teleporting an unknown quantum state via dual classical
  and Einstein-Podolsky-Rosen channels}},\ }\href
  {https://doi.org/10.1103/PhysRevLett.70.1895} {\bibfield  {journal} {\bibinfo
   {journal} {Phys. Rev. Lett.}\ }\textbf {\bibinfo {volume} {70}},\ \bibinfo
  {pages} {1895} (\bibinfo {year} {1993})}\BibitemShut {NoStop}%
\bibitem [{\citenamefont {Schlosshauer}(2019)}]{Q-Decoherence}%
  \BibitemOpen
  \bibfield  {author} {\bibinfo {author} {\bibfnamefont {M.}~\bibnamefont
  {Schlosshauer}},\ }\bibfield  {title} {\bibinfo {title} {Quantum
  decoherence},\ }\href
  {https://doi.org/https://doi.org/10.1016/j.physrep.2019.10.001} {\bibfield
  {journal} {\bibinfo  {journal} {Physics Reports}\ }\textbf {\bibinfo {volume}
  {831}},\ \bibinfo {pages} {1} (\bibinfo {year} {2019})}\BibitemShut {NoStop}%
\bibitem [{\citenamefont {Xu}\ \emph {et~al.}(2025)\citenamefont {Xu},
  \citenamefont {Han}, \citenamefont {Luo},\ and\ \citenamefont {Li}}]{Xu2025}%
  \BibitemOpen
  \bibfield  {author} {\bibinfo {author} {\bibfnamefont {Y.}~\bibnamefont
  {Xu}}, \bibinfo {author} {\bibfnamefont {X.}~\bibnamefont {Han}}, \bibinfo
  {author} {\bibfnamefont {R.}~\bibnamefont {Luo}},\ and\ \bibinfo {author}
  {\bibfnamefont {Z.}~\bibnamefont {Li}},\ }\bibfield  {title} {\bibinfo
  {title} {Hybrid quantum-classical stochastic programming for co-planning 5g
  base stations and photovoltaic power stations in urban communities},\ }\href
  {https://doi.org/10.1038/s41598-025-26699-2} {\bibfield  {journal} {\bibinfo
  {journal} {Scientific Reports}\ }\textbf {\bibinfo {volume} {15}},\ \bibinfo
  {pages} {42642} (\bibinfo {year} {2025})}\BibitemShut {NoStop}%
\bibitem [{\citenamefont {Moniruzzaman}\ \emph {et~al.}(2025)\citenamefont
  {Moniruzzaman}, \citenamefont {Lu},\ and\ \citenamefont
  {Hossain}}]{Moniruzzaman2025Hybrid}%
  \BibitemOpen
  \bibfield  {author} {\bibinfo {author} {\bibfnamefont {M.}~\bibnamefont
  {Moniruzzaman}}, \bibinfo {author} {\bibfnamefont {C.}~\bibnamefont {Lu}},\
  and\ \bibinfo {author} {\bibfnamefont {M.~S.}\ \bibnamefont {Hossain}},\
  }\bibfield  {title} {\bibinfo {title} {Hybrid quantum-classical machine
  learning for energy-efficient electric bus charging optimization},\
  }\href@noop {} {\bibfield  {journal} {\bibinfo  {journal} {SSRN}\ } (\bibinfo
  {year} {2025})}\BibitemShut {NoStop}%
\bibitem [{\citenamefont {Zimborás}\ \emph {et~al.}(2025)\citenamefont
  {Zimborás}, \citenamefont {Koczor}, \citenamefont {Holmes}, \citenamefont
  {Borrelli}, \citenamefont {Gilyén}, \citenamefont {Huang}, \citenamefont
  {Cai}, \citenamefont {Acín}, \citenamefont {Aolita}, \citenamefont {Banchi},
  \citenamefont {Brandão}, \citenamefont {Cavalcanti}, \citenamefont {Cubitt},
  \citenamefont {Filippov}, \citenamefont {García-Pérez}, \citenamefont
  {Goold}, \citenamefont {Kálmán}, \citenamefont {Kyoseva}, \citenamefont
  {Rossi}, \citenamefont {Sokolov}, \citenamefont {Tavernelli},\ and\
  \citenamefont {Maniscalco}}]{Zimborás2024QCmyths}%
  \BibitemOpen
  \bibfield  {author} {\bibinfo {author} {\bibfnamefont {Z.}~\bibnamefont
  {Zimborás}}, \bibinfo {author} {\bibfnamefont {B.}~\bibnamefont {Koczor}},
  \bibinfo {author} {\bibfnamefont {Z.}~\bibnamefont {Holmes}}, \bibinfo
  {author} {\bibfnamefont {E.-M.}\ \bibnamefont {Borrelli}}, \bibinfo {author}
  {\bibfnamefont {A.}~\bibnamefont {Gilyén}}, \bibinfo {author} {\bibfnamefont
  {H.-Y.}\ \bibnamefont {Huang}}, \bibinfo {author} {\bibfnamefont
  {Z.}~\bibnamefont {Cai}}, \bibinfo {author} {\bibfnamefont {A.}~\bibnamefont
  {Acín}}, \bibinfo {author} {\bibfnamefont {L.}~\bibnamefont {Aolita}},
  \bibinfo {author} {\bibfnamefont {L.}~\bibnamefont {Banchi}}, \bibinfo
  {author} {\bibfnamefont {F.~G. S.~L.}\ \bibnamefont {Brandão}}, \bibinfo
  {author} {\bibfnamefont {D.}~\bibnamefont {Cavalcanti}}, \bibinfo {author}
  {\bibfnamefont {T.}~\bibnamefont {Cubitt}}, \bibinfo {author} {\bibfnamefont
  {S.~N.}\ \bibnamefont {Filippov}}, \bibinfo {author} {\bibfnamefont
  {G.}~\bibnamefont {García-Pérez}}, \bibinfo {author} {\bibfnamefont
  {J.}~\bibnamefont {Goold}}, \bibinfo {author} {\bibfnamefont
  {O.}~\bibnamefont {Kálmán}}, \bibinfo {author} {\bibfnamefont
  {E.}~\bibnamefont {Kyoseva}}, \bibinfo {author} {\bibfnamefont {M.~A.~C.}\
  \bibnamefont {Rossi}}, \bibinfo {author} {\bibfnamefont {B.}~\bibnamefont
  {Sokolov}}, \bibinfo {author} {\bibfnamefont {I.}~\bibnamefont
  {Tavernelli}},\ and\ \bibinfo {author} {\bibfnamefont {S.}~\bibnamefont
  {Maniscalco}},\ }\href {https://doi.org/10.48550/ARXIV.2501.05694} {\bibinfo
  {title} {Myths around quantum computation before full fault tolerance: What
  no-go theorems rule out and what they don't}} (\bibinfo {year}
  {2025})\BibitemShut {NoStop}%
\bibitem [{\citenamefont {Patel}\ \emph {et~al.}(2020)\citenamefont {Patel},
  \citenamefont {Potharaju}, \citenamefont {Li}, \citenamefont {Roy},\ and\
  \citenamefont {Tiwari}}]{patel_experimental_2020}%
  \BibitemOpen
  \bibfield  {author} {\bibinfo {author} {\bibfnamefont {T.}~\bibnamefont
  {Patel}}, \bibinfo {author} {\bibfnamefont {A.}~\bibnamefont {Potharaju}},
  \bibinfo {author} {\bibfnamefont {B.}~\bibnamefont {Li}}, \bibinfo {author}
  {\bibfnamefont {R.~B.}\ \bibnamefont {Roy}},\ and\ \bibinfo {author}
  {\bibfnamefont {D.}~\bibnamefont {Tiwari}},\ }\bibfield  {title} {\bibinfo
  {title} {Experimental {Evaluation} of {NISQ} {Quantum} {Computers}: {Error}
  {Measurement}, {Characterization}, and {Implications}},\ }in\ \href
  {https://doi.org/10.1109/SC41405.2020.00050} {\emph {\bibinfo {booktitle}
  {{SC20}: {International} {Conference} for {High} {Performance} {Computing},
  {Networking}, {Storage} and {Analysis}}}}\ (\bibinfo {year} {2020})\ pp.\
  \bibinfo {pages} {1--15}\BibitemShut {NoStop}%
\bibitem [{\citenamefont {Fowler}\ \emph {et~al.}(2012)\citenamefont {Fowler},
  \citenamefont {Mariantoni}, \citenamefont {Martinis},\ and\ \citenamefont
  {Cleland}}]{fowler_surface_2012}%
  \BibitemOpen
  \bibfield  {author} {\bibinfo {author} {\bibfnamefont {A.~G.}\ \bibnamefont
  {Fowler}}, \bibinfo {author} {\bibfnamefont {M.}~\bibnamefont {Mariantoni}},
  \bibinfo {author} {\bibfnamefont {J.~M.}\ \bibnamefont {Martinis}},\ and\
  \bibinfo {author} {\bibfnamefont {A.~N.}\ \bibnamefont {Cleland}},\
  }\bibfield  {title} {\bibinfo {title} {Surface codes: {Towards} practical
  large-scale quantum computation},\ }\href
  {https://doi.org/10.1103/PhysRevA.86.032324} {\bibfield  {journal} {\bibinfo
  {journal} {Phys. Rev. A}\ }\textbf {\bibinfo {volume} {86}},\ \bibinfo
  {pages} {032324} (\bibinfo {year} {2012})},\ \bibinfo {note} {publisher:
  American Physical Society}\BibitemShut {NoStop}%
\bibitem [{\citenamefont {Cai}\ \emph {et~al.}(2023)\citenamefont {Cai},
  \citenamefont {Babbush}, \citenamefont {Benjamin}, \citenamefont {Endo},
  \citenamefont {Huggins}, \citenamefont {Li}, \citenamefont {McClean},\ and\
  \citenamefont {O’Brien}}]{cai_quantum_2023}%
  \BibitemOpen
  \bibfield  {author} {\bibinfo {author} {\bibfnamefont {Z.}~\bibnamefont
  {Cai}}, \bibinfo {author} {\bibfnamefont {R.}~\bibnamefont {Babbush}},
  \bibinfo {author} {\bibfnamefont {S.~C.}\ \bibnamefont {Benjamin}}, \bibinfo
  {author} {\bibfnamefont {S.}~\bibnamefont {Endo}}, \bibinfo {author}
  {\bibfnamefont {W.~J.}\ \bibnamefont {Huggins}}, \bibinfo {author}
  {\bibfnamefont {Y.}~\bibnamefont {Li}}, \bibinfo {author} {\bibfnamefont
  {J.~R.}\ \bibnamefont {McClean}},\ and\ \bibinfo {author} {\bibfnamefont
  {T.~E.}\ \bibnamefont {O’Brien}},\ }\bibfield  {title} {\bibinfo {title}
  {Quantum error mitigation},\ }\href
  {https://doi.org/10.1103/RevModPhys.95.045005} {\bibfield  {journal}
  {\bibinfo  {journal} {Rev. Mod. Phys.}\ }\textbf {\bibinfo {volume} {95}},\
  \bibinfo {pages} {045005} (\bibinfo {year} {2023})},\ \bibinfo {note}
  {publisher: American Physical Society}\BibitemShut {NoStop}%
\bibitem [{\citenamefont {Herbst}\ \emph {et~al.}(2024)\citenamefont {Herbst},
  \citenamefont {De~Maio},\ and\ \citenamefont {Brandic}}]{Herbst2024}%
  \BibitemOpen
  \bibfield  {author} {\bibinfo {author} {\bibfnamefont {S.}~\bibnamefont
  {Herbst}}, \bibinfo {author} {\bibfnamefont {V.}~\bibnamefont {De~Maio}},\
  and\ \bibinfo {author} {\bibfnamefont {I.}~\bibnamefont {Brandic}},\
  }\bibinfo {title} {Streaming iot data and the quantum edge: A classic/quantum
  machine learning use case},\ in\ \href
  {https://doi.org/10.1007/978-3-031-50684-0_14} {\emph {\bibinfo {booktitle}
  {Euro-Par 2023: Parallel Processing Workshops}}}\ (\bibinfo  {publisher}
  {Springer Nature Switzerland},\ \bibinfo {year} {2024})\ p.\ \bibinfo {pages}
  {177–188}\BibitemShut {NoStop}%
\bibitem [{\citenamefont {Degen}\ \emph {et~al.}(2017)\citenamefont {Degen},
  \citenamefont {Reinhard},\ and\ \citenamefont
  {Cappellaro}}]{Degen2017QuantumSensing}%
  \BibitemOpen
  \bibfield  {author} {\bibinfo {author} {\bibfnamefont {C.~L.}\ \bibnamefont
  {Degen}}, \bibinfo {author} {\bibfnamefont {F.}~\bibnamefont {Reinhard}},\
  and\ \bibinfo {author} {\bibfnamefont {P.}~\bibnamefont {Cappellaro}},\
  }\bibfield  {title} {\bibinfo {title} {Quantum sensing},\ }\bibfield
  {journal} {\bibinfo  {journal} {Rev. Mod. Phys.}\ }\textbf {\bibinfo {volume}
  {89}},\ \href {https://doi.org/10.1103/revmodphys.89.035002}
  {10.1103/revmodphys.89.035002} (\bibinfo {year} {2017})\BibitemShut {NoStop}%
\bibitem [{\citenamefont {Doherty}\ \emph {et~al.}(2014)\citenamefont
  {Doherty}, \citenamefont {Struzhkin}, \citenamefont {Simpson}, \citenamefont
  {McGuinness}, \citenamefont {Meng}, \citenamefont {Stacey}, \citenamefont
  {Karle}, \citenamefont {Hemley}, \citenamefont {Manson}, \citenamefont
  {Hollenberg},\ and\ \citenamefont {Prawer}}]{Doherty2014ElectronicPressure}%
  \BibitemOpen
  \bibfield  {author} {\bibinfo {author} {\bibfnamefont {M.~W.}\ \bibnamefont
  {Doherty}}, \bibinfo {author} {\bibfnamefont {V.~V.}\ \bibnamefont
  {Struzhkin}}, \bibinfo {author} {\bibfnamefont {D.~A.}\ \bibnamefont
  {Simpson}}, \bibinfo {author} {\bibfnamefont {L.~P.}\ \bibnamefont
  {McGuinness}}, \bibinfo {author} {\bibfnamefont {Y.}~\bibnamefont {Meng}},
  \bibinfo {author} {\bibfnamefont {A.}~\bibnamefont {Stacey}}, \bibinfo
  {author} {\bibfnamefont {T.~J.}\ \bibnamefont {Karle}}, \bibinfo {author}
  {\bibfnamefont {R.~J.}\ \bibnamefont {Hemley}}, \bibinfo {author}
  {\bibfnamefont {N.~B.}\ \bibnamefont {Manson}}, \bibinfo {author}
  {\bibfnamefont {L.~C.}\ \bibnamefont {Hollenberg}},\ and\ \bibinfo {author}
  {\bibfnamefont {S.}~\bibnamefont {Prawer}},\ }\bibfield  {title} {\bibinfo
  {title} {{Electronic properties and metrology applications of the diamond
  {NV}$^-$ Center under pressure}},\ }\bibfield  {journal} {\bibinfo  {journal}
  {Phys. Rev. Lett.}\ }\textbf {\bibinfo {volume} {112}},\ \href
  {https://doi.org/10.1103/PHYSREVLETT.112.047601}
  {10.1103/PHYSREVLETT.112.047601} (\bibinfo {year} {2014})\BibitemShut
  {NoStop}%
\bibitem [{\citenamefont {Fang}\ \emph {et~al.}(2024)\citenamefont {Fang},
  \citenamefont {Wang}, \citenamefont {Zhou}, \citenamefont {Li}, \citenamefont
  {Zhang}, \citenamefont {Tang}, \citenamefont {Zhong}, \citenamefont {Hu},
  \citenamefont {Zhou}, \citenamefont {Chen}, \citenamefont {Wang},\ and\
  \citenamefont {Zhan}}]{Fang2024}%
  \BibitemOpen
  \bibfield  {author} {\bibinfo {author} {\bibfnamefont {J.}~\bibnamefont
  {Fang}}, \bibinfo {author} {\bibfnamefont {W.}~\bibnamefont {Wang}}, \bibinfo
  {author} {\bibfnamefont {Y.}~\bibnamefont {Zhou}}, \bibinfo {author}
  {\bibfnamefont {J.}~\bibnamefont {Li}}, \bibinfo {author} {\bibfnamefont
  {D.}~\bibnamefont {Zhang}}, \bibinfo {author} {\bibfnamefont
  {B.}~\bibnamefont {Tang}}, \bibinfo {author} {\bibfnamefont {J.}~\bibnamefont
  {Zhong}}, \bibinfo {author} {\bibfnamefont {J.}~\bibnamefont {Hu}}, \bibinfo
  {author} {\bibfnamefont {F.}~\bibnamefont {Zhou}}, \bibinfo {author}
  {\bibfnamefont {X.}~\bibnamefont {Chen}}, \bibinfo {author} {\bibfnamefont
  {J.}~\bibnamefont {Wang}},\ and\ \bibinfo {author} {\bibfnamefont
  {M.}~\bibnamefont {Zhan}},\ }\bibfield  {title} {\bibinfo {title} {Classical
  and atomic gravimetry},\ }\href {https://doi.org/10.3390/rs16142634}
  {\bibfield  {journal} {\bibinfo  {journal} {Remote Sensing}\ }\textbf
  {\bibinfo {volume} {16}},\ \bibinfo {pages} {2634} (\bibinfo {year}
  {2024})}\BibitemShut {NoStop}%
\bibitem [{\citenamefont {Oh}\ \emph {et~al.}(2024)\citenamefont {Oh},
  \citenamefont {Gregoire}, \citenamefont {Black}, \citenamefont
  {Jeramy~Hughes}, \citenamefont {Kunz}, \citenamefont {Larsen}, \citenamefont
  {Lautier-Gaud}, \citenamefont {Lee}, \citenamefont {Schwindt}, \citenamefont
  {Mouradian}, \citenamefont {Narducci},\ and\ \citenamefont
  {Sackett}}]{Oh2024}%
  \BibitemOpen
  \bibfield  {author} {\bibinfo {author} {\bibfnamefont {E.}~\bibnamefont
  {Oh}}, \bibinfo {author} {\bibfnamefont {M.~D.}\ \bibnamefont {Gregoire}},
  \bibinfo {author} {\bibfnamefont {A.~T.}\ \bibnamefont {Black}}, \bibinfo
  {author} {\bibfnamefont {K.}~\bibnamefont {Jeramy~Hughes}}, \bibinfo {author}
  {\bibfnamefont {P.~D.}\ \bibnamefont {Kunz}}, \bibinfo {author}
  {\bibfnamefont {M.}~\bibnamefont {Larsen}}, \bibinfo {author} {\bibfnamefont
  {J.}~\bibnamefont {Lautier-Gaud}}, \bibinfo {author} {\bibfnamefont
  {J.}~\bibnamefont {Lee}}, \bibinfo {author} {\bibfnamefont {P.~D.~D.}\
  \bibnamefont {Schwindt}}, \bibinfo {author} {\bibfnamefont {S.~L.}\
  \bibnamefont {Mouradian}}, \bibinfo {author} {\bibfnamefont {F.~A.}\
  \bibnamefont {Narducci}},\ and\ \bibinfo {author} {\bibfnamefont {C.~A.}\
  \bibnamefont {Sackett}},\ }\bibfield  {title} {\bibinfo {title} {Perspective
  on quantum sensors from basic research to commercial applications},\ }\href
  {https://doi.org/10.2514/1.j062707} {\bibfield  {journal} {\bibinfo
  {journal} {AIAA Journal}\ }\textbf {\bibinfo {volume} {62}},\ \bibinfo
  {pages} {4029–4053} (\bibinfo {year} {2024})}\BibitemShut {NoStop}%
\bibitem [{\citenamefont {Casola}\ \emph {et~al.}(2018)\citenamefont {Casola},
  \citenamefont {van~der Sar},\ and\ \citenamefont
  {Yacoby}}]{Casola2018ProbingCondensed}%
  \BibitemOpen
  \bibfield  {author} {\bibinfo {author} {\bibfnamefont {F.}~\bibnamefont
  {Casola}}, \bibinfo {author} {\bibfnamefont {T.}~\bibnamefont {van~der
  Sar}},\ and\ \bibinfo {author} {\bibfnamefont {A.}~\bibnamefont {Yacoby}},\
  }\bibfield  {title} {\bibinfo {title} {Probing condensed matter physics with
  magnetometry based on nitrogen-vacancy centres in diamond},\ }\bibfield
  {journal} {\bibinfo  {journal} {Nature Reviews Materials}\ }\textbf {\bibinfo
  {volume} {3}},\ \href {https://doi.org/10.1038/natrevmats.2017.88}
  {10.1038/natrevmats.2017.88} (\bibinfo {year} {2018})\BibitemShut {NoStop}%
\bibitem [{\citenamefont {Hayashi}\ \emph {et~al.}(2018)\citenamefont
  {Hayashi}, \citenamefont {Matsuzaki}, \citenamefont {Taniguchi},
  \citenamefont {Shimo-Oka}, \citenamefont {Nakamura}, \citenamefont {Onoda},
  \citenamefont {Ohshima}, \citenamefont {Morishita}, \citenamefont {Fujiwara},
  \citenamefont {Saito},\ and\ \citenamefont
  {Mizuochi}}]{Hayashi2018OptimizationEnsemble}%
  \BibitemOpen
  \bibfield  {author} {\bibinfo {author} {\bibfnamefont {K.}~\bibnamefont
  {Hayashi}}, \bibinfo {author} {\bibfnamefont {Y.}~\bibnamefont {Matsuzaki}},
  \bibinfo {author} {\bibfnamefont {T.}~\bibnamefont {Taniguchi}}, \bibinfo
  {author} {\bibfnamefont {T.}~\bibnamefont {Shimo-Oka}}, \bibinfo {author}
  {\bibfnamefont {I.}~\bibnamefont {Nakamura}}, \bibinfo {author}
  {\bibfnamefont {S.}~\bibnamefont {Onoda}}, \bibinfo {author} {\bibfnamefont
  {T.}~\bibnamefont {Ohshima}}, \bibinfo {author} {\bibfnamefont
  {H.}~\bibnamefont {Morishita}}, \bibinfo {author} {\bibfnamefont
  {M.}~\bibnamefont {Fujiwara}}, \bibinfo {author} {\bibfnamefont
  {S.}~\bibnamefont {Saito}},\ and\ \bibinfo {author} {\bibfnamefont
  {N.}~\bibnamefont {Mizuochi}},\ }\bibfield  {title} {\bibinfo {title}
  {{Optimization of Temperature Sensitivity Using the Optically Detected
  Magnetic-Resonance Spectrum of a Nitrogen-Vacancy Center Ensemble}},\ }\href
  {https://doi.org/10.1103/PhysRevApplied.10.034009} {\bibfield  {journal}
  {\bibinfo  {journal} {Phys. Rev. Appl.}\ }\textbf {\bibinfo {volume} {10}},\
  \bibinfo {pages} {034009} (\bibinfo {year} {2018})}\BibitemShut {NoStop}%
\bibitem [{\citenamefont {Alam}\ \emph {et~al.}(2024)\citenamefont {Alam},
  \citenamefont {Gorrini}, \citenamefont {Gawe\l{}czyk}, \citenamefont
  {Wigger}, \citenamefont {Coccia}, \citenamefont {Guo}, \citenamefont
  {Shahbazi}, \citenamefont {Bharadwaj}, \citenamefont {Kubanek}, \citenamefont
  {Ramponi}, \citenamefont {Barclay}, \citenamefont {Bennett}, \citenamefont
  {Hadden}, \citenamefont {Bifone}, \citenamefont {Eaton},\ and\ \citenamefont
  {Machnikowski}}]{Alam2024Determining}%
  \BibitemOpen
  \bibfield  {author} {\bibinfo {author} {\bibfnamefont {M.~S.}\ \bibnamefont
  {Alam}}, \bibinfo {author} {\bibfnamefont {F.}~\bibnamefont {Gorrini}},
  \bibinfo {author} {\bibfnamefont {M.}~\bibnamefont {Gawe\l{}czyk}}, \bibinfo
  {author} {\bibfnamefont {D.}~\bibnamefont {Wigger}}, \bibinfo {author}
  {\bibfnamefont {G.}~\bibnamefont {Coccia}}, \bibinfo {author} {\bibfnamefont
  {Y.}~\bibnamefont {Guo}}, \bibinfo {author} {\bibfnamefont {S.}~\bibnamefont
  {Shahbazi}}, \bibinfo {author} {\bibfnamefont {V.}~\bibnamefont {Bharadwaj}},
  \bibinfo {author} {\bibfnamefont {A.}~\bibnamefont {Kubanek}}, \bibinfo
  {author} {\bibfnamefont {R.}~\bibnamefont {Ramponi}}, \bibinfo {author}
  {\bibfnamefont {P.~E.}\ \bibnamefont {Barclay}}, \bibinfo {author}
  {\bibfnamefont {A.~J.}\ \bibnamefont {Bennett}}, \bibinfo {author}
  {\bibfnamefont {J.~P.}\ \bibnamefont {Hadden}}, \bibinfo {author}
  {\bibfnamefont {A.}~\bibnamefont {Bifone}}, \bibinfo {author} {\bibfnamefont
  {S.~M.}\ \bibnamefont {Eaton}},\ and\ \bibinfo {author} {\bibfnamefont
  {P.}~\bibnamefont {Machnikowski}},\ }\bibfield  {title} {\bibinfo {title}
  {Determining strain components in a diamond waveguide from zero-field
  optically detected magnetic resonance spectra of negatively charged
  nitrogen-vacancy-center ensembles},\ }\href
  {https://doi.org/10.1103/PhysRevApplied.22.024055} {\bibfield  {journal}
  {\bibinfo  {journal} {Phys. Rev. Appl.}\ }\textbf {\bibinfo {volume} {22}},\
  \bibinfo {pages} {024055} (\bibinfo {year} {2024})}\BibitemShut {NoStop}%
\bibitem [{\citenamefont {Hatano}\ \emph {et~al.}(2021)\citenamefont {Hatano},
  \citenamefont {Shin}, \citenamefont {Nishitani}, \citenamefont {Iwatsuka},
  \citenamefont {Masuyama}, \citenamefont {Sugiyama}, \citenamefont {Ishii},
  \citenamefont {Onoda}, \citenamefont {Ohshima}, \citenamefont {Arai},
  \citenamefont {Iwasaki},\ and\ \citenamefont
  {Hatano}}]{Hatano2021Simultaneous}%
  \BibitemOpen
  \bibfield  {author} {\bibinfo {author} {\bibfnamefont {Y.}~\bibnamefont
  {Hatano}}, \bibinfo {author} {\bibfnamefont {J.}~\bibnamefont {Shin}},
  \bibinfo {author} {\bibfnamefont {D.}~\bibnamefont {Nishitani}}, \bibinfo
  {author} {\bibfnamefont {H.}~\bibnamefont {Iwatsuka}}, \bibinfo {author}
  {\bibfnamefont {Y.}~\bibnamefont {Masuyama}}, \bibinfo {author}
  {\bibfnamefont {H.}~\bibnamefont {Sugiyama}}, \bibinfo {author}
  {\bibfnamefont {M.}~\bibnamefont {Ishii}}, \bibinfo {author} {\bibfnamefont
  {S.}~\bibnamefont {Onoda}}, \bibinfo {author} {\bibfnamefont
  {T.}~\bibnamefont {Ohshima}}, \bibinfo {author} {\bibfnamefont
  {K.}~\bibnamefont {Arai}}, \bibinfo {author} {\bibfnamefont {T.}~\bibnamefont
  {Iwasaki}},\ and\ \bibinfo {author} {\bibfnamefont {M.}~\bibnamefont
  {Hatano}},\ }\bibfield  {title} {\bibinfo {title} {Simultaneous thermometry
  and magnetometry using a fiber-coupled quantum diamond sensor},\ }\bibfield
  {journal} {\bibinfo  {journal} {Applied Physics Letters}\ }\textbf {\bibinfo
  {volume} {118}},\ \href {https://doi.org/10.1063/5.0031502}
  {10.1063/5.0031502} (\bibinfo {year} {2021})\BibitemShut {NoStop}%
\bibitem [{\citenamefont {Pogorzelski}\ \emph {et~al.}(2024)\citenamefont
  {Pogorzelski}, \citenamefont {Horsthemke}, \citenamefont {Homrighausen},
  \citenamefont {Stiegek\"{o}tter}, \citenamefont {Gregor},\ and\ \citenamefont
  {Gl\"{o}sek\"{o}tter}}]{Pogorzelski2024Compact}%
  \BibitemOpen
  \bibfield  {author} {\bibinfo {author} {\bibfnamefont {J.}~\bibnamefont
  {Pogorzelski}}, \bibinfo {author} {\bibfnamefont {L.}~\bibnamefont
  {Horsthemke}}, \bibinfo {author} {\bibfnamefont {J.}~\bibnamefont
  {Homrighausen}}, \bibinfo {author} {\bibfnamefont {D.}~\bibnamefont
  {Stiegek\"{o}tter}}, \bibinfo {author} {\bibfnamefont {M.}~\bibnamefont
  {Gregor}},\ and\ \bibinfo {author} {\bibfnamefont {P.}~\bibnamefont
  {Gl\"{o}sek\"{o}tter}},\ }\bibfield  {title} {\bibinfo {title} {Compact and
  fully integrated led quantum sensor based on nv centers in diamond},\ }\href
  {https://doi.org/10.3390/s24030743} {\bibfield  {journal} {\bibinfo
  {journal} {Sensors}\ }\textbf {\bibinfo {volume} {24}},\ \bibinfo {pages}
  {743} (\bibinfo {year} {2024})}\BibitemShut {NoStop}%
\bibitem [{\citenamefont {IBM}(2025)}]{IBM-QC-Roadmap}%
  \BibitemOpen
  \bibfield  {author} {\bibinfo {author} {\bibnamefont {IBM}},\ }\href
  {https://web.archive.org/web/20250404135913/https://www.ibm.com/quantum/blog/quantum-development-roadmap}
  {\bibinfo {title} {{IBM’s roadmap for building an open quantum software
  ecosystem}}} (\bibinfo {year} {2025}),\ \bibinfo {note} {archived / last
  accessed 2025-04-04}\BibitemShut {NoStop}%
\bibitem [{\citenamefont {Barker}\ and\ \citenamefont
  {Roginsky}(2019)}]{Nist-2048bit-RSA}%
  \BibitemOpen
  \bibfield  {author} {\bibinfo {author} {\bibfnamefont {E.}~\bibnamefont
  {Barker}}\ and\ \bibinfo {author} {\bibfnamefont {A.}~\bibnamefont
  {Roginsky}},\ }\href {https://doi.org/10.6028/NIST.SP.800-131Ar2} {\emph
  {\bibinfo {title} {Transitioning the Use of Cryptographic Algorithms and Key
  Lengths}}},\ \bibinfo {type} {Tech. Rep.}\ \bibinfo {number} {NIST Special
  Publication (SP) 800-131A, Rev. 2}\ (\bibinfo  {institution} {National
  Institute of Standards and Technology},\ \bibinfo {address} {Gaithersburg,
  MD},\ \bibinfo {year} {2019})\BibitemShut {NoStop}%
\bibitem [{\citenamefont {for Information~Security}(2025)}]{BSI-3000bit-RSA}%
  \BibitemOpen
  \bibfield  {author} {\bibinfo {author} {\bibfnamefont {G.~F.~O.}\
  \bibnamefont {for Information~Security}},\ }\href
  {https://doi.org/10.6028/NIST.SP.800-131Ar2} {\emph {\bibinfo {title}
  {Cryptographic Mechanisms: Recommendations and Key Lengths, Version:
  2025-1}}},\ \bibinfo {type} {Tech. Rep.}\ \bibinfo {number} {BSI TR-02102-1}\
  (\bibinfo  {institution} {German Federal Office for Information Security},\
  \bibinfo {address} {Bonn, Germany},\ \bibinfo {year} {2025})\BibitemShut
  {NoStop}%
\bibitem [{\citenamefont {Gidney}\ and\ \citenamefont
  {Eker{\aa{}}}(2021)}]{rsa-2024-in-8-hours}%
  \BibitemOpen
  \bibfield  {author} {\bibinfo {author} {\bibfnamefont {C.}~\bibnamefont
  {Gidney}}\ and\ \bibinfo {author} {\bibfnamefont {M.}~\bibnamefont
  {Eker{\aa{}}}},\ }\bibfield  {title} {\bibinfo {title} {{How to factor 2048
  bit {RSA} integers in 8 hours using 20 million noisy qubits}},\ }\href
  {https://doi.org/10.22331/q-2021-04-15-433} {\bibfield  {journal} {\bibinfo
  {journal} {{Quantum}}\ }\textbf {\bibinfo {volume} {5}},\ \bibinfo {pages}
  {433} (\bibinfo {year} {2021})}\BibitemShut {NoStop}%
\bibitem [{\citenamefont {Boudot}\ \emph {et~al.}(2020)\citenamefont {Boudot},
  \citenamefont {Gaudry}, \citenamefont {Guillevic}, \citenamefont {Heninger},
  \citenamefont {Thom{\'e}},\ and\ \citenamefont
  {Zimmermann}}]{RSA-Thousands-of-Core-Years}%
  \BibitemOpen
  \bibfield  {author} {\bibinfo {author} {\bibfnamefont {F.}~\bibnamefont
  {Boudot}}, \bibinfo {author} {\bibfnamefont {P.}~\bibnamefont {Gaudry}},
  \bibinfo {author} {\bibfnamefont {A.}~\bibnamefont {Guillevic}}, \bibinfo
  {author} {\bibfnamefont {N.}~\bibnamefont {Heninger}}, \bibinfo {author}
  {\bibfnamefont {E.}~\bibnamefont {Thom{\'e}}},\ and\ \bibinfo {author}
  {\bibfnamefont {P.}~\bibnamefont {Zimmermann}},\ }\bibfield  {title}
  {\bibinfo {title} {Comparing the difficulty of factorization and discrete
  logarithm: A 240-digit experiment},\ }in\ \href@noop {} {\emph {\bibinfo
  {booktitle} {Advances in Cryptology -- CRYPTO 2020}}},\ \bibinfo {editor}
  {edited by\ \bibinfo {editor} {\bibfnamefont {D.}~\bibnamefont {Micciancio}}\
  and\ \bibinfo {editor} {\bibfnamefont {T.}~\bibnamefont {Ristenpart}}}\
  (\bibinfo  {publisher} {Springer International Publishing},\ \bibinfo
  {address} {Cham},\ \bibinfo {year} {2020})\ pp.\ \bibinfo {pages}
  {62--91}\BibitemShut {NoStop}%
\bibitem [{\citenamefont {Pirandola}\ \emph {et~al.}(2020)\citenamefont
  {Pirandola}, \citenamefont {Andersen}, \citenamefont {Banchi}, \citenamefont
  {Berta}, \citenamefont {Bunandar}, \citenamefont {Colbeck}, \citenamefont
  {Englund}, \citenamefont {Gehring}, \citenamefont {Lupo}, \citenamefont
  {Ottaviani}, \citenamefont {Pereira}, \citenamefont {Razavi}, \citenamefont
  {Shamsul~Shaari}, \citenamefont {Tomamichel}, \citenamefont {Usenko},
  \citenamefont {Vallone}, \citenamefont {Villoresi},\ and\ \citenamefont
  {Wallden}}]{pirandola}%
  \BibitemOpen
  \bibfield  {author} {\bibinfo {author} {\bibfnamefont {S.}~\bibnamefont
  {Pirandola}}, \bibinfo {author} {\bibfnamefont {U.~L.}\ \bibnamefont
  {Andersen}}, \bibinfo {author} {\bibfnamefont {L.}~\bibnamefont {Banchi}},
  \bibinfo {author} {\bibfnamefont {M.}~\bibnamefont {Berta}}, \bibinfo
  {author} {\bibfnamefont {D.}~\bibnamefont {Bunandar}}, \bibinfo {author}
  {\bibfnamefont {R.}~\bibnamefont {Colbeck}}, \bibinfo {author} {\bibfnamefont
  {D.}~\bibnamefont {Englund}}, \bibinfo {author} {\bibfnamefont
  {T.}~\bibnamefont {Gehring}}, \bibinfo {author} {\bibfnamefont
  {C.}~\bibnamefont {Lupo}}, \bibinfo {author} {\bibfnamefont {C.}~\bibnamefont
  {Ottaviani}}, \bibinfo {author} {\bibfnamefont {J.~L.}\ \bibnamefont
  {Pereira}}, \bibinfo {author} {\bibfnamefont {M.}~\bibnamefont {Razavi}},
  \bibinfo {author} {\bibfnamefont {J.}~\bibnamefont {Shamsul~Shaari}},
  \bibinfo {author} {\bibfnamefont {M.}~\bibnamefont {Tomamichel}}, \bibinfo
  {author} {\bibfnamefont {V.~C.}\ \bibnamefont {Usenko}}, \bibinfo {author}
  {\bibfnamefont {G.}~\bibnamefont {Vallone}}, \bibinfo {author} {\bibfnamefont
  {P.}~\bibnamefont {Villoresi}},\ and\ \bibinfo {author} {\bibfnamefont
  {P.}~\bibnamefont {Wallden}},\ }\bibfield  {title} {\bibinfo {title}
  {Advances in quantum cryptography},\ }\href
  {https://doi.org/10.1364/AOP.361502} {\bibfield  {journal} {\bibinfo
  {journal} {Advances in Optics and Photonics}\ }\textbf {\bibinfo {volume}
  {12}},\ \bibinfo {pages} {1012} (\bibinfo {year} {2020})}\BibitemShut
  {NoStop}%
\bibitem [{\citenamefont {Sharma}\ \emph {et~al.}(2021)\citenamefont {Sharma},
  \citenamefont {Agrawal}, \citenamefont {Bhatia}, \citenamefont {Prakash},\
  and\ \citenamefont {Mishra}}]{Sharma2021}%
  \BibitemOpen
  \bibfield  {author} {\bibinfo {author} {\bibfnamefont {P.}~\bibnamefont
  {Sharma}}, \bibinfo {author} {\bibfnamefont {A.}~\bibnamefont {Agrawal}},
  \bibinfo {author} {\bibfnamefont {V.}~\bibnamefont {Bhatia}}, \bibinfo
  {author} {\bibfnamefont {S.}~\bibnamefont {Prakash}},\ and\ \bibinfo {author}
  {\bibfnamefont {A.~K.}\ \bibnamefont {Mishra}},\ }\bibfield  {title}
  {\bibinfo {title} {Quantum key distribution secured optical networks: A
  survey},\ }\href {https://doi.org/10.1109/ojcoms.2021.3106659} {\bibfield
  {journal} {\bibinfo  {journal} {IEEE Open Journal of the Communications
  Society}\ }\textbf {\bibinfo {volume} {2}},\ \bibinfo {pages} {2049–2083}
  (\bibinfo {year} {2021})}\BibitemShut {NoStop}%
\bibitem [{\citenamefont {Cao}\ \emph {et~al.}(2022)\citenamefont {Cao},
  \citenamefont {Zhao}, \citenamefont {Wang}, \citenamefont {Zhang},
  \citenamefont {Ng},\ and\ \citenamefont {Hanzo}}]{Cao2022}%
  \BibitemOpen
  \bibfield  {author} {\bibinfo {author} {\bibfnamefont {Y.}~\bibnamefont
  {Cao}}, \bibinfo {author} {\bibfnamefont {Y.}~\bibnamefont {Zhao}}, \bibinfo
  {author} {\bibfnamefont {Q.}~\bibnamefont {Wang}}, \bibinfo {author}
  {\bibfnamefont {J.}~\bibnamefont {Zhang}}, \bibinfo {author} {\bibfnamefont
  {S.~X.}\ \bibnamefont {Ng}},\ and\ \bibinfo {author} {\bibfnamefont
  {L.}~\bibnamefont {Hanzo}},\ }\bibfield  {title} {\bibinfo {title} {The
  evolution of quantum key distribution networks: On the road to the
  qinternet},\ }\href {https://doi.org/10.1109/COMST.2022.3144219} {\bibfield
  {journal} {\bibinfo  {journal} {{IEEE Communications Surveys \& Tutorials}}\
  }\textbf {\bibinfo {volume} {24}},\ \bibinfo {pages} {839} (\bibinfo {year}
  {2022})}\BibitemShut {NoStop}%
\bibitem [{\citenamefont {Davidson}\ \emph {et~al.}(2024)\citenamefont
  {Davidson}, \citenamefont {Hugues-Salas}, \citenamefont {Bonner},
  \citenamefont {Jones}, \citenamefont {Prentice}, \citenamefont {Kariappa},
  \citenamefont {Fowler}, \citenamefont {Oliveira}, \citenamefont {Zhang},
  \citenamefont {Andersson}, \citenamefont {Kosmatos}, \citenamefont
  {Stavdas},\ and\ \citenamefont {Lord}}]{10713122}%
  \BibitemOpen
  \bibfield  {author} {\bibinfo {author} {\bibfnamefont {Z.~C.~M.}\
  \bibnamefont {Davidson}}, \bibinfo {author} {\bibfnamefont {E.}~\bibnamefont
  {Hugues-Salas}}, \bibinfo {author} {\bibfnamefont {G.~M.}\ \bibnamefont
  {Bonner}}, \bibinfo {author} {\bibfnamefont {B.~E.}\ \bibnamefont {Jones}},
  \bibinfo {author} {\bibfnamefont {J.}~\bibnamefont {Prentice}}, \bibinfo
  {author} {\bibfnamefont {S.}~\bibnamefont {Kariappa}}, \bibinfo {author}
  {\bibfnamefont {D.~S.}\ \bibnamefont {Fowler}}, \bibinfo {author}
  {\bibfnamefont {R.~D.}\ \bibnamefont {Oliveira}}, \bibinfo {author}
  {\bibfnamefont {P.}~\bibnamefont {Zhang}}, \bibinfo {author} {\bibfnamefont
  {Y.}~\bibnamefont {Andersson}}, \bibinfo {author} {\bibfnamefont {E.~A.}\
  \bibnamefont {Kosmatos}}, \bibinfo {author} {\bibfnamefont {A.}~\bibnamefont
  {Stavdas}},\ and\ \bibinfo {author} {\bibfnamefont {A.}~\bibnamefont
  {Lord}},\ }\bibfield  {title} {\bibinfo {title} {Airqkd: The role of
  free-space optics quantum key distribution enabling pragmatic secure and
  scalable communications},\ }\href {https://doi.org/10.1109/MCOM.001.2300813}
  {\bibfield  {journal} {\bibinfo  {journal} {IEEE Communications Magazine}\
  }\textbf {\bibinfo {volume} {62}},\ \bibinfo {pages} {40} (\bibinfo {year}
  {2024})}\BibitemShut {NoStop}%
\bibitem [{\citenamefont {Bedington}\ \emph {et~al.}(2017)\citenamefont
  {Bedington}, \citenamefont {Arrazola},\ and\ \citenamefont
  {Ling}}]{Bedington2017}%
  \BibitemOpen
  \bibfield  {author} {\bibinfo {author} {\bibfnamefont {R.}~\bibnamefont
  {Bedington}}, \bibinfo {author} {\bibfnamefont {J.~M.}\ \bibnamefont
  {Arrazola}},\ and\ \bibinfo {author} {\bibfnamefont {A.}~\bibnamefont
  {Ling}},\ }\bibfield  {title} {\bibinfo {title} {Progress in satellite
  quantum key distribution},\ }\href
  {https://doi.org/10.1038/s41534-017-0031-5} {\bibfield  {journal} {\bibinfo
  {journal} {npj Quantum Information}\ }\textbf {\bibinfo {volume} {3}},\
  \bibinfo {pages} {30} (\bibinfo {year} {2017})}\BibitemShut {NoStop}%
\bibitem [{\citenamefont {Dai}\ \emph {et~al.}(2024)\citenamefont {Dai},
  \citenamefont {Sun}, \citenamefont {Lv}, \citenamefont {Wang},\ and\
  \citenamefont {Tong}}]{10845578}%
  \BibitemOpen
  \bibfield  {author} {\bibinfo {author} {\bibfnamefont {H.}~\bibnamefont
  {Dai}}, \bibinfo {author} {\bibfnamefont {X.}~\bibnamefont {Sun}}, \bibinfo
  {author} {\bibfnamefont {B.}~\bibnamefont {Lv}}, \bibinfo {author}
  {\bibfnamefont {H.}~\bibnamefont {Wang}},\ and\ \bibinfo {author}
  {\bibfnamefont {L.}~\bibnamefont {Tong}},\ }\bibfield  {title} {\bibinfo
  {title} {Multi-scenario quantum key distribution mechanism for power grid
  terminals},\ }in\ \href {https://doi.org/10.1109/ISCTech63666.2024.10845578}
  {\emph {\bibinfo {booktitle} {2024 12th International Conference on
  Information Systems and Computing Technology (ISCTech)}}}\ (\bibinfo {year}
  {2024})\ pp.\ \bibinfo {pages} {1--5}\BibitemShut {NoStop}%
\bibitem [{\citenamefont {Li}\ \emph {et~al.}(2023)\citenamefont {Li},
  \citenamefont {Xue}, \citenamefont {Li}, \citenamefont {Chen}, \citenamefont
  {Li}, \citenamefont {Wang}, \citenamefont {Yu}, \citenamefont {Wei},
  \citenamefont {Sun},\ and\ \citenamefont {Lu}}]{Li2023}%
  \BibitemOpen
  \bibfield  {author} {\bibinfo {author} {\bibfnamefont {Z.}~\bibnamefont
  {Li}}, \bibinfo {author} {\bibfnamefont {K.}~\bibnamefont {Xue}}, \bibinfo
  {author} {\bibfnamefont {J.}~\bibnamefont {Li}}, \bibinfo {author}
  {\bibfnamefont {L.}~\bibnamefont {Chen}}, \bibinfo {author} {\bibfnamefont
  {R.}~\bibnamefont {Li}}, \bibinfo {author} {\bibfnamefont {Z.}~\bibnamefont
  {Wang}}, \bibinfo {author} {\bibfnamefont {N.}~\bibnamefont {Yu}}, \bibinfo
  {author} {\bibfnamefont {D.~S.~L.}\ \bibnamefont {Wei}}, \bibinfo {author}
  {\bibfnamefont {Q.}~\bibnamefont {Sun}},\ and\ \bibinfo {author}
  {\bibfnamefont {J.}~\bibnamefont {Lu}},\ }\bibfield  {title} {\bibinfo
  {title} {Entanglement-assisted quantum networks: Mechanics, enabling
  technologies, challenges, and research directions},\ }\href
  {https://doi.org/10.1109/comst.2023.3294240} {\bibfield  {journal} {\bibinfo
  {journal} {IEEE Communications Surveys \& Tutorials}\ }\textbf {\bibinfo
  {volume} {25}},\ \bibinfo {pages} {2133–2189} (\bibinfo {year}
  {2023})}\BibitemShut {NoStop}%
\bibitem [{\citenamefont {Dolphin}\ \emph {et~al.}(2023)\citenamefont
  {Dolphin}, \citenamefont {Para{\"i}so}, \citenamefont {Du}, \citenamefont
  {Woodward}, \citenamefont {Marangon},\ and\ \citenamefont
  {Shields}}]{Dolphin2023}%
  \BibitemOpen
  \bibfield  {author} {\bibinfo {author} {\bibfnamefont {J.~A.}\ \bibnamefont
  {Dolphin}}, \bibinfo {author} {\bibfnamefont {T.~K.}\ \bibnamefont
  {Para{\"i}so}}, \bibinfo {author} {\bibfnamefont {H.}~\bibnamefont {Du}},
  \bibinfo {author} {\bibfnamefont {R.~I.}\ \bibnamefont {Woodward}}, \bibinfo
  {author} {\bibfnamefont {D.~G.}\ \bibnamefont {Marangon}},\ and\ \bibinfo
  {author} {\bibfnamefont {A.~J.}\ \bibnamefont {Shields}},\ }\bibfield
  {title} {\bibinfo {title} {A hybrid integrated quantum key distribution
  transceiver chip},\ }\href {https://doi.org/10.1038/s41534-023-00751-3}
  {\bibfield  {journal} {\bibinfo  {journal} {npj Quantum Information}\
  }\textbf {\bibinfo {volume} {9}},\ \bibinfo {pages} {84} (\bibinfo {year}
  {2023})}\BibitemShut {NoStop}%
\bibitem [{\citenamefont {Alshowkan}\ \emph {et~al.}(2022)\citenamefont
  {Alshowkan}, \citenamefont {Evans}, \citenamefont {Starke}, \citenamefont
  {Earl},\ and\ \citenamefont {Peters}}]{alshowkan2022authentication}%
  \BibitemOpen
  \bibfield  {author} {\bibinfo {author} {\bibfnamefont {M.}~\bibnamefont
  {Alshowkan}}, \bibinfo {author} {\bibfnamefont {P.~G.}\ \bibnamefont
  {Evans}}, \bibinfo {author} {\bibfnamefont {M.}~\bibnamefont {Starke}},
  \bibinfo {author} {\bibfnamefont {D.}~\bibnamefont {Earl}},\ and\ \bibinfo
  {author} {\bibfnamefont {N.~A.}\ \bibnamefont {Peters}},\ }\bibfield  {title}
  {\bibinfo {title} {Authentication of smart grid communications using quantum
  key distribution},\ }\href@noop {} {\bibfield  {journal} {\bibinfo  {journal}
  {Scientific Reports}\ }\textbf {\bibinfo {volume} {12}},\ \bibinfo {pages}
  {12731} (\bibinfo {year} {2022})}\BibitemShut {NoStop}%
\bibitem [{\citenamefont {Evans}\ \emph {et~al.}(2021)\citenamefont {Evans},
  \citenamefont {Alshowkan}, \citenamefont {Earl}, \citenamefont {Mulkey},
  \citenamefont {Newell}, \citenamefont {Peterson}, \citenamefont {Safi},
  \citenamefont {Tripp},\ and\ \citenamefont {Peters}}]{9405393}%
  \BibitemOpen
  \bibfield  {author} {\bibinfo {author} {\bibfnamefont {P.~G.}\ \bibnamefont
  {Evans}}, \bibinfo {author} {\bibfnamefont {M.}~\bibnamefont {Alshowkan}},
  \bibinfo {author} {\bibfnamefont {D.}~\bibnamefont {Earl}}, \bibinfo {author}
  {\bibfnamefont {D.~D.}\ \bibnamefont {Mulkey}}, \bibinfo {author}
  {\bibfnamefont {R.}~\bibnamefont {Newell}}, \bibinfo {author} {\bibfnamefont
  {G.}~\bibnamefont {Peterson}}, \bibinfo {author} {\bibfnamefont
  {C.}~\bibnamefont {Safi}}, \bibinfo {author} {\bibfnamefont {J.~L.}\
  \bibnamefont {Tripp}},\ and\ \bibinfo {author} {\bibfnamefont {N.~A.}\
  \bibnamefont {Peters}},\ }\bibfield  {title} {\bibinfo {title} {Trusted node
  qkd at an electrical utility},\ }\href
  {https://doi.org/10.1109/ACCESS.2021.3070222} {\bibfield  {journal} {\bibinfo
   {journal} {IEEE Access}\ }\textbf {\bibinfo {volume} {9}},\ \bibinfo {pages}
  {105220} (\bibinfo {year} {2021})}\BibitemShut {NoStop}%
\bibitem [{\citenamefont {{AZO Quantum}}(2022)}]{Mozi_Satellite}%
  \BibitemOpen
  \bibfield  {author} {\bibinfo {author} {\bibnamefont {{AZO Quantum}}},\
  }\href {www.azoquantum.com/Article.aspx?ArticleID=308} {\bibinfo {title} {How
  the’mozi’satellite grants quantum security from space}} (\bibinfo {year}
  {2022})\BibitemShut {NoStop}%
\bibitem [{\citenamefont {Tang}\ \emph {et~al.}(2021)\citenamefont {Tang},
  \citenamefont {Zhang},\ and\ \citenamefont {Krawec}}]{tang_quantum_2021}%
  \BibitemOpen
  \bibfield  {author} {\bibinfo {author} {\bibfnamefont {Z.}~\bibnamefont
  {Tang}}, \bibinfo {author} {\bibfnamefont {P.}~\bibnamefont {Zhang}},\ and\
  \bibinfo {author} {\bibfnamefont {W.~O.}\ \bibnamefont {Krawec}},\ }\bibfield
   {title} {\bibinfo {title} {A quantum leap in microgrids security: The
  prospects of quantum-secure microgrids},\ }\href
  {https://doi.org/10.1109/mele.2020.3047167} {\bibfield  {journal} {\bibinfo
  {journal} {IEEE Electrification Magazine}\ }\textbf {\bibinfo {volume} {9}},\
  \bibinfo {pages} {66–73} (\bibinfo {year} {2021})}\BibitemShut {NoStop}%
\bibitem [{\citenamefont {Tang}\ \emph {et~al.}(2020)\citenamefont {Tang},
  \citenamefont {Qin}, \citenamefont {Jiang}, \citenamefont {Krawec},\ and\
  \citenamefont {Zhang}}]{tang_quantum-secure_2020}%
  \BibitemOpen
  \bibfield  {author} {\bibinfo {author} {\bibfnamefont {Z.}~\bibnamefont
  {Tang}}, \bibinfo {author} {\bibfnamefont {Y.}~\bibnamefont {Qin}}, \bibinfo
  {author} {\bibfnamefont {Z.}~\bibnamefont {Jiang}}, \bibinfo {author}
  {\bibfnamefont {W.~O.}\ \bibnamefont {Krawec}},\ and\ \bibinfo {author}
  {\bibfnamefont {P.}~\bibnamefont {Zhang}},\ }\bibfield  {title} {\bibinfo
  {title} {Quantum-secure networked microgrids},\ }in\ \href
  {https://doi.org/10.1109/PESGM41954.2020.9281884} {\emph {\bibinfo
  {booktitle} {2020 {IEEE} Power \& Energy Society General Meeting
  ({PESGM})}}}\ (\bibinfo {year} {2020})\ pp.\ \bibinfo {pages} {1--5},\
  \bibinfo {note} {{ISSN}: 1944-9933}\BibitemShut {NoStop}%
\bibitem [{\citenamefont {Grice}\ \emph {et~al.}(2025)\citenamefont {Grice},
  \citenamefont {Olama}, \citenamefont {Lee},\ and\ \citenamefont
  {Evans}}]{10852309}%
  \BibitemOpen
  \bibfield  {author} {\bibinfo {author} {\bibfnamefont {W.}~\bibnamefont
  {Grice}}, \bibinfo {author} {\bibfnamefont {M.}~\bibnamefont {Olama}},
  \bibinfo {author} {\bibfnamefont {A.}~\bibnamefont {Lee}},\ and\ \bibinfo
  {author} {\bibfnamefont {P.~G.}\ \bibnamefont {Evans}},\ }\bibfield  {title}
  {\bibinfo {title} {Quantum key distribution applicability to smart grid
  cybersecurity systems},\ }\href {https://doi.org/10.1109/ACCESS.2025.3533942}
  {\bibfield  {journal} {\bibinfo  {journal} {IEEE Access}\ }\textbf {\bibinfo
  {volume} {13}},\ \bibinfo {pages} {17398} (\bibinfo {year}
  {2025})}\BibitemShut {NoStop}%
\bibitem [{\citenamefont {Diamanti}\ \emph {et~al.}(2016)\citenamefont
  {Diamanti}, \citenamefont {Lo}, \citenamefont {Qi},\ and\ \citenamefont
  {Yuan}}]{Diamanti2016}%
  \BibitemOpen
  \bibfield  {author} {\bibinfo {author} {\bibfnamefont {E.}~\bibnamefont
  {Diamanti}}, \bibinfo {author} {\bibfnamefont {H.-K.}\ \bibnamefont {Lo}},
  \bibinfo {author} {\bibfnamefont {B.}~\bibnamefont {Qi}},\ and\ \bibinfo
  {author} {\bibfnamefont {Z.}~\bibnamefont {Yuan}},\ }\bibfield  {title}
  {\bibinfo {title} {Practical challenges in quantum key distribution},\ }\href
  {https://doi.org/10.1038/npjqi.2016.25} {\bibfield  {journal} {\bibinfo
  {journal} {npj Quantum Information}\ }\textbf {\bibinfo {volume} {2}},\
  \bibinfo {pages} {16025} (\bibinfo {year} {2016})}\BibitemShut {NoStop}%
\bibitem [{\citenamefont {Gado}\ \emph {et~al.}(2024)\citenamefont {Gado},
  \citenamefont {Ismail},\ and\ \citenamefont {Krawec}}]{gado_upgrading_2024}%
  \BibitemOpen
  \bibfield  {author} {\bibinfo {author} {\bibfnamefont {M.}~\bibnamefont
  {Gado}}, \bibinfo {author} {\bibfnamefont {M.}~\bibnamefont {Ismail}},\ and\
  \bibinfo {author} {\bibfnamefont {W.~O.}\ \bibnamefont {Krawec}},\ }\bibfield
   {title} {\bibinfo {title} {Upgrading the cyber layer of power systems to
  support semi-quantum key distribution},\ }in\ \href
  {https://doi.org/10.1109/ISGT59692.2024.10454162} {\emph {\bibinfo
  {booktitle} {2024 {IEEE} Power \& Energy Society Innovative Smart Grid
  Technologies Conference ({ISGT})}}}\ (\bibinfo {year} {2024})\ pp.\ \bibinfo
  {pages} {1--5},\ \bibinfo {note} {{ISSN}: 2472-8152}\BibitemShut {NoStop}%
\bibitem [{\citenamefont {Beutel}\ \emph {et~al.}(2021)\citenamefont {Beutel},
  \citenamefont {Gehring}, \citenamefont {Wolff}, \citenamefont {Schuck},\ and\
  \citenamefont {Pernice}}]{Beutel2021}%
  \BibitemOpen
  \bibfield  {author} {\bibinfo {author} {\bibfnamefont {F.}~\bibnamefont
  {Beutel}}, \bibinfo {author} {\bibfnamefont {H.}~\bibnamefont {Gehring}},
  \bibinfo {author} {\bibfnamefont {M.~A.}\ \bibnamefont {Wolff}}, \bibinfo
  {author} {\bibfnamefont {C.}~\bibnamefont {Schuck}},\ and\ \bibinfo {author}
  {\bibfnamefont {W.}~\bibnamefont {Pernice}},\ }\bibfield  {title} {\bibinfo
  {title} {Detector-integrated on-chip qkd receiver for ghz clock rates},\
  }\href {https://doi.org/10.1038/s41534-021-00373-7} {\bibfield  {journal}
  {\bibinfo  {journal} {npj Quantum Information}\ }\textbf {\bibinfo {volume}
  {7}},\ \bibinfo {pages} {40} (\bibinfo {year} {2021})}\BibitemShut {NoStop}%
\bibitem [{\citenamefont {Astaburuaga}\ \emph {et~al.}(2025)\citenamefont
  {Astaburuaga}, \citenamefont {Rath},\ and\ \citenamefont
  {Sengupta}}]{QKD-Eavesdropping-DOS-1}%
  \BibitemOpen
  \bibfield  {author} {\bibinfo {author} {\bibfnamefont {I.}~\bibnamefont
  {Astaburuaga}}, \bibinfo {author} {\bibfnamefont {S.}~\bibnamefont {Rath}},\
  and\ \bibinfo {author} {\bibfnamefont {S.}~\bibnamefont {Sengupta}},\
  }\bibfield  {title} {\bibinfo {title} {Unveiling bb84 vulnerabilities: A
  quantum key distribution simulation library},\ }in\ \href
  {https://doi.org/10.1109/CCNC54725.2025.10975888} {\emph {\bibinfo
  {booktitle} {2025 IEEE 22nd Consumer Communications \& Networking Conference
  (CCNC)}}}\ (\bibinfo {year} {2025})\ pp.\ \bibinfo {pages} {1--8}\BibitemShut
  {NoStop}%
\bibitem [{\citenamefont {Price}\ \emph {et~al.}(2020)\citenamefont {Price},
  \citenamefont {Rarity},\ and\ \citenamefont
  {Erven}}]{QKD-Eavesdropping-DOS-2}%
  \BibitemOpen
  \bibfield  {author} {\bibinfo {author} {\bibfnamefont {A.~B.}\ \bibnamefont
  {Price}}, \bibinfo {author} {\bibfnamefont {J.~G.}\ \bibnamefont {Rarity}},\
  and\ \bibinfo {author} {\bibfnamefont {C.}~\bibnamefont {Erven}},\ }\bibfield
   {title} {\bibinfo {title} {A quantum key distribution protocol for rapid
  denial of service detection},\ }\href
  {https://doi.org/10.1140/epjqt/s40507-020-00084-6} {\bibfield  {journal}
  {\bibinfo  {journal} {EPJ Quantum Technology}\ }\textbf {\bibinfo {volume}
  {7}},\ \bibinfo {pages} {8} (\bibinfo {year} {2020})}\BibitemShut {NoStop}%
\bibitem [{\citenamefont {Apostolov}(2021)}]{R-GOOSE}%
  \BibitemOpen
  \bibfield  {author} {\bibinfo {author} {\bibfnamefont {A.}~\bibnamefont
  {Apostolov}},\ }\bibfield  {title} {\bibinfo {title} {{Wide Area GOOSE and
  Its Applications to System Integrity Protection Schemes}},\ }in\ \href
  {https://doi.org/10.1109/CPRE48231.2021.9429849} {\emph {\bibinfo {booktitle}
  {2021 74th Conference for Protective Relay Engineers (CPRE)}}}\ (\bibinfo
  {year} {2021})\ pp.\ \bibinfo {pages} {1--5}\BibitemShut {NoStop}%
\bibitem [{\citenamefont {Buechel}\ \emph {et~al.}(2025)\citenamefont
  {Buechel}, \citenamefont {Zimmer}, \citenamefont {Carta},\ and\ \citenamefont
  {Benigni}}]{Buechel}%
  \BibitemOpen
  \bibfield  {author} {\bibinfo {author} {\bibfnamefont {M.}~\bibnamefont
  {Buechel}}, \bibinfo {author} {\bibfnamefont {M.}~\bibnamefont {Zimmer}},
  \bibinfo {author} {\bibfnamefont {D.}~\bibnamefont {Carta}},\ and\ \bibinfo
  {author} {\bibfnamefont {A.}~\bibnamefont {Benigni}},\ }\bibfield  {title}
  {\bibinfo {title} {Distribution grid voltage estimation by correlated
  gaussian processes},\ }in\ \href
  {https://doi.org/10.1109/AMPS66841.2025.11219969} {\emph {\bibinfo
  {booktitle} {2025 IEEE 15th International Workshop on Applied Measurements
  for Power Systems (AMPS)}}}\ (\bibinfo {year} {2025})\ pp.\ \bibinfo {pages}
  {1--6}\BibitemShut {NoStop}%
\bibitem [{\citenamefont {Xie}\ \emph {et~al.}(2024)\citenamefont {Xie},
  \citenamefont {Rahman},\ and\ \citenamefont {Sun}}]{Xie}%
  \BibitemOpen
  \bibfield  {author} {\bibinfo {author} {\bibfnamefont {J.}~\bibnamefont
  {Xie}}, \bibinfo {author} {\bibfnamefont {A.}~\bibnamefont {Rahman}},\ and\
  \bibinfo {author} {\bibfnamefont {W.}~\bibnamefont {Sun}},\ }\bibfield
  {title} {\bibinfo {title} {Bayesian gan-based false data injection attack
  detection in active distribution grids with ders},\ }\href
  {https://doi.org/10.1109/TSG.2023.3337340} {\bibfield  {journal} {\bibinfo
  {journal} {IEEE Transactions on Smart Grid}\ }\textbf {\bibinfo {volume}
  {15}},\ \bibinfo {pages} {3223} (\bibinfo {year} {2024})}\BibitemShut
  {NoStop}%
\bibitem [{\citenamefont {ETSI}(2020)}]{etsi-qkd-interfaces}%
  \BibitemOpen
  \bibfield  {author} {\bibinfo {author} {\bibnamefont {ETSI}},\ }\href@noop {}
  {\emph {\bibinfo {title} {GS QKD 004; Version: 2.1.1, Quantum Key
  Distribution (QKD); Application Interface}}},\ \bibinfo {type} {Tech. Rep.}\
  (\bibinfo  {institution} {ETSI},\ \bibinfo {year} {2020})\BibitemShut
  {NoStop}%
\bibitem [{\citenamefont {ETSI}(2025)}]{etsi-qkd-architectures}%
  \BibitemOpen
  \bibfield  {author} {\bibinfo {author} {\bibnamefont {ETSI}},\ }\href@noop {}
  {\emph {\bibinfo {title} {Draft GR QKD 017; Version: 0.1.13, Quantum Key
  Distribution (QKD); Analysis of QKD Network architectures}}},\ \bibinfo
  {type} {Tech. Rep.}\ (\bibinfo  {institution} {ETSI},\ \bibinfo {year}
  {2025})\BibitemShut {NoStop}%
\bibitem [{\citenamefont {Peres}\ and\ \citenamefont
  {Terno}(2004)}]{No-SOL-Q-Communication}%
  \BibitemOpen
  \bibfield  {author} {\bibinfo {author} {\bibfnamefont {A.}~\bibnamefont
  {Peres}}\ and\ \bibinfo {author} {\bibfnamefont {D.~R.}\ \bibnamefont
  {Terno}},\ }\bibfield  {title} {\bibinfo {title} {Quantum information and
  relativity theory},\ }\href {https://doi.org/10.1103/RevModPhys.76.93}
  {\bibfield  {journal} {\bibinfo  {journal} {Rev. Mod. Phys.}\ }\textbf
  {\bibinfo {volume} {76}},\ \bibinfo {pages} {93} (\bibinfo {year}
  {2004})}\BibitemShut {NoStop}%
\bibitem [{\citenamefont {Eberhard}(1978)}]{No-Communication}%
  \BibitemOpen
  \bibfield  {author} {\bibinfo {author} {\bibfnamefont {P.~H.}\ \bibnamefont
  {Eberhard}},\ }\bibfield  {title} {\bibinfo {title} {Bell’s theorem and the
  different concepts of locality},\ }\href {https://doi.org/10.1007/bf02728628}
  {\bibfield  {journal} {\bibinfo  {journal} {Il Nuovo Cimento B Series 11}\
  }\textbf {\bibinfo {volume} {46}},\ \bibinfo {pages} {392–419} (\bibinfo
  {year} {1978})}\BibitemShut {NoStop}%
\bibitem [{\citenamefont {Babahajiani}\ and\ \citenamefont
  {Zhang}(2022)}]{MG-Control-With-Q-Teleportation}%
  \BibitemOpen
  \bibfield  {author} {\bibinfo {author} {\bibfnamefont {P.}~\bibnamefont
  {Babahajiani}}\ and\ \bibinfo {author} {\bibfnamefont {P.}~\bibnamefont
  {Zhang}},\ }\bibfield  {title} {\bibinfo {title} {Quantum distributed
  microgrid control},\ }in\ \href
  {https://doi.org/10.1109/pesgm48719.2022.9916709} {\emph {\bibinfo
  {booktitle} {2022 IEEE Power \& amp; Energy Society General Meeting
  (PESGM)}}}\ (\bibinfo  {publisher} {IEEE},\ \bibinfo {year} {2022})\ pp.\
  \bibinfo {pages} {1--5}\BibitemShut {NoStop}%
\bibitem [{\citenamefont {Babahajiani}\ \emph {et~al.}(2022)\citenamefont
  {Babahajiani}, \citenamefont {Zhang}, \citenamefont {Wei}, \citenamefont
  {Liu},\ and\ \citenamefont {Lu}}]{MG-Control-With-Q-Teleportation-2}%
  \BibitemOpen
  \bibfield  {author} {\bibinfo {author} {\bibfnamefont {P.}~\bibnamefont
  {Babahajiani}}, \bibinfo {author} {\bibfnamefont {P.}~\bibnamefont {Zhang}},
  \bibinfo {author} {\bibfnamefont {T.-C.}\ \bibnamefont {Wei}}, \bibinfo
  {author} {\bibfnamefont {J.}~\bibnamefont {Liu}},\ and\ \bibinfo {author}
  {\bibfnamefont {X.}~\bibnamefont {Lu}},\ }\bibfield  {title} {\bibinfo
  {title} {Employing interacting qubits for distributed microgrid control},\
  }\href {https://doi.org/10.1109/tpwrs.2022.3196608} {\bibfield  {journal}
  {\bibinfo  {journal} {IEEE Transactions on Power Systems}\ ,\ \bibinfo
  {pages} {1}} (\bibinfo {year} {2022})}\BibitemShut {NoStop}%
\bibitem [{\citenamefont {Ahmad}\ \emph {et~al.}(2025)\citenamefont {Ahmad},
  \citenamefont {Hassan}, \citenamefont {Islam}, \citenamefont {Shafiullah},
  \citenamefont {Abido},\ and\ \citenamefont
  {Al-Dhaifallah}}]{Distributed-Control-MG-Review}%
  \BibitemOpen
  \bibfield  {author} {\bibinfo {author} {\bibfnamefont {G.}~\bibnamefont
  {Ahmad}}, \bibinfo {author} {\bibfnamefont {A.}~\bibnamefont {Hassan}},
  \bibinfo {author} {\bibfnamefont {A.}~\bibnamefont {Islam}}, \bibinfo
  {author} {\bibfnamefont {M.}~\bibnamefont {Shafiullah}}, \bibinfo {author}
  {\bibfnamefont {M.~A.}\ \bibnamefont {Abido}},\ and\ \bibinfo {author}
  {\bibfnamefont {M.}~\bibnamefont {Al-Dhaifallah}},\ }\bibfield  {title}
  {\bibinfo {title} {{Distributed Control Strategies for Microgrids: A Critical
  Review of Technologies and Challenges}},\ }\href
  {https://doi.org/10.1109/ACCESS.2025.3552940} {\bibfield  {journal} {\bibinfo
   {journal} {IEEE Access}\ }\textbf {\bibinfo {volume} {13}},\ \bibinfo
  {pages} {60702} (\bibinfo {year} {2025})}\BibitemShut {NoStop}%
\bibitem [{\citenamefont {Wootters}\ and\ \citenamefont
  {Zurek}(1982)}]{No-Cloning-1}%
  \BibitemOpen
  \bibfield  {author} {\bibinfo {author} {\bibfnamefont {W.~K.}\ \bibnamefont
  {Wootters}}\ and\ \bibinfo {author} {\bibfnamefont {W.~H.}\ \bibnamefont
  {Zurek}},\ }\bibfield  {title} {\bibinfo {title} {A single quantum cannot be
  cloned},\ }\href@noop {} {\bibfield  {journal} {\bibinfo  {journal} {Nature}\
  }\textbf {\bibinfo {volume} {299}},\ \bibinfo {pages} {802} (\bibinfo {year}
  {1982})}\BibitemShut {NoStop}%
\bibitem [{\citenamefont {Dieks}(1982)}]{No-Cloning2}%
  \BibitemOpen
  \bibfield  {author} {\bibinfo {author} {\bibfnamefont {D.}~\bibnamefont
  {Dieks}},\ }\bibfield  {title} {\bibinfo {title} {{Communication by EPR
  devices}},\ }\href
  {https://doi.org/https://doi.org/10.1016/0375-9601(82)90084-6} {\bibfield
  {journal} {\bibinfo  {journal} {Physics Letters A}\ }\textbf {\bibinfo
  {volume} {92}},\ \bibinfo {pages} {271} (\bibinfo {year} {1982})}\BibitemShut
  {NoStop}%
\bibitem [{\citenamefont {Zhou}\ \emph {et~al.}(2022)\citenamefont {Zhou},
  \citenamefont {Tang}, \citenamefont {Nikmehr}, \citenamefont {Babahajiani},
  \citenamefont {Feng}, \citenamefont {Wei}, \citenamefont {Zheng},\ and\
  \citenamefont {Zhang}}]{qc-power-ystems-zhou}%
  \BibitemOpen
  \bibfield  {author} {\bibinfo {author} {\bibfnamefont {Y.}~\bibnamefont
  {Zhou}}, \bibinfo {author} {\bibfnamefont {Z.}~\bibnamefont {Tang}}, \bibinfo
  {author} {\bibfnamefont {N.}~\bibnamefont {Nikmehr}}, \bibinfo {author}
  {\bibfnamefont {P.}~\bibnamefont {Babahajiani}}, \bibinfo {author}
  {\bibfnamefont {F.}~\bibnamefont {Feng}}, \bibinfo {author} {\bibfnamefont
  {T.-C.}\ \bibnamefont {Wei}}, \bibinfo {author} {\bibfnamefont
  {H.}~\bibnamefont {Zheng}},\ and\ \bibinfo {author} {\bibfnamefont
  {P.}~\bibnamefont {Zhang}},\ }\bibfield  {title} {\bibinfo {title} {Quantum
  computing in power systems},\ }\href {https://doi.org/10.23919/IEN.2022.0021}
  {\bibfield  {journal} {\bibinfo  {journal} {iEnergy}\ }\textbf {\bibinfo
  {volume} {1}},\ \bibinfo {pages} {170} (\bibinfo {year} {2022})}\BibitemShut
  {NoStop}%
\bibitem [{\citenamefont {Golestan}\ \emph {et~al.}(2023)\citenamefont
  {Golestan}, \citenamefont {Habibi}, \citenamefont {{Mousazadeh Mousavi}},
  \citenamefont {Guerrero},\ and\ \citenamefont
  {Vasquez}}]{qc-power-systems-golestan}%
  \BibitemOpen
  \bibfield  {author} {\bibinfo {author} {\bibfnamefont {S.}~\bibnamefont
  {Golestan}}, \bibinfo {author} {\bibfnamefont {M.}~\bibnamefont {Habibi}},
  \bibinfo {author} {\bibfnamefont {S.}~\bibnamefont {{Mousazadeh Mousavi}}},
  \bibinfo {author} {\bibfnamefont {J.}~\bibnamefont {Guerrero}},\ and\
  \bibinfo {author} {\bibfnamefont {J.}~\bibnamefont {Vasquez}},\ }\bibfield
  {title} {\bibinfo {title} {Quantum computation in power systems: An overview
  of recent advances},\ }\href
  {https://doi.org/https://doi.org/10.1016/j.egyr.2022.11.185} {\bibfield
  {journal} {\bibinfo  {journal} {Energy Reports}\ }\textbf {\bibinfo {volume}
  {9}},\ \bibinfo {pages} {584} (\bibinfo {year} {2023})}\BibitemShut {NoStop}%
\bibitem [{\citenamefont {Eskandarpour}\ \emph {et~al.}(2020)\citenamefont
  {Eskandarpour}, \citenamefont {Bahadur~Ghosh}, \citenamefont {Khodaei},
  \citenamefont {Paaso},\ and\ \citenamefont {Zhang}}]{qgrid-eskandarpour}%
  \BibitemOpen
  \bibfield  {author} {\bibinfo {author} {\bibfnamefont {R.}~\bibnamefont
  {Eskandarpour}}, \bibinfo {author} {\bibfnamefont {K.~J.}\ \bibnamefont
  {Bahadur~Ghosh}}, \bibinfo {author} {\bibfnamefont {A.}~\bibnamefont
  {Khodaei}}, \bibinfo {author} {\bibfnamefont {A.}~\bibnamefont {Paaso}},\
  and\ \bibinfo {author} {\bibfnamefont {L.}~\bibnamefont {Zhang}},\ }\bibfield
   {title} {\bibinfo {title} {Quantum-enhanced grid of the future: A primer},\
  }\href {https://doi.org/10.1109/ACCESS.2020.3031595} {\bibfield  {journal}
  {\bibinfo  {journal} {IEEE Access}\ }\textbf {\bibinfo {volume} {8}},\
  \bibinfo {pages} {188993} (\bibinfo {year} {2020})}\BibitemShut {NoStop}%
\bibitem [{\citenamefont {Liu}\ \emph {et~al.}(2018)\citenamefont {Liu},
  \citenamefont {Choo},\ and\ \citenamefont
  {Grossschadl}}]{liu2018postquantumiot}%
  \BibitemOpen
  \bibfield  {author} {\bibinfo {author} {\bibfnamefont {Z.}~\bibnamefont
  {Liu}}, \bibinfo {author} {\bibfnamefont {K.-K.~R.}\ \bibnamefont {Choo}},\
  and\ \bibinfo {author} {\bibfnamefont {J.}~\bibnamefont {Grossschadl}},\
  }\bibfield  {title} {\bibinfo {title} {Securing edge devices in the
  post-quantum internet of things using lattice-based cryptography},\ }\href
  {https://doi.org/10.1109/MCOM.2018.1700330} {\bibfield  {journal} {\bibinfo
  {journal} {IEEE Communications Magazine}\ }\textbf {\bibinfo {volume} {56}},\
  \bibinfo {pages} {158} (\bibinfo {year} {2018})}\BibitemShut {NoStop}%
\bibitem [{\citenamefont {Ebrahimi}\ \emph {et~al.}(2019)\citenamefont
  {Ebrahimi}, \citenamefont {Bayat-Sarmadi},\ and\ \citenamefont
  {Mosanaei-Boorani}}]{ebrahimi2019cryptoprocessor}%
  \BibitemOpen
  \bibfield  {author} {\bibinfo {author} {\bibfnamefont {S.}~\bibnamefont
  {Ebrahimi}}, \bibinfo {author} {\bibfnamefont {S.}~\bibnamefont
  {Bayat-Sarmadi}},\ and\ \bibinfo {author} {\bibfnamefont {H.}~\bibnamefont
  {Mosanaei-Boorani}},\ }\bibfield  {title} {\bibinfo {title} {Post-quantum
  cryptoprocessors optimized for edge and resource-constrained devices in
  iot},\ }\href {https://doi.org/10.1109/JIOT.2019.2903082} {\bibfield
  {journal} {\bibinfo  {journal} {IEEE Internet of Things Journal}\ }\textbf
  {\bibinfo {volume} {6}},\ \bibinfo {pages} {5500} (\bibinfo {year}
  {2019})}\BibitemShut {NoStop}%
\bibitem [{\citenamefont {Zhuang}\ \emph {et~al.}(2018)\citenamefont {Zhuang},
  \citenamefont {Zhang},\ and\ \citenamefont {Shapiro}}]{Zhuang2018}%
  \BibitemOpen
  \bibfield  {author} {\bibinfo {author} {\bibfnamefont {Q.}~\bibnamefont
  {Zhuang}}, \bibinfo {author} {\bibfnamefont {Z.}~\bibnamefont {Zhang}},\ and\
  \bibinfo {author} {\bibfnamefont {J.~H.}\ \bibnamefont {Shapiro}},\
  }\bibfield  {title} {\bibinfo {title} {Distributed quantum sensing using
  continuous-variable multipartite entanglement},\ }\bibfield  {journal}
  {\bibinfo  {journal} {Physical Review A}\ }\textbf {\bibinfo {volume} {97}},\
  \href {https://doi.org/10.1103/physreva.97.032329}
  {10.1103/physreva.97.032329} (\bibinfo {year} {2018})\BibitemShut {NoStop}%
\bibitem [{\citenamefont {Schöffel}\ \emph {et~al.}(2022)\citenamefont
  {Schöffel}, \citenamefont {Lauer}, \citenamefont {Rheinländer},\ and\
  \citenamefont {Wehn}}]{schöffel2022secureiotq}%
  \BibitemOpen
  \bibfield  {author} {\bibinfo {author} {\bibfnamefont {M.}~\bibnamefont
  {Schöffel}}, \bibinfo {author} {\bibfnamefont {F.}~\bibnamefont {Lauer}},
  \bibinfo {author} {\bibfnamefont {C.~C.}\ \bibnamefont {Rheinländer}},\ and\
  \bibinfo {author} {\bibfnamefont {N.}~\bibnamefont {Wehn}},\ }\bibfield
  {title} {\bibinfo {title} {Secure iot in the era of quantum computers—where
  are the bottlenecks?},\ }\bibfield  {journal} {\bibinfo  {journal} {Sensors}\
  }\textbf {\bibinfo {volume} {22}},\ \href {https://doi.org/10.3390/s22072484}
  {10.3390/s22072484} (\bibinfo {year} {2022})\BibitemShut {NoStop}%
\bibitem [{\citenamefont {Duan}\ and\ \citenamefont
  {Dinavahi}(2023)}]{SDN-FastFailover}%
  \BibitemOpen
  \bibfield  {author} {\bibinfo {author} {\bibfnamefont {T.}~\bibnamefont
  {Duan}}\ and\ \bibinfo {author} {\bibfnamefont {V.}~\bibnamefont
  {Dinavahi}},\ }\bibfield  {title} {\bibinfo {title} {Dataplane-based fast
  failover in sdn-enabled wide area measurement system of smart grid},\ }\href
  {https://doi.org/10.1109/TII.2022.3216568} {\bibfield  {journal} {\bibinfo
  {journal} {IEEE Transactions on Industrial Informatics}\ }\textbf {\bibinfo
  {volume} {19}},\ \bibinfo {pages} {8148} (\bibinfo {year}
  {2023})}\BibitemShut {NoStop}%
\bibitem [{\citenamefont {Goodney}\ \emph {et~al.}(2013)\citenamefont
  {Goodney}, \citenamefont {Kumar}, \citenamefont {Ravi},\ and\ \citenamefont
  {Cho}}]{SDN-for-PMUs}%
  \BibitemOpen
  \bibfield  {author} {\bibinfo {author} {\bibfnamefont {A.}~\bibnamefont
  {Goodney}}, \bibinfo {author} {\bibfnamefont {S.}~\bibnamefont {Kumar}},
  \bibinfo {author} {\bibfnamefont {A.}~\bibnamefont {Ravi}},\ and\ \bibinfo
  {author} {\bibfnamefont {Y.~H.}\ \bibnamefont {Cho}},\ }\bibfield  {title}
  {\bibinfo {title} {Efficient pmu networking with software defined networks},\
  }in\ \href {https://doi.org/10.1109/SmartGridComm.2013.6687987} {\emph
  {\bibinfo {booktitle} {2013 IEEE International Conference on Smart Grid
  Communications (SmartGridComm)}}}\ (\bibinfo {year} {2013})\ pp.\ \bibinfo
  {pages} {378--383}\BibitemShut {NoStop}%
\bibitem [{\citenamefont {Kunze}\ \emph {et~al.}(2021)\citenamefont {Kunze},
  \citenamefont {Glebke}, \citenamefont {Scheiper}, \citenamefont
  {Bodenbenner}, \citenamefont {Schmitt},\ and\ \citenamefont
  {Wehrle}}]{COMSYS-In-Network-Computing}%
  \BibitemOpen
  \bibfield  {author} {\bibinfo {author} {\bibfnamefont {I.}~\bibnamefont
  {Kunze}}, \bibinfo {author} {\bibfnamefont {R.}~\bibnamefont {Glebke}},
  \bibinfo {author} {\bibfnamefont {J.}~\bibnamefont {Scheiper}}, \bibinfo
  {author} {\bibfnamefont {M.}~\bibnamefont {Bodenbenner}}, \bibinfo {author}
  {\bibfnamefont {R.~H.}\ \bibnamefont {Schmitt}},\ and\ \bibinfo {author}
  {\bibfnamefont {K.}~\bibnamefont {Wehrle}},\ }\bibfield  {title} {\bibinfo
  {title} {Investigating the applicability of in-network computing to
  industrial scenarios},\ }in\ \href
  {https://doi.org/10.1109/ICPS49255.2021.9468247} {\emph {\bibinfo {booktitle}
  {2021 4th IEEE International Conference on Industrial Cyber-Physical Systems
  (ICPS)}}}\ (\bibinfo {year} {2021})\ pp.\ \bibinfo {pages}
  {334--340}\BibitemShut {NoStop}%
\bibitem [{\citenamefont {Azuma}\ \emph {et~al.}(2023)\citenamefont {Azuma},
  \citenamefont {Economou}, \citenamefont {Elkouss}, \citenamefont {Hilaire},
  \citenamefont {Jiang}, \citenamefont {Lo},\ and\ \citenamefont
  {Tzitrin}}]{Quantum-Repeaters}%
  \BibitemOpen
  \bibfield  {author} {\bibinfo {author} {\bibfnamefont {K.}~\bibnamefont
  {Azuma}}, \bibinfo {author} {\bibfnamefont {S.~E.}\ \bibnamefont {Economou}},
  \bibinfo {author} {\bibfnamefont {D.}~\bibnamefont {Elkouss}}, \bibinfo
  {author} {\bibfnamefont {P.}~\bibnamefont {Hilaire}}, \bibinfo {author}
  {\bibfnamefont {L.}~\bibnamefont {Jiang}}, \bibinfo {author} {\bibfnamefont
  {H.-K.}\ \bibnamefont {Lo}},\ and\ \bibinfo {author} {\bibfnamefont
  {I.}~\bibnamefont {Tzitrin}},\ }\bibfield  {title} {\bibinfo {title} {Quantum
  repeaters: From quantum networks to the quantum internet},\ }\href
  {https://doi.org/10.1103/RevModPhys.95.045006} {\bibfield  {journal}
  {\bibinfo  {journal} {Rev. Mod. Phys.}\ }\textbf {\bibinfo {volume} {95}},\
  \bibinfo {pages} {045006} (\bibinfo {year} {2023})}\BibitemShut {NoStop}%
\bibitem [{\citenamefont {{Linux Foundation / SOGNO Members \&
  Contributors}}(2025)}]{SOGNO}%
  \BibitemOpen
  \bibfield  {author} {\bibinfo {author} {\bibnamefont {{Linux Foundation /
  SOGNO Members \& Contributors}}},\ }\href
  {https://lfenergy.org/projects/sogno/} {\bibinfo {title} {{SOGNO - LF
  Energy}}} (\bibinfo {year} {2025})\BibitemShut {NoStop}%
\bibitem [{\citenamefont {{50Hertz}}(2025)}]{MCCS}%
  \BibitemOpen
  \bibfield  {author} {\bibinfo {author} {\bibnamefont {{50Hertz}}},\ }\href
  {https://www.mccs.com/} {\bibinfo {title} {{MCCS}}} (\bibinfo {year}
  {2025})\BibitemShut {NoStop}%
\bibitem [{\citenamefont {Malevannaya}\ \emph {et~al.}(2025)\citenamefont
  {Malevannaya}, \citenamefont {Polozov}, \citenamefont {Andriyash},\ and\
  \citenamefont {Rodionov}}]{ESQMalevannaya}%
  \BibitemOpen
  \bibfield  {author} {\bibinfo {author} {\bibfnamefont {E.~I.}\ \bibnamefont
  {Malevannaya}}, \bibinfo {author} {\bibfnamefont {I.~A.}\ \bibnamefont
  {Polozov}, \bibfnamefont {Vikto...ov}}, \bibinfo {author} {\bibfnamefont
  {A.~V.}\ \bibnamefont {Andriyash}},\ and\ \bibinfo {author} {\bibfnamefont
  {I.~A.}\ \bibnamefont {Rodionov}},\ }\bibfield  {title} {\bibinfo {title} {An
  engineering guide to superconducting quantum circuit shielding},\ }\href
  {https://doi.org/10.1063/5.0250262} {\bibfield  {journal} {\bibinfo
  {journal} {Applied Physics Reviews}\ }\textbf {\bibinfo {volume} {12}},\
  \bibinfo {pages} {031334} (\bibinfo {year} {2025})},\ \Eprint
  {https://arxiv.org/abs/https://pubs.aip.org/aip/apr/article-pdf/doi/10.1063/5.0250262/20708867/031334\_1\_5.0250262.pdf}
  {https://pubs.aip.org/aip/apr/article-pdf/doi/10.1063/5.0250262/20708867/031334\_1\_5.0250262.pdf}
  \BibitemShut {NoStop}%
\bibitem [{\citenamefont {Le}\ \emph {et~al.}(2025)\citenamefont {Le},
  \citenamefont {Mayer}, \citenamefont {Magaletti}, \citenamefont {Schmidt},
  \citenamefont {Roch},\ and\ \citenamefont {Debuisschert}}]{Le2025}%
  \BibitemOpen
  \bibfield  {author} {\bibinfo {author} {\bibfnamefont {X.~P.}\ \bibnamefont
  {Le}}, \bibinfo {author} {\bibfnamefont {L.}~\bibnamefont {Mayer}}, \bibinfo
  {author} {\bibfnamefont {S.}~\bibnamefont {Magaletti}}, \bibinfo {author}
  {\bibfnamefont {M.}~\bibnamefont {Schmidt}}, \bibinfo {author} {\bibfnamefont
  {J.-F.}\ \bibnamefont {Roch}},\ and\ \bibinfo {author} {\bibfnamefont
  {T.}~\bibnamefont {Debuisschert}},\ }\bibfield  {title} {\bibinfo {title}
  {Field-effect detected magnetic resonance of nitrogen-vacancy centers in
  diamond based on all-carbon schottky contacts},\ }\href
  {https://doi.org/10.1038/s44172-025-00541-z} {\bibfield  {journal} {\bibinfo
  {journal} {Communications Engineering}\ }\textbf {\bibinfo {volume} {4}},\
  \bibinfo {pages} {209} (\bibinfo {year} {2025})}\BibitemShut {NoStop}%
\bibitem [{\citenamefont {Crawford}\ \emph {et~al.}(2021)\citenamefont
  {Crawford}, \citenamefont {Shugayev}, \citenamefont {Paudel}, \citenamefont
  {Lu}, \citenamefont {Syamlal}, \citenamefont {Ohodnicki}, \citenamefont
  {Chorpening}, \citenamefont {Gentry},\ and\ \citenamefont
  {Duan}}]{Crawford2021QuantumReview}%
  \BibitemOpen
  \bibfield  {author} {\bibinfo {author} {\bibfnamefont {S.~E.}\ \bibnamefont
  {Crawford}}, \bibinfo {author} {\bibfnamefont {R.~A.}\ \bibnamefont
  {Shugayev}}, \bibinfo {author} {\bibfnamefont {H.~P.}\ \bibnamefont
  {Paudel}}, \bibinfo {author} {\bibfnamefont {P.}~\bibnamefont {Lu}}, \bibinfo
  {author} {\bibfnamefont {M.}~\bibnamefont {Syamlal}}, \bibinfo {author}
  {\bibfnamefont {P.~R.}\ \bibnamefont {Ohodnicki}}, \bibinfo {author}
  {\bibfnamefont {B.}~\bibnamefont {Chorpening}}, \bibinfo {author}
  {\bibfnamefont {R.}~\bibnamefont {Gentry}},\ and\ \bibinfo {author}
  {\bibfnamefont {Y.}~\bibnamefont {Duan}},\ }\bibfield  {title} {\bibinfo
  {title} {Quantum sensing for energy applications: Review and perspective},\
  }\href {https://doi.org/https://doi.org/10.1002/qute.202100049} {\bibfield
  {journal} {\bibinfo  {journal} {Advanced Quantum Technologies}\ }\textbf
  {\bibinfo {volume} {4}},\ \bibinfo {pages} {2100049} (\bibinfo {year}
  {2021})},\ \Eprint
  {https://arxiv.org/abs/https://advanced.onlinelibrary.wiley.com/doi/pdf/10.1002/qute.202100049}
  {https://advanced.onlinelibrary.wiley.com/doi/pdf/10.1002/qute.202100049}
  \BibitemShut {NoStop}%
\bibitem [{\citenamefont {Barenco}\ \emph {et~al.}(1995)\citenamefont
  {Barenco}, \citenamefont {Bennett}, \citenamefont {Cleve}, \citenamefont
  {DiVincenzo}, \citenamefont {Margolus}, \citenamefont {Shor}, \citenamefont
  {Sleator}, \citenamefont {Smolin},\ and\ \citenamefont
  {Weinfurter}}]{barenco_elementary_1995}%
  \BibitemOpen
  \bibfield  {author} {\bibinfo {author} {\bibfnamefont {A.}~\bibnamefont
  {Barenco}}, \bibinfo {author} {\bibfnamefont {C.~H.}\ \bibnamefont
  {Bennett}}, \bibinfo {author} {\bibfnamefont {R.}~\bibnamefont {Cleve}},
  \bibinfo {author} {\bibfnamefont {D.~P.}\ \bibnamefont {DiVincenzo}},
  \bibinfo {author} {\bibfnamefont {N.}~\bibnamefont {Margolus}}, \bibinfo
  {author} {\bibfnamefont {P.}~\bibnamefont {Shor}}, \bibinfo {author}
  {\bibfnamefont {T.}~\bibnamefont {Sleator}}, \bibinfo {author} {\bibfnamefont
  {J.~A.}\ \bibnamefont {Smolin}},\ and\ \bibinfo {author} {\bibfnamefont
  {H.}~\bibnamefont {Weinfurter}},\ }\bibfield  {title} {\bibinfo {title}
  {Elementary gates for quantum computation},\ }\href
  {https://doi.org/10.1103/PhysRevA.52.3457} {\bibfield  {journal} {\bibinfo
  {journal} {Phys. Rev. A}\ }\textbf {\bibinfo {volume} {52}},\ \bibinfo
  {pages} {3457} (\bibinfo {year} {1995})},\ \bibinfo {note} {number:
  5}\BibitemShut {NoStop}%
\bibitem [{\citenamefont {Deutsch}\ and\ \citenamefont
  {Jozsa}(1992)}]{deutsch_rapid_1992}%
  \BibitemOpen
  \bibfield  {author} {\bibinfo {author} {\bibfnamefont {D.}~\bibnamefont
  {Deutsch}}\ and\ \bibinfo {author} {\bibfnamefont {R.}~\bibnamefont
  {Jozsa}},\ }\bibfield  {title} {\bibinfo {title} {Rapid solution of problems
  by quantum computation},\ }\href {https://doi.org/10.1098/rspa.1992.0167}
  {\bibfield  {journal} {\bibinfo  {journal} {Proc. R. Soc. Lond. A}\ }\textbf
  {\bibinfo {volume} {439}},\ \bibinfo {pages} {553} (\bibinfo {year}
  {1992})},\ \bibinfo {note} {publisher: Royal Society}\BibitemShut {NoStop}%
\bibitem [{\citenamefont {Knill}\ and\ \citenamefont {Laflamme}(1997)}]{QECC}%
  \BibitemOpen
  \bibfield  {author} {\bibinfo {author} {\bibfnamefont {E.}~\bibnamefont
  {Knill}}\ and\ \bibinfo {author} {\bibfnamefont {R.}~\bibnamefont
  {Laflamme}},\ }\bibfield  {title} {\bibinfo {title} {{Theory of quantum
  error-correcting codes}},\ }\href {https://doi.org/10.1103/PhysRevA.55.900}
  {\bibfield  {journal} {\bibinfo  {journal} {Phys. Rev. A}\ }\textbf {\bibinfo
  {volume} {55}},\ \bibinfo {pages} {900} (\bibinfo {year} {1997})}\BibitemShut
  {NoStop}%
\bibitem [{\citenamefont {Dür}\ and\ \citenamefont
  {Briegel}(2007)}]{Entanglement-Purification}%
  \BibitemOpen
  \bibfield  {author} {\bibinfo {author} {\bibfnamefont {W.}~\bibnamefont
  {Dür}}\ and\ \bibinfo {author} {\bibfnamefont {H.~J.}\ \bibnamefont
  {Briegel}},\ }\bibfield  {title} {\bibinfo {title} {Entanglement purification
  and quantum error correction},\ }\href
  {https://doi.org/10.1088/0034-4885/70/8/R03} {\bibfield  {journal} {\bibinfo
  {journal} {Reports on Progress in Physics}\ }\textbf {\bibinfo {volume}
  {70}},\ \bibinfo {pages} {1381} (\bibinfo {year} {2007})}\BibitemShut
  {NoStop}%
\end{thebibliography}
\end{document}